\documentclass[a4paper,10pt]{article}

\usepackage{jheppub} % for details on the use of the package, please
\usepackage[T1]{fontenc} % if needed%%%%%%%%%%%%%%%%%%%%%%%%%%%

\usepackage{bm}
\usepackage{latexsym}
\usepackage{dcolumn}
\usepackage{amsmath,amsfonts,amssymb, amsthm}
\usepackage{graphicx,epsfig}
\usepackage{environ}
\usepackage{mathtools}
\usepackage{braket}
\usepackage{float}
\usepackage{mathrsfs}
\usepackage{esint}
\usepackage{amsmath}
\usepackage{tensor}
\usepackage{fancyhdr}
\usepackage{subcaption}
\usepackage{hyperref}
\usepackage{graphicx,epstopdf}
\usepackage{capt-of}
\usepackage{tikz}

\hypersetup{
	colorlinks   = true, %Colours links instead of ugly boxes
	urlcolor     = blue, %Colour for external hyperlinks
	linkcolor    = blue, %Colour of internal links
	citecolor   = red %Colour of citations
}

\def\b{\beta}

\def\d{\delta}

\def\p{\partial}

\def\k{\kappa}

\def\be{\begin{equation}}
\def\ee{\end{equation}}
\def\ba{\begin{align}}
\def\ea{\end{align}}

\def\G{\Gamma}

\def\Y{\mathcal{Y}}

\newcommand{\levicivita}{}% initialize
\def\levicivita#1#{\tensor#1{\epsilon}}
\def\tbf{\textbf}
\def\mcl{\mathcal}
\def\sptm{(\mathcal{M}, \textbf{g}, \hat{\bm{\nabla}})}
\def\sptml{(\mathcal{M}, \textbf{g}, {\bm{\nabla}})}
\def\hech{\mathcal{H}}
\def\udl{\underline}
\def\udll{\underline{\bm{l}}}

\def\hcvd{\hat{\nabla}}
\def\cvd{\nabla}
\def\ct{\mathcal{T}}
\def\thht{\hat{\Theta}}
\def\ti{\mathbb{T}}
\def\tpm{\ct_P(\mathcal{M})}

\def\tpss{\ct^{*}_{P}({S_t})}
\def\hw{\hat{\omega}}
\def\hhw{\hat{\Omega}}
\def\dxa{\bm{d} x^a}
\def\dxb{\bm{d} x^b}
\def\dl{\bm{d}\underline{\bm{l}}}
\def\hchi{\hat{\chi}}
\def\lil{\pounds_{\bm{l}}}
\def\lik{\pounds_{\bm{k}}}
\def\thdl{\overset{(d)}{\hat{\theta_{\bm{l}}}}}
\def\thdk{\overset{(d)}{\hat{\theta_{\bm{k}}}}}
\def\shdl{{^{(\bm{l},d)}{\sigma_{ab}}}}

\def\shdk{{^{(\bm{k},d)}{\sigma_{ab}}}}

\def\thel{\overset{(e)}{\hat{\theta_{\bm{l}}}}}
\def\shel{{^{(\bm{l},e)}{\sigma_{ab}}}}

\def\whel{{^{(\bm{l},e)}{\omega_{ab}}}}

\def\xih{\hat{\Xi}}
\def\fr{\frac{1}{2}}
\def\hb{\hat{\mathcal{B}}}
\def\hbr{\tilde{\mathcal{B}}}
\def\thbl{\overset{(B)}{\hat{\theta_{\bm{l}}}}}
\def\shbl{{^{(\bm{l},B)}{\hat{\sigma_{ab}}}}}
\def\shblr{{^{(\bm{l},B)}{{\sigma_{ab}}}}}
\def\shblu{{^{(\bm{l},B)}{\hat{\sigma^{ab}}}}}
\def\shblru{{^{(\bm{l},B)}{{\sigma^{ab}}}}}
\def\whbl{{^{(\bm{l},B)}{\hat{\omega_{ab}}}}}

\def\whblu{{^{(\bm{l},B)}{\hat{\omega^{ab}}}}}

\def\hr{\hat{R}}
\def\sptq{(S_t, \bm{q}, \bm{\hat{\mathcal{D}}})}
\def\hsvd{\hat{\mathcal{D}}}
\def\twr{{^{(2)}\hat{R}}}
\def\twt{{^{(2)}T}}
\def\bmx{\bm{X}}
\def\bmy{\bm{Y}}
\def\bmt{^{(2)}\bm{T}}
\def\hr{\hat{R}}
\def\hq{\hat{\mathcal{Q}}}
\def\hp{\hat{\mathscr{P}}}
\def\mt{\overset{(m)}{T}}
\def\lk{\lambda_{(k)}}
\def\ll{\lambda_{(l)}}
\def\dlk{\delta \lambda_{(k)}}
\def\ints{\int_{S_t}}
\def\intsd{\int_{S_t} d^2 x \sqrt{q}}
\def\dellk{\delta_{\lambda_{(k)}}}
\def\snu{s_{\text{null}}}
\def\stor{s_{\text{tor}}}
\def\twor{{{^{(2)}} \hr}}
\def\bthk{\Big(\thdk - q^{ij}T_{ihj}k^h\Big)}
\def\mt{T^{(m)}}
\def\hg{\hat{G}}
\def\hgm{\hat{\Gamma}}
\def\aec{\mathcal{A}_{\text{EC}}}
\def\aem{\mathcal{A}_{\text{m}}}
\title{Kinematics and dynamics of null hypersurfaces in the Einstein-Cartan spacetime and related thermodynamic interpretation}
\author[a]{Sumit Dey,}
 \author[a]{and Bibhas Ranjan Majhi}
 \affiliation[a]{Department of Physics, Indian Institute of Technology Guwahati, Guwahati 781039, Assam, India.}
 \emailAdd{dey18@iitg.ac.in}
\emailAdd{bibhas.majhi@iitg.ac.in}\date{\today}
\abstract{A general geometric construction of a generic null hypersurface in presence of torsion in the spacetime (Riemann-Cartan background), generated by a null vector $l^a$, is being developed here. We then explicitly define and structure various corresponding kinematical quantities. The dynamics of the null surface, particularly given by $\hat{G}_{ab}k^al^b$, is also discussed. The later one is constructed under the {\it geodesic constraint} condition. This yields a relation among the rate of change of expansion scalar corresponding to auxiliary null vector $k^a$ and various kinematical entities on the null surface. Using this relation we show that the Einstein-Cartan-Kibble-Sciama equation (which provides the dynamics of the metric and the torsion tensor) on this null hypersurface acquires a thermodynamic interpretation. The thermodynamic entities like temperature, entropy density, energy and pressure are properly identified. In the whole analysis we adopt the geometrical field interpretation of torsion and all discussions are done in a covariant manner. 
}

\begin{document}
\maketitle
\flushbottom%%%%%%%%%%%%%%%%%%%%%%%%%%%%%%%%%%%%%%%%%%%%%%%%%%%%%%%%%%%%%%%%%%%%%%%%%%%%%%%
%\tableofcontents
\noindent

%%%%%%%%%%%%%%%%%%%%%%%%%%%%%%%%%%%%%%%%%%%%%%%%%%%%%%%%%%%%%%%%%%%%%%%%%%%%%%%%%%%%%%%%%%%%%%%%%%%%%%%%%%%%%%%%%%%%%%
\section{Introduction and motivation}
In the context Einstein gravity, the fascinating analogy between the well known laws of black-hole mechanics and classical thermodynamics came about through the works of Bekenstein, Hawking and others \cite{Bekenstein:1973ur, Bardeen:1973gs, Hawking:1971vc, Hawking:1976de, Davies:1978mf} (for a review see \cite{Wald:1999vt, Carlip:2014pma, Wall:2018ydq}). This connection allowed the conception of entropy and temperature to be assigned to stationary black-hole event horizons. Such possible connections to thermodynamics have also been explored on stationary event horizons in modified theories of gravity \cite{Kolekar:2011bb, Kolekar:2012tq, Mureika:2015sda, Soroushfar:2016nbu, Aros:2019quj, Sarkar:2019xfd}. Generalizations to first and second laws of black hole mechanics in the context of dynamical horizons have also been explored \cite{Hayward:1997jp, Hayward:1998ee, Ashtekar:2003hk, Majhi:2014hpa, Majhi:2014lka, Majhi:2015tpa, Bhattacharya:2016kbm}. Moving away from global event horizons, black hole thermodynamics for \textit{quasi-local} horizons have also been studied \cite{Nielsen:2008cr, Cai:2008mh}. 

A clear indication of the underlying connection between gravitational dynamics and thermodynamics came about from the work of Jacobson \cite{Jacobson:1995ab}. He derived the Einstein field equations from the underlying Clausius identity $\d S = \frac{\d Q}{T}$ as applied to local Rindler horizons in equilibrium constructed at any point in the spacetime. Here, $T$ and $\d Q$ are interpreted as the Unruh temperature and the energy-momentum flux crossing the Rindler horizon as observed by the accelerated observer. The entropy change $\d S$ is proportional to the change of cross-sectional area of the null curves generating the Rindler horizon. Moreover, it has been shown by Padmanabhan {\it et. al.} \cite{Padmanabhan:2002sha, Padmanabhan:2003gd, Padmanabhan:2009vy, Kothawala:2007em, Paranjape:2006ca} that gravitational field equations near static as well as stationary event horizons in general relativity, Lanczos-Lovelock and other modified theories of gravity assume a thermodynamic identity $T \d_{\lambda} S = \d_{\lambda} E + P \d_{\lambda}V$. For a more complete set of references towards this thermodynamic identity, see \cite{Chakraborty:2015aja}. Such structure of the thermodynamic identity analogous to the first law of thermodynamics (where the symbols have their usual meanings) is defined for {\it virtual displacement} $\delta \lambda$ along an affine parameter $\lambda$ off the event horizon. 

It might appear that such connections between field equations and thermodynamics only applies to special spacetime solutions having event horizons. However, it has been shown that any generic null hypersurface constructed at any point in spacetime acts as a local Rindler horizon for a specific accelerated observer \cite{Padmanabhan:2009vy, Jacobson:1995ab}. Unruh effect \cite{unruh1976notes,unruh1984happens} provides a clear concept of temperature assigned to the Rindler horizon by this class of observers. This renders a thermodynamical structure to this local null surface \cite{Parikh:2018anm, Adami:2021kvx} which may not necessarily be a black hole horizon. Particularly the same \textit{observer-dependent} program (following Jacobson) of deriving the field equations via the Clausius identity or (following Padmanabhan)  interpreting the gravitational field equations near the null surface in a thermodynamic form can be followed. It has also been shown \cite{Padmanabhan:2007en, Padmanabhan:2007xy} that the extremization of the sum of gravitational heat density and matter density of a generic null surface in the spacetime produces the given gravitational field equations.
%These connections have however since been extended to any generic null hypersurface $\hech$. Null hypersurfaces act as event horizons for a given class of observers. Hence it is possible to attribute a heat density $T s$ to a given null surface.

%\textcolor{blue}{These vivid connections between gravitational dynamics and thermodynamics established in the context of generic null hypersurfaces form the motivation for \textit{``emergent gravity''} paradigm}. In a paradigm shift, the `emergent gravity" program considers gravitational dynamics to be not fundamental. Rather it considers gravity to be \textit{emergent} from fundamental degrees of freedom associated with the gravitational field \cite{Padmanabhan:2009kr, Padmanabhan:2010xh, Padmanabhan:2013nxa, Chakraborty:2014rga, Parattu:2013gwa, Chakraborty:2014joa}. That is, gravity and its dynamics emerges \textcolor{blue}{much like thermodynamics of matter as an effective field theory arises from the statistical mechanics of its constituent atoms.} 

Infact, it has been shown in the literature that three particular projections of the Einstein tensor $G_{ab}$ on a generic null surface $\hech$ upon use of the gravitational field equations lead to fluid-dynamical and thermodynamic interpretations. The relevant projections are $G_{ab}l^a q^b_{~c}$, $G_{ab}k^a l^b$ and $G_{ab}l^a l^b$. Here $l^a$ and $k^a$ are the null generator and the auxiliary null vector field respectively of $\hech$, while $q^a_{~b}$ is the induced metric on a transverse spacelike cross-section $S_t$ of $\hech$. The specific observations are as follows :
\begin{itemize}
	\item It was shown by Damour \cite{Damour:1979wya, damour1982surface} that the projection component $G_{ab}l^a q^b_{~c}$ on a null surface in Einstein gravity leads to an equation which is similar to Navier-Stokes (NS) equation, known as the Damour-Navier-Stokes (DNS) equation. The DNS equation when viewed from local inertial frames leads to the non-relativistic NS equation \cite{Padmanabhan:2010rp}. This connection between gravitational dynamics and fluid dynamical NS equation was also explored through the extremization of entropy functional defined on a generic null hypersurface \cite{Kolekar:2011gw}.
	\item{As mentioned above, Jacobson \cite{Jacobson:1995ab} used the null Raychaudhuri equation (NRE) \cite{Poisson:2009pwt} to a system of non-expanding congruence of null curves to derive the Einstein field equations from the Clausis identity $\d S = \frac{\d Q}{T}$ as applied to a local causal horizon \textit{in equilibrium}. The NRE relates $G_{ab}l^a l^b$ with the evolution dynamics of the outgoing expansion scalar of the null generators. Infact, this procedure can be reverse engineered. It has been shown \cite{Mohd:2013jva, Parattu:2017cjd} that the field equations for any diffeomorphism invariant theory of gravity via the component $G_{ab}l^a l^b$ on a non-expanding general null surface leads to the Clausius identity. Jacobson's formalism was later extended to local causal horizons in the non-equilibrium case \cite{Eling:2006aw, Chirco:2009dc, Dey:2017fld} and also for modified theories of gravity \cite{Chirco:2009dc, Dey:2017fld}.}
	\item{Padmanabhan and his collaborators \cite{Chakraborty:2015aja, Padmanabhan:2002sha, Kothawala:2007em} have shown that the projection component $G_{ab}k^a l^b$ of the Einstein field equation on the generic null surface $\hech$ leads to a thermodynamic interpretation that is structurally similar to the first law of thermodynamics. The formalism was later extended to Lanczos-Lovelock theories of gravity \cite{Paranjape:2006ca, Kothawala:2009kc, Chakraborty:2015wma}. However such interpretational analogy had been done so by adapting Gaussian null coordinate system on the null surface {\footnote{Recently justification on assigning temperature on the generic null surface is addressed in \cite{Dalui:2021sme}.}}. As a result, the thermodynamic parameters had been coordinate dependent. A completely covariant approach to the thermodynamic structure provided by $G_{ab}k^a l^b$ on the null surface was given in \cite{Dey:2020tkj} and was also applied in the case of Scalar-Tensor theory of gravity \cite{Dey:2021rke}. It was pointed out in \cite{Chakraborty:2015aja} that $G_{ab}k^a l^b$ is more natural a projection component than $G_{ab}l^a l^b$ for the thermodynamic interpretation. This has to do with the fact that $G_{ab}k^a l^b$ is the projection component of the vector field $-G^a_{~b}l^b$ along the null generators $l^a$ as opposed to $G_{ab}l^a l^b$ which represents the projection component along the auxiliary null field $k^a$. Hence $G_{ab}k^a l^b$ is a quantity intrinsic to $\hech$ as compared to $G_{ab}l^a l^b$.}
\end{itemize}

These vivid connections of gravitational dynamics with thermodynamics and fluid equation established in the context of generic null hypersurfaces form the motivation for thinking about gravity as an \textit{``emergent phenomenon''} \footnote{Emergent nature of gravity was initially introduced in $1967$ by Sakharov \cite{sakharov1967vacuum}.}.
 In a paradigm shift, the ``emergent gravity'' program considers gravitational dynamics to be not fundamental. Rather it considers gravity to be \textit{emergent} from fundamental degrees of freedom associated with the gravitational field \cite{Padmanabhan:2009kr, Padmanabhan:2010xh, Padmanabhan:2013nxa, Chakraborty:2014rga, Parattu:2013gwa, Chakraborty:2014joa}. That is, gravity and its dynamics emerge much like thermodynamics of matter arises as an effective theory from the statistical mechanics of its constituent atoms.
 
Under this point of view, if gravity is indeed emergent (as seen especially for Einstein gravity), then the connections between gravitational dynamics and thermodynamics should indeed transcend to other theories of gravity. Here, in our case we take the example of Einstein-Cartan (EC) theory \cite{ASENS_1923_3_40__325_0}. The EC theory is built in the geometrical backdrop of the Riemann-Cartan (RC) spacetime. The EC theory is a natural extension of Einsten gravity obtained by including the intrinsic spin of the particle(s) in the geometrization of spacetime \cite{Hehl:1976kj}. The presence of intrinsic spin causes non-zero torsion in the spacetime geometry and the relevant gravitational field equations are the Einstein-Cartan-Kibble-Sciama (ECKS) equations \cite{kibble1961lorentz, sciama1962, sciama1964physical, sciama1964erratum, hehl1971does} (for textbook expositions and reviews see \cite{de1986introduction, de1994spin, kleinert1989gauge, Poplawski:2009fb, Shapiro:2001rz, Obukhov:2006gea}). The presence of spin allows for a non-zero spin angular momentum tensor in addition to the energy-momentum tensor. In the macroscopic classical domain, the spin degrees of freedom cancel out due to their dipole nature and hence dynamics of macroscopic bodies are characterized by the energy-momentum tensor alone. However, in the microscopic regime, one cannot ignore the spin angular momentum which actually  ``sources'' torsion as being a geometric field in the spacetime in addition to the metric tensor.

In this paper, we aim to address exclusively the question whether the projection component $\hg_{ab}k^a l^b$ on $\hech$ in the EC theory can be provided a thermodynamic interpretation in a completely covariant way. Here, $\hg_{ab}$ is the analogue (not symmteric) of the Einstein tensor in the RC spacetime. Let us pause to mention that the component $\hg_{ab}l^a l^b$ is related to the NRE determining the dynamical evolution of the outgoing expansion scalar. The evolution equations, corresponding to the expansion, shear and vorticity, for congruences of both timelike as well as null curves in spacetimes with torsion (under different assumptions on the nature of the torsion) have been provided in \cite{PhysRevD.104.084073, PhysRevD.96.024021, PhysRevD.58.044021, Dey:2017fld}. The thermodynamic interpretation for $\hg_{ab}l^a l^b$ has been provided in \cite{Dey:2017fld} and hence will not be pursued here. In absence of torsion, it can be shown \cite{Dey:2020tkj} that $G_{ab}k^a l^b$ is related to the dynamics of the ingoing expansion scalar as opposed to the outgoing one in spacetimes without torsion. This can be achieved by taking the trace of the evolution equation of the transversal deformation rate tensor \cite{Gourgoulhon:2005ng}.  In order to arrive at such a covariant evolution equation, we would require to foliate the spacetime in the neighbourhood of $\hech$ by a family of null hypersurfaces. We would then foliate this null family by a stack of spacelike surfaces in the spirit of $3+1$ decomposition. This allows us to unambiguously construct the relevant kinematics and dynamics of the general null surface \cite{Gourgoulhon:2005ng}. Upon using the Einstein field equations and combining the evolution equation of the ingoing expansion scalar with the process of a virtual displacement leads to a covariant structurization of the thermodynamics attested to $G_{ab}k^a l^b$. 

Our aim, hence in this paper is two fold. Firstly, under the framework of the $3+1$ null foliation of $\hech$, we construct all the relevant kinematics and dynamics of $\hech$ in the RC spacetime \footnote{As far we know this has not been dealt explicitly in the literature.}. However, especially while studying the dynamics, we will impose a particular constraint which we name as the \textit{geodesic constraint}. The geodesic constraint is the choice that restricts the null generators of $\hech$ in the RC spacetime to be simultaneously auto-parallel as well as geodesic curves. We are hence led to the evolution equation of the transversal deformation rate tensor and the evolution equation of the ingoing expansion scalar of $\hech$ in RC spacetime under the geodesic constraint. As usual, we will see that the dynamics of the ingoing expansion scalar is related to $\hg_{ab}k^a l^b$ in the RC spacetime. Secondly, with the help of the notion of virtual displacement \footnote{In this context the concept of virtual displacement was initially introduced in \cite{padmanabhan2002classical}.} applied to this evolution equation and the relevant field equations, we provide a covariant thermodynamic interpretation to $\hg_{ab}k^a l^b$. In doing so, we will be able to access how the thermodynamic parameters and their interpretation are affected by the inclusion of torsion under the geodesic constraint. 

The organization of the paper is as follows. In Sec. \ref{spacetimegeom}, we very briefly discuss about the geometric properties of the RC spacetime and the corresponding ECKS field equations. In Sec. \ref{nullhypgeom}, we discuss in detail about the structure of a generic null hypersurface $\hech$ in the RC spacetime. In Sec. \ref{kinematicsh} and Sec. \ref{dynamicsh}, we explore respectively the kinematics and dynamics of $\hech$ in such a spacetime. In Sec. \ref{section6}, we begin our in-depth study of the thermodynamic interpretation provided to $\hg_{ab}k^a l^b$ and also discuss some special cases. Finally, we conclude in Sec. \ref{conclusion}. At the end, in Appendix \ref{NRE3+1} we rederive for completeness, the NRE corresponding to the outgoing expansion scalar of the null generators in the framework of the $3+1$ null foliation. We also provide four other appendixes for calculational details.

Let us clarify our position on notations and dimensions. We work in $d=4$ spacetime dimensions and use the metric signature $(-,+,+,+)$. We employ geometrized unit system where we set $c$, $\hbar$ and $G$ to unity. The lowercase Latin alphabets $a,b, \cdot \cdot \cdot$ are for the bulk spacetime and run from $0$ to $3$. The spatial coordinate indices on the transverse two dimensional subspace $S_t$ of $\hech$ are denoted by uppercase Latin alphabets $A,B, \cdot \cdot \cdot$ and run from $2$ to $3$.
%%%%%%%%%%%%%%%%%%%%%%%%%%%%%%%%%%%%%%%%%%%%%%%%%%%%%%%%%%%%%%%%%%%%%%%%%%%%%%%%%%%%%%%%%%%%%%%%%%%%%%%%%%%%%%%%%%%%%%%%
\section{A brief review of spacetime with torsion and gravitational field equations}\label{spacetimegeom}
Our objective is to provide a thermodynamical interpretation to the ECKS field equations. We are hence interested in the spacetime with torsion in the background. Our ambient spacetime $(\mathcal{M},\tbf{g}, \hat{\bm{\nabla}})$ is provided with a metric compatible affine connection,
\begin{equation}
	\hat{\nabla}_a g_{bc} = 0 ~.
	\label{metricity}
\end{equation}
Such a spacetime is designated as the \textit{Riemann-Cartan spacetime}. 
\subsection{Geometrical properties of $\sptm$ :}
Let us now very briefly review the geometrical properties of such a spacetime. For details refer to \cite{Poplawski:2009fb}. Just to set the convention straight, we define the covariant derivative of a $(r,s)$ rank tensor $T^{a_1 \cdot \cdot \cdot a_r}_{~~b_1 \cdot \cdot \cdot b_s}$ to be,
\begin{align}
\hcvd_a T^{a_1 \cdot \cdot \cdot a_r}_{~~b_1 \cdot \cdot \cdot b_s} &\equiv \p_a T^{a_1 \cdot \cdot \cdot a_r}_{~~b_1 \cdot \cdot \cdot b_s} + \hgm^{a_1}_{~a i_1} T^{i_1 \cdot \cdot \cdot a_r}_{~~b_1 \cdot \cdot \cdot b_s} + \cdot \cdot \cdot + \hgm^{a_r}_{~a i_r} T^{a_1 \cdot \cdot \cdot i_r}_{~~b_1 \cdot \cdot \cdot b_s} - \hgm^{j_1}_{~a b_1} T^{a_1 \cdot \cdot \cdot a_r}_{~~j_1 \cdot \cdot \cdot b_s}\cdot \cdot \cdot \nonumber \\
&-\hgm^{j_s}_{~a b_s} T^{a_1 \cdot \cdot \cdot a_r}_{~~b_1 \cdot \cdot \cdot j_s}~,
\label{manip38}
\end{align}
where notice that the differentiating index $``a"$ sits at the first position in the subscript indices of affine connections. 
The torsion tensor is basically defined as,
\begin{align}
T^a_{~bc} \equiv \hgm^a_{~bc} - \hgm^a_{~cb}~,
\label{torsion}
\end{align}
implying that the torsion tensor is antisymmetric in the last two indices.
This general metric compatible affine connection is related to the symmetrical Levi-Civita connection $\G^a_{~bc}$ via the contorsion tensor $K^a_{~bc}$, 
\begin{align}
\hgm^a_{~bc} = \G^a_{bc} + K^a_{~bc}~.
\label{contorsion}
\end{align}
The contorsion tensor can be expressed in terms of the torsion tensor as,
\begin{align}
K^a_{~bc} = \fr \Big(T^a_{~bc} + T_{b~~c}^{~a} + T_{c~~b}^{~a}\Big) ~.
\label{contorsion1}
\end{align}
The above relation can be obtained by the familiar trick of setting $(-\hcvd_a g_{bc} + \hcvd_b g_{ca} + \hcvd_c g_{ab}) = 0$ via \eqref{metricity} and then using \eqref{manip38}.
It is quite easy to verify (according to the convention followed in \eqref{torsion}) that the contorsion tensor $K_{abc}$ is antisymmetric in the first and last indices.
Here, we use the metric tensor $g_{ab}$ to raise and lower all spacetime indices of a tensorial quantity. Let us now look at the trace of the torsion tensor defined via contracting the first and the third index. Following properties can then be easily deduced from the definition,
\begin{align}
T_b \equiv g^{ac}T_{abc} = T^a_{~ba} = -T^a_{~ab} ~; \nonumber \\
K^a_{~ab} = -T_b ; ~~~~ K^a_{~ba} = 0;~~~~ K_{b~~a}^{~a} = T_b ~.
\label{tracetorsion}
\end{align}
Another quantity of interest that comes into play is the modified torsion tensor $S^a_{~bc}$ defined as,
\begin{align}
S^a_{~bc}  \equiv T^a_{~bc} + \d^a_{~b}T_c - \d^a_{~c}T_b ~,\label{modifiedtorsion}
\end{align}
which like the torsion tensor is antisymmetric in the last two indices.

In general, the existence of torsion in the spacetime $\sptm$ can be characterized from the fact that the action of the commutator of the covariant derivatives $\bm{\hcvd}$ on any scalar field does not vanish,
\begin{align}
[\hcvd_a, \hcvd_b] \Phi = - T^d_{~ab} (\hcvd_d \Phi) ~.
\label{torsiondef}
\end{align}
The corresponding action on contravariant and covariant vectors are summarized below,
\begin{align}
&[\hcvd_a, \hcvd_b]A^i = \hr^i_{~kab} A^k - T^d_{~ab} (\hcvd_d A^i)~; \\
&[\hcvd_a, \hcvd_b] \omega_c = - \hr^d_{~cab} \omega_d - T^d_{~ab} (\hcvd_d \omega_c) ~.
\label{ricciidenx}
\end{align}
The Riemann tensor in the spacetime $\sptm$ follows the usual definition in terms of the affine connection $\hgm^a_{~bc}$ and as per our convention is,
\begin{align}
	\hr^a_{~bcd} \equiv \p_c \hgm^a_{~db} - \p_d \hgm^a_{~cb} + \hgm^a_{~ci} \hgm^i_{~db} - \hgm^a_{~di}\hgm^i_{~cb} ~.
\end{align}
 The following symmetries of the Riemann tensor are then quite evident,
\begin{align}
\hr_{abcd} = - \hr_{abdc} ~~~ \text{and}~~~ \hr_{abcd} = -\hr_{bacd} ~.
\label{Riemsym}
\end{align}
However, the usual symmetry under pairwise exchange of the indices does not follow through over here, as $\hr_{cdab} = \hr_{abcd} + \hq_{abcd}$ where,
\begin{align}
 \hq_{abcd} &= -\frac{3}{2}\Big(\hcvd_{[b} T_{|a| cd]} - \hcvd_{[a}T_{|b|cd]} - \hcvd_{[d}T_{|c|ab]} + \hcvd_{[c}T_{|d|ab]} \nonumber \\
&+T_{ae[b}T^e_{~cd]} - T_{be[a}T^{e}_{~cd]} - T_{ce[d}T^e_{~ab]} + T_{de[c}T^e_{~ab]}\Big) ~.
\label{riemann1}
\end{align}
In the above equation, $||$ indicates the enclosed index barred from antisymmetrization. Similarly, the usual first and the second Bianchi identities do not follow: 
\begin{align}
&\hr^d_{[cab]} = \hcvd_{[a} T^d_{~bc]} - T^f_{~[ab}  T^d_{~~c]f}~ ; \\
&\hcvd_{[a}\hr^f_{~|d|bc]} = -T^k_{~[ab} \hr^f_{~|dk|c]} ~.
\end{align} 
Here, we have used the convention that,
\begin{align}
	A_{[i_1 \cdot \cdot \cdot i_n]} &= \frac{1}{n !} \underset{\sigma}{\Sigma} (-1)^{\epsilon_{\sigma}} A_{i_{\sigma(1)} \cdot \cdot \cdot i_{\sigma(n)}} ~, \nonumber \\
	A_{(i_1 \cdot \cdot \cdot i_n)} &= \frac{1}{n !} \underset{\sigma}{\Sigma}  A_{i_{\sigma(1)} \cdot \cdot \cdot i_{\sigma(n)}} ~, 
\end{align}
where the summation is over all possible permutations $\{\sigma\}$ of the set $\{1,2, \cdot \cdot \cdot n\}$ and 
\begin{align}
	\epsilon_{\sigma} &= 0, ~~~~~\text{when $\sigma$ is an even permuation of $\{1,2, \cdot \cdot \cdot n\}$ } \nonumber \\
	\epsilon_{\sigma} &= 1, ~~~~~\text{when $\sigma$ is an odd permuation of $\{1,2, \cdot \cdot \cdot n\}$ } ~.
\end{align}
The Ricci tensor is no longer symmetric owing to the presence of torsion,
\begin{align}
\hr_{[ab]} = -\fr (\hcvd_i+ T_i) S^i_{~ab} ~.
\label{antisymmetryRicci}
\end{align}
In analogy with the Einstein tensor of the usual Riemannian geometry, we introduce the tensor $\hg^a_{~b} \equiv \hr^a_{~b} - \fr \d^a_{~b} \hr$ in the spacetime $\sptm$. As anticipated, due to the presence of torsion, the tensor $\hg^a_{~b}$ fails to be divergenceless,
\begin{align}
\hcvd_a \hg^a_{~b} = - T^k_{~ab} \hr^a_{~k} + \fr T^{kad} \hr_{adkb} ~.
\end{align}
The Lie derivative of the metric tensor of $\sptm$ along a given vector field $\bm{v}$ is,
\begin{align}
\pounds_{\bm{v}} g_{ab} = \hcvd_a v_b + \hcvd_b v_a + (T_{acb} + T_{bca})v^c ~.
\label{Lie}
\end{align}
%%%%%%%%%%%%%%%%%%%%%%%%%%%%%%%%%%%%%%%%%%%%%%%%%%%%%%%%%%%%%%%%%%%%%%%%%%%%%%%%%%%%%%%%%%%%%%%%%%%%%%%%%%%%%%%%%%%%%
\subsection{The gravitational field equations}
The gravitational action in the spacetime $\sptm$ will henceforth be referred to as the Einstein-Cartan action $\mathcal{A}_{\text{EC}}$. For details refer to \cite{Poplawski:2009fb, kleinert1989gauge, Pilling:2002dz}. In this theory both the metric and the torsion tensors are treated as independent dynamical variables. The total action for the theory is, 
\begin{align}
\mathcal{A}_{\text{tot}} = \aec + \aem = \frac{1}{16 \pi} \int_{\mathcal{V}} d^4 x \sqrt{-g} \hr + \aem ~,
\label{totalaction}
\end{align} 
where, $\aem$ is the corresponding matter action. Obviously, the above action is extremized by varying with respect to (w.r.t) both the metric and the torsion (preferably here the contorsion tensor) to yield the field equations. The Einstein-Cartan-Sciama-Kibble field equation (by varying w.r.t the metric) is \cite{Poplawski:2009fb, kleinert1989gauge, Pilling:2002dz},
\begin{align}
\hg_{ab}  +\fr (\hcvd_c + T_c) \Big(-S^c_{~ab} + S_{ab}^{~~c} + S_{ba}^{~~c}\Big) = 8 \pi \mt_{ab} ~,
\label{ECKSeom}
\end{align}
where $\mt_{ab}$ is the matter stress energy-momentum tensor. The field equation obtained by extremizing the total action w.r.t the contorsion tensor is, 
\begin{align}
S^a_{~bc} = 8 \pi \tau^a_{~bc} ~,
\label{spindensity}
\end{align}
where $\tau^a_{~bc}$ is the spin angular momentum tensor. Hence given a matter Lagrangian depending upon the metric, the matter field and its first derivative, the variation of the matter action is given as,
\begin{align}
\d \aem \equiv - \fr \int_{\mathcal{V}} d^4 x \sqrt{-g} \Big[\mt_{ab} \d g^{ab} + \tau^{b~c}_{~a}~ \d K^a_{~bc}\Big] ~.
\end{align} 
This indicates that the matter energy-momentum tensor $\mt_{ab}$ is symmetric whereas the spin angular momentum tensor $\tau^a_{~bc}$ is antisymmetric in the last two indices.
In anticipation of the result we are trying to achieve, let us state the following identity,
\begin{align}
\Big(\hcvd_a T_b - \hcvd_b T_a\Big) + \Big(\hcvd_i + T_i\Big)T^i_{~ab} = \Big(\hcvd_i + T_i\Big)S^i_{~ab} ~.
\label{manip33}
\end{align}
The above result can quite easily be verified by using the definition of the modified torsion tensor \eqref{modifiedtorsion}. Upon using \eqref{manip33} in \eqref{ECKSeom}, we obtain,
\begin{align}
\hg_{ab} + \Big(\hcvd_a T_b - \hcvd_b T_a\Big) + \Big(\hcvd_i + T_i\Big)T^i_{~ab} = 8 \pi \mt_{ab} + \fr (\hcvd_i + T_i) \Big[3 S^i_{~ab} + S_{a~b}^{~i} + S_{b~a}^{~i}\Big] ~.
\label{manip37}
\end{align}
Using \eqref{spindensity}, the last term on the right hand side (R.H.S) of the above equation can be expressed in terms of the spin angular momentum tensor. This form of the gravitational field equation \eqref{manip37} will be used later in our analysis.
%%%%%%%%%%%%%%%%%%%%%%%%%%%%%%%%%%%%%%%%%%%%%%%%%%%%%%%%%%%
%%%%%%%%%%%%%%%%%%%%%%%%%%%%%%%%%%%%%%%%%%%%%%%%%%%%%%%%%%%%%%%%%%%%%%%%%%%%%%%%%%%%%%%%%%%%%%%%%%%%%%%%%%%%%%%%%%%%%%
\section{Geometry of a generic null hypersurface in the Riemann-Cartan spacetime}\label{nullhypgeom} 
In this introduction to the structure of a generic null hypersurface in the RC spacetime $\sptm$ and its associated kinematics and dynamics (to be introduced in the next sections), we will stick to the notations and formalism introduced in \cite{Gourgoulhon:2005ng}, which provided the formulation for torsionless spacetime. Infact part of our objective is to see what modifications do the kinematics and dynamics of a general null surface incur provided our ambient spacetime has non-trivial torsion present in it. It is to this respect that we adopt the formalism introduced in \cite{Gourgoulhon:2005ng} and follow the notions.

A null hypersurface is basically a surface of codimension one, such that it can be described by specifying its corresponding induced metric and second fundamental form (characterizing the extrinsic curvature). These relevant quantities need to described properly for a null hypersurface in $\sptm$ i.e when there is non-trivial torsion in the spacetime.
We consider the existence of a generic null hypersurface $\mcl{H}$ in the spacetime $\sptm$, defined via the scalar field $u(x^a) = 0$. The surface $\hech$ is integrable into a hypersurface-orthogonal null surface such that its null normal $\underline{\bm{l}}$ is given by,
\begin{equation}
	l_a = - e^{\rho} \partial_a u = -e^{\rho} \hat{\nabla}_a u ~,
	\label{nullnormal}
\end{equation}
where $\rho$ is some smooth scalar field on $\hech$.
The co-efficient $e^{\rho}$ that relates the null normal $l_a$ with the gradient of the scalar field $\partial_a u$ is chosen to be negative such that $l^a$ is future pointing. This can be done by a suitable choice of the scalar field $u(x^a)$. Notice that the null normal cannot be provided a unique normalization on account of the fact that $\bm{l} \cdot \bm{l} = 0$. Next, we postulate the existence of an auxiliary null foliation in the neighbourhood of our null hypersurface $\mcl{H}$. In order to have well defined operations valid in the spacetime $\sptm$ like the covariant derivative $\hat{\bm{\nabla}}$, we cannot be only constrained on the single null hypersurface $u(x^a) = 0$. The support of the null vector field $\bm{l}$ needs to be extended from the null surface to at least in its vicinity. Following Carter \cite{carter1997extended}, this is facilitated by considering not just a single hypersurface $u = 0$, but by rather a family of null hypersurfaces $u(x^a) = c$, where $c$ is a constant. Hence the spacetime is foliated by a family of null hypersurfaces $\hech_u$ out which our particular chosen surface $\hech_{u=0} = \hech$ is just an element with $c=0$. This null foliation of $\sptm$ in the neighbourhood of $\hech$ extends the validity of the scalar field $\rho$ and hence $\bm{l}$ to not just on $\hech$, but to an open neighbourhood of $\sptm$ about $\hech$. This finally allows us to perform operations that are valid on the ambient spacetime rather than on the hypersurface. Even though such a foliation is non-unique, yet the geometrical quantities of interest that will be introduced and evaluated on our chosen hypersurface $\hech$ does not depend on the choice of foliation. 

Next, we proceed to a discussion of the Frobenius identity on our null surface. The fact that we have been able to write our null generator in the form of \eqref{nullnormal} means that the exterior derivative of the null normal to the hypersurface satisfies,
\begin{equation}
\bm{d}\underline{\bm{l}} = \bm{d} \rho \wedge \underline{\bm{l}} ~.
\label{frobeniusdual}
\end{equation}
This represents the Frobenius theorem in its dual formulation \cite{wald2010general}. The Frobenius identity quantifies the fact that the hyperplane, normal to $\underline{\bm{l}}$, is integrable into our null hypersurface $\hech$ and is hence hypersurface orthogonal. As a consequence of the dual formulation of the Frobenius identity we have,
\begin{equation}
\underline{\bm{l}} \wedge \bm{d} \underline{l} = \udl{\bm{l}} \wedge \bm{d} \rho \wedge \udl{\bm{l}} = -\bm{d} \rho \wedge \udll \wedge \udll  = 0~. 
\end{equation}
In the index notation, the above implies,
\begin{equation}
l_{[a}\partial_b l_{c]} = 0 ~.
\end{equation}
Converting to the spacetime covariant derivatives, the above formula translates to,
\begin{equation}
 \omega_{abc} \equiv l_{[a} \hat{\nabla}_b l_{c]} = -l_a \Big(T^d_{~~bc}~l_d \Big) - l_b \Big(T^d_{~~ca}~l_d\Big) - l_c \Big(T^d_{~~ab}l_d\Big) ~.
\end{equation}
Thus we see that due to the presence of torsion in the spacetime, hypersurface orthogonality does not imply a zero twist $\omega_{abc}$.
\subsection{Hypersurface orthogonal null geodesic congruence }
As a consequence of the Frobenius identity \eqref{frobeniusdual}, we have,
\begin{equation}
\p_a l_b - \p_b l_a = \Big(\hcvd_al_b - \hcvd_b l_a\Big) + T^{c}_{~ab} ~ l_c = \Big(\nabla_a l_b - \nabla_b l_a\Big) = (\p_a \rho)l_b - (\p_b \rho)l_a ~.
\end{equation}
Contracting the above formula with $l^a$, we obtain,
\begin{equation}
l^a  \hcvd_a l_b + T_{cab} ~l^a l^c = l^a \cvd_a l_b = (l^a \p_a \rho) l_b ~.
\end{equation}
The operator $\cvd$ is the covariant derivative of the spacetime taken w.r.t the Levi-Civita connection. Defining the directional rate of change of the scalar field $\rho(x^a)$ along the null generators $l^a$ to be $\kappa$, i.e $l^a \p_a \rho = \kappa$ and using the antisymmetry of the torsion tensor in its last two indices \eqref{torsion}, we hence have,
\begin{equation}
l^a \hcvd_a l_b - T_{abc}~l^a l^c = l^a \cvd_a l_b = \kappa l_b ~.
\end{equation}
The above equation indicates that even though the vector field $l^a$ is the null generator of $\hech$, yet it is not an auto-parallel vector field i.e. $l^a$ does not satisfy the parallel transport equation w.r.t to the spacetime connection $\hcvd$,
\begin{equation}
	l^a \hcvd_a l_b = \kappa l_b + T_{abc}l^a l^c  = \kappa l_b + \ti_b~,
	\label{notautoparallel}
\end{equation}
where $\ti_b \equiv T_{abc}~l^a l^c$.
Notice that even though $\bm{l}$ is not an auto-parallel vector field in the spacetime $\sptm$, yet $\bm{l}$ is extremal in the sense that they are null geodesic curves of extremal length. This is because the notion of extremal curves is defined only w.r.t. the Levi-Civita connection. 
%%%%%%%%%%%%%%%%%%%%%%%%%%%%%%%%%%%%%%%%%%%%%%%%%%%%%%%%%%%%
\subsection{Extrinsic geometry of the null hypersurface}
For any hypersurface of codimension one embedded  in the spacetime $\sptm$, the extrinsic curvature captures the notion of bending of the hypersurface. The extrinsic curvature is quantified by the Weingarten map (also known as the shape operator). The Weingarten map at a point $P$ on the hypersurface measures how its normal changes as we move along a vector on the tangent space established on the hypersurface at $P$. Let us now focus on our null surface $\hech$. For any vector $\bm{v}$ $\in$  $T_{P}(\hech)$, we have the definition of the Weingarten map $^{\hech}\chi^a_{~b}$ as follows,
\begin{equation}
^{\hech}\chi^a_{~b} v^b \equiv \bm{\hcvd_{\bm{v}}} ~l^a ~. 
\label{weingartenmap}
\end{equation} Now, since the notion of the Weingarten map involves taking the covariant derivative of the null generator along a tangent vector to $\hech$, the quantity $^{\hech}\chi^a_{~b} v^b$ is independent of the null foliation. The quantity $^{\hech}\chi^a_{~b}v^b$ again itself lies on the tangent space of $\hech$ established on $P$ as verified via the following,
\begin{equation}
l_a (^{\hech}\chi^a_{~b} v^b) = l_a \bm{\hcvd_v}~l^a = l_a v^b \hcvd_b l^a = 0 \implies (^{\hech}\chi^a_{~b} v^b)\bm{\partial_a} \in \ct_{P} (\hech) ~.  
\end{equation}
In contrast to Riemannian spacetimes, the presence of non trivial torsion in the Riemann-Cartan spacetime forces the Weingarten map to not be self-adjoint. This can be shown along the following lines. Consider two vectors $\bm{u}$ and $\bm{v}$ established on the tangent space of $\hech$ at the point $P$ i.e $l_a u^a = 0 = l_a v^a$. Then obviously the Lie commutator of these two vectors again lies on the tangent space of $\hech$ at $P$ i.e. $[\bm{u}, \bm{v}]$ $\in$ $\ct_{P}(\hech)$ as can be verified by showing  that,
\begin{align}
l_a [\bm{u}, \bm{v}]^a &= l_a u^b \hcvd_b v^a - l_a v^b \hcvd_b u^a - l_a T^{a}_{~bc} ~u^b v^c \nonumber \\
&= -u^b v^a \hcvd_b l_a + u^a v^b \hcvd_b l_a - T^{a}_{~bc}~l_a u^b v^c = u^b v^a \Big(\hcvd_a l_b- \hcvd_b l_a \Big)  - T^a_{~bc} ~l_a u^b v^c \nonumber \\
&= u^b v^a \Big((\partial_a \rho)~l_b - (\partial_b \rho)~l_a - T^c_{~ab}~l_c\Big) - T^a_{~bc}~l_a u^b v^c = -l_c T^c_{ba}~u^a v^b - l_c T^c_{~ab} u^a v^b = 0 ~.
\end{align}   
Getting back to our reasoning and using the definition of the torsion tensor we have the following,
\begin{align}
u_a ~( ^{\hech}\chi^a_{~b} v^b) &= u_a (v^b \hcvd_b l^a) = -l_a v^b \hcvd_b u^a \nonumber \\
&= -l_a u^b \hcvd_b v^a + l_a [\bm{u}, \bm{v}]^a + T_{abc}l^a u^b v^c \nonumber \\
&= v_a (u^b \hcvd_b l^a) + T_{abc}l^a u^b v^c = v_a ~ (^{\hech}\chi^a_{~b} u^b) + T_{abc}l^a u^b v^c ~.
\label{selfadjWein}
\end{align}
This shows that the presence of non-zero torsion in the Riemann-Cartan spacetime accounts for the fact that $u_a ~ ^{\hech}\chi^a_{~b} v^b \neq v_a ~ ^{\hech}\chi^a_{~b} u^b$ and hence the Weingarten map is in general not self-adjoint.

Let us then move on to the concept of the second fundamental form  of $\hech$. The second fundamental form $^{\hech}\Theta_{ab}$ is a second rank tensor belonging to the cotangent space of the hypersurface $\hech$ defined as the following. Consider any two vectors $\bm{u}$ and $\bm{v}$ belonging to the tangent space of $\hech$, then,
\begin{equation}
^{\hech}\Theta_{ab} u^a v^b \equiv u_a ~(^{\hech}\chi^a_{~b} v^b) ~~ \implies ^{\hech}\Theta_{ab} \bm{d}x^a \otimes \bm{d}x^b \in \ct^{*}_P{(\hech)} \otimes  \ct^{*}_P{(\hech)} ~.
\end{equation}
Notice that since the Weingarten map is not self-adjoint, the second fundamental form is not symmetric in its two indices. It can very easily be verified that,
\begin{equation}
^{\hech}\Theta_{ab}v^a u^b = {^{\hech}\Theta_{ab}} u^a v^b + T_{abc}l^a v^b u^c ~.
\end{equation}
Our aim is to finally consider a $3+1$ induced foliation of the null family of the hypersurfaces. For that we consider in our spacetime, a stack of spacelike hypersurfaces defined via $\Sigma_t \equiv t(x^a) = \text{constant}$ slices. The timelike normal to these spacelike slices is defined via,
\begin{equation}
n_a = -N \partial_a t = -N \hcvd_a t ~.
\label{normaltosigmat}
\end{equation}
The timelike normal is normalized i.e. $n_a n^a = -1$ and the proportionality factor $N$ is called the lapse function. The orthogonal projection tensor onto the surfaces $\Sigma_t$ defined as,
\begin{equation}
\gamma^a_{~b} = \delta^a_{~b} + n^a n_b ~,
\label{projectiontensorspacelike}
\end{equation}
projects any vector in $\sptm$ onto the surface $\Sigma_t$. Now with the help of this spacelike foliation we can introduce an adapted coordinate system for $\sptm$. Let the independent coordinates parametrizing the $t(x^a) = c ~(\text{constant})$ surface be $y^{\mu}$. This allows a locally well defined coordinate system to be established in an open neighbourhood of $\sptm$ given via $x^a = (t, y^{\mu})$. The coordinate time vector $\bm{t}$ along its flow basically joins the points having the same spatial coordinates $y^{\mu}$ for the different time slices and is defined as,
\begin{equation}
\bm{t} \equiv \bm{\frac{\partial}{\partial t}} ~~ \text{and} ~~ t^a \partial_a t = 1 ~.
\end{equation}
This allows an orthogonal decomposition of the time vector as follows,
\begin{equation}
t^a = N n^a + \b^a ~.
\label{timevectordecomp}
\end{equation}
The vector $\b^a$ is basically the projection of the time vector onto the $t(x^a) = c$ slice and is known as the shift vector.

Now, we finally come to the topic of foliating our family of null hypersurfaces by a stack of spacelike hypersurfaces. The reason for doing this is as follows. As of yet, precisely because of the unique structure of our null hypersurface $\hech$, we do not have any notion of a vector that is transverse to $\hech$. Hence we do not have any well defined projection tensor onto the null surface. To define a projector onto $\ct_{P}(\hech)$, we need to have a direction that is transverse to the hypersurface so that we may define a projector along this direction. There exists no unique transverse direction to $\hech$. The auxiliary null vector $\bm{k}_{\text{(au)}}$ $\in$ $\ct_{P}{(\mathcal{M})}$ defined via the relations,
\begin{equation}
\bm{k}_{\text{(au)}} \cdot \bm{k}_{\text{(au)}}= 0 ~~ \text{and} ~~ \bm{k}_{\text{(au)}} \cdot \bm{l} \equiv -1 ~,
\end{equation}
defines a notion of a vector transverse to the null surface $\hech$. However such a vector is non-unique. Hence the idea behind foliating our null surface by the family of spacelike $t(x^a) = \text{constant}$ slices is to provide some extra structure onto $\hech$ so that we can unambigously define a unique transverse direction to our null surface. This extra structure will provide two immediate benefits. One is that it allows a unique projection tensor onto $\hech$ and that we can in a sense normalize our null generators by choosing a specific parameter made possible via this extra structure.

The spacelike slices $\Sigma_t$ cut our generic null surface into a stack of $2$-dimensional cross-sections $S_t$ defined as,
\begin{equation}
S_t \equiv \hech \cap \Sigma_t ~.
\label{Stdefn}
\end{equation}
As we can visualize, the family of these transverse spacelike $2$-dimensional cross-sections $S_t$ provide a foliation of  $\hech$. The value of the scalar field $t(x^a) = \text{constant}$ can be chosen as a parameter along each null generator $l^a$ of $\hech$. With this (in general) non-affine parametrization $t$, we can in essence normalize \cite{Gourgoulhon:2005ng} the null normal by demanding that,
\begin{equation}
	l^a = \frac{d x^a}{d t} ~.
	\label{nullgeneratorpar}
\end{equation}
Geometrically this means that the $2$-surfaces $S_t$ can be Lie dragged along the null generators $l^a$ thus forming our null hypersurface $\hech$. It is in this sense that the null vector $\bm{l}$ is called the null generator of $\hech$. Thus following $\eqref{Stdefn}$, the transverse $2$-surfaces $S_t$ can be defined as level sets of the scalar field $u(x^a) = 0$ such that it is the set of all the points $P$,
\begin{equation}
	S_t \equiv \{P \in \mathcal{M}, ~~P \in S_t : u(P) = 0  \text{~~and~~} t(P) = t\} ~.
	\label{Stdefn1}
\end{equation}
Let us consider a unit spacelike vector field $\bm{s}$ belonging to the tangent space of $\Sigma_t$ and which is directed outward from the transverse $2$-surface $S_t$. This vector field follows the properties given as,
\begin{align}
\bm{s} \cdot \bm{s} = 1, ~~~ \bm{n} \cdot \bm{s} = 0, ~~~ s^a \partial_a u < 0 \nonumber \\
\forall \bm{v} \in \ct_P{(\Sigma_t)}, ~~ \bm{v} \in \ct_P (S_t) ~~ \iff ~~ \bm{s} \cdot \bm{v} = 0 ~.
\label{defnoutwards}
\end{align}
Then it can be quite easily shown \cite{Gourgoulhon:2005ng} that, the $3+1$ decomposition of the null generator is, 
\begin{equation}
\bm{l} = N(\bm{n} + \bm{s}) ~.
\label{l3+1}
\end{equation}
Obviously it can be seen that the projection of the null generator onto the spacelike slices $\Sigma_t$  given by $\gamma^a_{~b} l^b  = N s^a$ with $N >0$ points in the exterior direction w.r.t. $S_t$. It is in this respect that the null generators are outgoing w.r.t. the $2$-surface $S_t$. Using \eqref{nullnormal} and \eqref{normaltosigmat}, it can very easily be shown that,
\begin{equation}
  s_a = \Big(\frac{1}{N}\Big) l_a -n_a ~=~ N \p_a t - \Big(\frac{e^{\rho}}{N}\Big) \p_a u ~=~ N\p_a t + M \p_a u ~,
  \label{outwardtoSt}
\end{equation}
where $M = -e^{\rho}/N$. Now once we have provided a null normal $n_a$ to the spacelike slices $\Sigma_t$ and the unit outward spacelike vector field $s_a$ to the transverse cross-sections $S_t$, we can define an orthogonal projection tensor $q_{ab}$ onto the spacelike $2$-surface $S_t$ as,
\begin{equation}
q_{ab} = g_{ab} + n_a n_b -s_a s_b ~.
\label{Stprojectiontonsor1}
\end{equation}
Now let us come to the discussion of the construction of an ingoing transverse auxiliary null vector field to $\hech$. Any null vector pointing in the direction $(\bm{n}- \bm{s})$ points in the ingoing direction w.r.t. $S_t$. We can normalize this ingoing auxiliary null field \cite{Gourgoulhon:2005ng} by taking it as,
\begin{equation}
\bm{k} = \frac{1}{2 N} (\bm{n}-\bm{s} ) ~.
\label{auxiliaryk1}
\end{equation}
The normalization has been chosen so that $\bm{k}$ is consistent with a unique definition of the auxiliary null vector field thanks to the foliation of the null surface $\hech$ by the stack of spacelike slices $\Sigma_t$,
\begin{equation}
\bm{l} \cdot \bm{k} = -1, ~~~ \bm{k} \cdot \bm{k} = 0 ~~ \text{and} ~~ \bm{k} \cdot \bm{e_{{A}}} = 0 ~,
\label{defnauxiliaryk}
\end{equation}
where $\{\bm{e_{{A}}}\}$ denotes the set of basis vectors lying on the tangent space $\ct_P (S_t)$ of $S_t$. By using the Eqs. \eqref{auxiliaryk1}, \eqref{outwardtoSt} and \eqref{normaltosigmat} we have,
\begin{equation}
	k_a  = -\p_a t - \Big(\frac{M}{2 N}\Big) \p_a u ~.
	\label{auxiliarynullnormal}
\end{equation}
The unique orthogonal projection tensor onto the transverse $2$-surface $S_t$ can also be defined in terms of the null normal $l_a$ and the auxiliary null normal $k_a$ as,
\begin{equation}
q_{ab} = g_{ab} + l_a k_b + k_a l_b ~.
\label{projectiontensorSt1}
\end{equation} 
Its trivial to verify that $q^a_{~b} l^b = 0$ and $q^a_{~b} k^b = 0$.
Finally, owing to the foliation of $\hech$ by the stack of spacelike $\Sigma_t$ we have at our disposal both a unique normalized null normal $l_a$ and a unique auxiliary transverse null vector field $k_a$. This enables us to define a unique projection tensor onto $\hech$ as, 
\begin{equation}
\Pi^a_{~b} = \delta^a_{~b} + k^a l_b  = q^a_{~b} - l^a k_b~.
\end{equation}
This projection tensor basically projects any vector belonging to $\ct_{P}(\mathcal{M})$ to only the part belonging to $\ct_P(\hech)$. The projection tensor satisfies the following properties as can be easily verified,
\begin{align}
&\Pi^a_{~b} l^b = l^a, ~~ \Pi^a_{~b} k^b = 0,\nonumber \\
&\Pi^a_{~b} l_a = 0 ~~ \text{and} ~~ \Pi^a_{~b} k_a = k_b ~.
\end{align}
%%%%%%%%%%%%%%%%%%%%%%%%%%%%%%%%%%%%%%%%%%%%%%%%%%%%%%%%%%%%%%%%%%%%%%%%%%%%%%%%%%%%%%%%%%%%%%%%%%%%%%%%%%%%%%%%%%%%%%
\section{Kinematics of the null hypersurface $\hech$}\label{kinematicsh}
{Following \cite{Gourgoulhon:2005ng}, what we imply by kinematical quantities are all those geometrical entities that have first order derivatives of the null vector fields $\bm{l}$ and $\bm{k}$, their associated $1$-forms $\underline{\bm{l}}$ and $\underline{\bm{k}}$ and the metric fields $\bm{g}$ and $\bm{q}$ as well. By first order derivatives, we mean spacetime covariant derivatives $\bm{\hcvd}$ as well as the Lie derivatives along $\bm{l}$ and $\bm{k}$.} The extension of such kinematical quantities (to be described in detail in this section) to case of the Riemann-Cartan spacetime $\sptm$ is quite necessary for our analysis. In doing so, we will keep track of the modifications that arise in these kinematical quantities when we consider torsion in the spacetime.
\subsection{The extended second fundamental form}\label{kinema1}
Previously we have defined our Weingarten map or the shape operator corresponding to vectors constrained only on the hypersurface $\hech$. Now, as we have a unique projection tensor onto the tangent space of $\hech$, we can extend the definition of the Weingarten map to vectors living in the tangent space $\ct_P(\mathcal{M})$. The extended Weingarten map $\chi^a_{~b}$ is defined for vectors living in $\ct_P(\mathcal{M})$ as,
\begin{equation}
\chi^a_{~b} v^b \equiv {^{\hech}\chi^a_{~b}} \Big(\Pi^b_{~c} v^c\Big) = \bm{\hcvd}_{\bm{\Pi(v)}} l^a   = \Big(\delta^b_{~c} + k^b l_c\Big) v^c \hcvd_c l^a~,
\label{extendedWeingarten}
\end{equation}
where contrary to the earlier section, now $\bm{v} \in \ct_P(\mathcal{M})$.
From the above, it follows that,
\begin{align}
\chi^a_{~b} = \hcvd_b l^a + (k^c \hcvd_c l^a) l_b ~.
\label{extendedWeingarten1}
\end{align}
Then one finds that,
\begin{align}
\chi^a_{~b} l^b = l^b \hcvd_b l^a = \kappa l^a + T_{b~~c}^{~a} ~l^b l^c = \kappa l^a + \mathbb{T}^a ~, \\
\chi^a_{~b} k^b = 0 ~,
\end{align}
where $\mathbb{T}^a \equiv T_{b~~c}^{~a} ~l^b l^c $. It is worthwhile to notice that the action of the extended Weingarten map onto any spacetime vector $\bm{v} ~ \in ~ \ct_{P}(\mathcal{M})$ is to effectively map it to another vector belonging to the tangent space of $\hech$. This can quite easily seen by,
\begin{equation}
l_a \chi^a_{~b} v^b = l_a \bm{\hcvd_{\Pi(v)}}l^a = \frac{1}{2} \Pi^a_{~b} v^b \hcvd_a (l_d l^d) = 0 ~.
\end{equation}
This clearly shows that $\chi^a_{~b} v^b$ belongs to the tangent space $\ct_P(\hech)$ for any vector $\bm{v} ~ \in ~ \ct_{P}(\mathcal{M})$. For this reason the action of the extended Weingarten map onto any spacetime vector is the same as its action on the projected part of the vector onto the tangent space of $\hech$ i.e. 
\begin{equation}
	\chi^a_{~b} (\Pi^b_{~c}v^c) = \chi^a_{~b} v^b ~.
	\label{extendedWeinprop2}
\end{equation}
Notice that $\bm{l}$ is not an eigenvector of the extended Weingarten map exclusively due to the presence of torsion in $\sptm$. In the same vein, we can extend the notion of the second fundamental form $^{\hech}\Theta_{ab}$ to its action over vectors living in $\ct_P(\mathcal{M})$ rather than $\ct_P(\hech)$. Hence for any two vectors $(\bm{u}, \bm{v}) ~ \in ~ \ct_P(\mathcal{M}) \times \ct_P(\mathcal{M}) $ we define the extended second fundamental form $\thht_{ab}$ as the following,
\begin{align}
	\thht_{ab} u^a v^b &\equiv {^{\hech}\Theta_{ab}} \Big(\Pi^a_{~c} u^c ~\Pi^b_{~d} v^d\Big)  = (\Pi^a_{~c} u^c) (\Pi^b_{~d} v^d) \hcvd_b l_a \nonumber \\
	&= \Big((q^a_{~c}-l^a k_c)u^c\Big) \Big((q^b_{~d} - l^b k_d) v^d\Big) \hcvd_b l_a \nonumber \\
	&= \Big(q^a_{~c}q^b_{~d} u^c v^d \hcvd_b l_a\Big) - (k_d v^d) q^a_{~c}u^c \Big(\kappa l_a + \mathbb{T}_a\Big) + (k_c u^c) (k_d v^d) l^a \Big(\kappa l_a + \mathbb{T}_a\Big) \nonumber \\
	&= \Big(q^c_{~a} q^d_{~b} \hcvd_d l_c - q^c_{~a} k_b \mathbb{T}_c\Big) u^a v^b ~.
	\label{secondfundform1}
\end{align}
This naturally allows us to define the extended second fundamental form as a bilinear,
\begin{equation}
\thht_{ab} = \Big(q^c_{~a} q^d_{~b} \hcvd_d l_c\Big) - \Big(q^c_{~a} k_b \ti_c\Big) ~.
\label{extsecondfundform}
\end{equation} 
Naturally by inspection it can be observed that,
\begin{equation}
\thht_{ab} l^a = 0, ~~ \thht_{ab}l^b = q^c_{~a} \ti_c , ~~ \thht_{ab} k^a = 0 ~~ \text{and} ~~ \thht_{ab} k^b = 0 ~.
\label{extfundformprops}
\end{equation}
The above Eq. \eqref{extfundformprops} very clearly shows that in the presence of torsion in $\sptm$, the extended second fundamental form $\thht_{ab}$ is not a second rank tensor lying only in the space $\ct^{*}(S_t) \otimes \ct^{*}(S_t)$; rather it lies in the space $\ct^{*}(\hech) \otimes \ct^{*}(\hech)$. We can compute the trace of this extended second fundamental form,
\begin{align}
\overset{(e)}{\hat{\theta_{\bm{l}}}} &\equiv g^{ab} \thht_{ab} = g^{ab} \Big[\Big(q^c_{~a} q^d_{~b} \hcvd_d l_c\Big) - \Big(q^c_{~a} k_b \ti_c\Big)\Big] = q^{cd} \hcvd_d l_c \nonumber \\
&=(g^{cd} +l^ck^d +k^c l^d) \hcvd_d l_c = \hcvd_a l^a -\kappa + k_a \ti^a ~.
\label{trace2ndfundamentalform}
\end{align}

It maybe noted that if we impose the constraint, 
\begin{equation}
\ti_b \equiv T_{abc}~l^a l^c = 0 ~,
\label{geodesicconstraint}
\end{equation} 
in the Riemann-Cartan spacetime, we then have, via \eqref{notautoparallel}, that the null generators are parallel-transported along themselves with $\kappa$ being the non-affinity parameter,
\begin{equation}
l^a \hcvd_a l_b \overset{\ti_b = 0}{=} \kappa l_b ~.
\end{equation}
The condition \eqref{geodesicconstraint} will be termed as the geodesic constraint. Application of the geodesic constraint implies,
\begin{equation}
l^a \hcvd_a l_b = l^a \cvd_a l_b - K^c_{~ab}~l^a l_c \overset{\ti_b = 0}{=} l^a \cvd_{a} l_b = \kappa l_b ~,
\end{equation}
thus verifying the fact that if the null generators $\bm{l}$ of $\hech$ satisfy the parallel-transport equation w.r.t. the connection $\hcvd$, then they also satisfy the geodesic equation w.r.t. the Levi-Civita connection $\cvd$. The above geodesic constraint \eqref{geodesicconstraint} represents the vanishing of the torsion current $T_{abc}l^a l^c$. In the context of Killing horizons established in the spacetime $\sptm$ it has been shown in \cite{Dey:2017fld} that the above condition of the vanishing of such a torsion current is necessary to establish the zeroth law. This allows a notion of equilibrium to be defined for such a horizon. We will have much to say about this later.
%\textcolor{red}{Provide the physical interpretation of the geodesic constraint following from Ramit et-al. Note that Ramit et-al defines the geodesic constraint as providing a notion of \textbf{equilibrium} at least for a Killing horizon where the notion of surface gravity and the non-affine parameter needs to coincide for the equilibrium case. However here we are not specifying to the case of Killing horizons and hence our generic null surface should not have a notion of equilibrium built into it. Think then what the corresponding motivation should be in order to set the quantity $\ti^a =0$}.

Imposing the geodesic constraint i.e. $\ti^a = 0$, we notice that the extended second fundamental form lies in the orthogonal transverse space (to $\bm{l}$ and $\bm{k}$) $\ct^{*}(S_t) \otimes \ct^*(S_t)$ and hence is orthogonal to both $\bm{l}$ and $\bm{k}$. Note that again due to the presence of torsion in the spacetime, the extended second fundamental form is not symmetric in its indices i.e. $\thht_{ab} \neq \thht_{ba}$. 
For the specific case of the geodesic constraint $\ti_a = 0$, the extended second fundamental form being a bilinear established on the $2$-dimensional transverse space orthogonal to both $\bm{l}$ and $\bm{k}$ can be provided an irreducible decomposition into a symmetric trace part $\thel$, a traceless symmetric part $\shel$ and an antisymmetric traceless part $\whel$,
\begin{align}
\thht_{ab} &= \frac{1}{2} q_{ab}~ \thel + \shel + \whel ~.
\label{irred2ndfundform2} \\
\thel &\overset{\ti^a = 0}{=} (\hcvd_a l^a - \k),  ~~ ~ \shel \overset{\ti^a =0}{ = } \thht_{(ab)} - \frac{1}{2} q_{ab} \thel ~~ \text{and} ~~ \whel \overset{\ti^a =0}{ = } \thht_{[ab]} ~. 
\label{irred2ndfundformprops2}
\end{align}
Note that the trace of the extended second fundamental form is not to be designated as the expansion scalar corresponding to the null congruence $\bm{l}$ since it does not quantify the fractional rate of change of the area element $\sqrt{q}$ of $S_t$ along $\bm{l}$. We will soon develop the proper notion of an expansion scalar corresponding to null congruences in $\sptm$ for our purpose.
\subsection{The Rotation $1$-form and the Hajicek $1$-form}
A quantity of great interest and utility for practical calculation dealing with this topic is the spacetime covariant derivative of the null normal $l_a$. Now having foliated the spacetime in the neighbourhood of our generic null surface $\hech$ by a family of null hypersurfaces and the slicing of this null family by a stack of spacelike slices $\Sigma_t$, we have now a well defined notion of $\hcvd_a l_b$. According to \eqref{secondfundform1}, we have for any $(\bm{u}, \bm{v}) ~ \in ~ \ct_P(\mathcal{M}) \times \ct_P(\mathcal{M}) $,
\begin{align}
\thht_{ab} u^a v^b &= (\Pi^a_{~c}u^c) \bm{\hcvd_{\Pi(v)}} l_a = \Big(u^a + (l_c u^c)k^a\Big) \bm{\hcvd_{\Pi(v)}} l_a \nonumber \\
&= u^a (\Pi^b_{~d} v^d) \hcvd_b l_a + (l_c u^c) k^a \bm{\hcvd_{\Pi(v)}} l_a \nonumber \\
& =u^a \Big(v^b + (l_d v^d)k^b\Big) \hcvd_b l_a + (l_c u^c) k_a \Big(\chi^a _{~d} v^d\Big) \nonumber \\
& = (\hcvd_b l_a) u^a v^b + u^a (l_d v^d) (k^b \hcvd_b l_a) + (l_a u^a )k_b \Big(\chi^b_{~d} v^d\Big) ~. 
\label{rotn1form1}
\end{align}
The rotation $1$-form $\underline{\hat{\bm{\omega}}} ~ \in \ct^{*}_{P} (\mathcal{M})$ is defined in such a way that its action on any vector $\bm{f} ~ \in \tpm$ is given via, \begin{equation}
	\hw_a f^a \equiv - k_a ~ (\chi^a_{~b}f^b) ~.
	\label{rotn1form2}
\end{equation}
Then using \eqref{rotn1form2} in \eqref{rotn1form1}, we obtain,
\begin{align}
&\thht_{ab} u^a v^b = (\hcvd_b l_a) u^a v^b + (l_b k^c \hcvd_c l_a) u^a v^b - (l_a u^a)(\hw_b v^b) ~.
\end{align}
Now, since $u^a$ and $v^a$ are arbitrary, one finds after rearranging,
\begin{align}
\hcvd_a l_b = \thht_{ba} + \hw_a l_b - l_a (\bm{\hcvd_{k}}l_b ) ~.
\label{nablaalbexpansion}
\end{align}
The above expansion \eqref{nablaalbexpansion} of the spacetime covariant derivative of the null normal is going to be of significant practical interest to us. 

Let us look at some properties of the rotation $1$- form. Using the basic definition \eqref{rotn1form2}, it can very easily verified that, 
\begin{equation}
\hw_a k^a = 0 ~~ \text{and} ~~ \hw_a l^a = \kappa - k_a \ti^a ~.
\label{rotn1formprops}
\end{equation}
Combining \eqref{extendedWeingarten1} and \eqref{nablaalbexpansion}, we have,
\begin{equation}
	\chi^a_{~b} = \thht^a_{~b} + \hw_b l^a ~.
\end{equation}
Let us now proceed to obtain an expression of the rotation $1$-form. We begin by noticing that for any vector $\bm{f} \in \tpm$, we have,
\begin{align}
\hw_a f^a &= - k_a \chi^a_{~b}f^b =-k_a \Pi^b_{~d}f^d \hcvd_b l^a = -k_a \Big((\delta^b_{d} + k^b l_d)f^d\Big) \hcvd_b l^a\nonumber \\
& = - k_a \Big(f^b + k^b(l_d f^d)\Big)\hcvd_b l^a = -(k_b \hcvd_a l^b)f^a - \Big(k_b (k^c \hcvd_c l^b)l_a\Big)f^a ~.
\end{align}
This allows us to have,
\begin{equation}
\hw_a = -(k_b \hcvd_a l^b) - l_a k_b (\bm{\hcvd_{k}}l^b) ~.
\label{omega1defn}
\end{equation}
The above equation stands as a working definition of the rotation $1$-form. However, we will soon come up with other expressions of this rotation $1$-form that will be useful later on.

The Hajicek $1$-form $\bm{\underline{\hhw}} \in \tpss$ is defined as the projection of the rotation $1$-form onto the cotangent space of $S_t$.
\begin{equation}
	\hhw_a \equiv q^b_{~a} \hw_b ~.
	\label{Omegadefn1}
\end{equation}
Following the above definition, we have for any vector $\bm{v} ~\in ~ \tpm$, 
\begin{equation}
\hhw_a v^a \equiv (q^b_{~a} \hw_b) v^a = \hw^a (q^a_{~b}v^b) ~.
\end{equation}
Using the fact that $q^a_{~b} = \Pi^a_{~b} + l^a k_b$, we have,
\begin{align}
\hhw_a v^a &= \hw_a \Big((\Pi^a_{~b} + l^a k_b) v^b\Big) = \hw_a (\Pi^a_{~b}v^b) + (\hw_a l^a) k_b v^b \nonumber \\
& =\hw_a v^a + (\kappa - k_a \ti^a) (k_b v^b) ~.
\end{align}
In the above, we have used the fact that $\hw_a (\Pi^a_{~b}v^b)  = \hw_a v^a$ precisely via the property of the extended Weingarten map as shown in \eqref{extendedWeinprop2}.
This allows us to have a relationship between the rotation $1$-form and the Hajicek $1$-form as,
\begin{equation}
\hw_a = \hhw_a - \kappa k_a + (k_b \ti^b)k_a ~.
\label{rotnhajireln}
\end{equation}

From the dual formulation of the Frobenius theorem for the null normal $\underline{\bm{l}}$, we have upon using \eqref{nablaalbexpansion},
\begin{align}
\dl &= \bm{d} \rho \wedge \underline{\bm{l}}  =(\p_a l_b - \p_b l_a) \dxa \otimes \dxb  \nonumber \\
& =  \Big(\hcvd_a l_b - \hcvd_b l_a + T_{cab}l^c\Big) \dxa \otimes \dxb \nonumber \\
& = \Big(\thht_{ba} - \thht_{ab} + \hw_a l_b - \hw_b l_a + (\bm{\hcvd_{k}} l_a) l_b - (\bm{\hcvd_{k}} l_b)l_a + T_{cab}l^c\Big) \dxa \otimes \dxb~.
\end{align}
Using \eqref{extsecondfundform}, we have the antisymmetric part of the extended second fundamental form as,
\begin{align}
	\thht_{ba}- \thht_{ab} = q^c_{~a} q^d_{~b} \Big((\hcvd_c l_d - \hcvd_d l_c) + (k_b q^c_{~a} - k_a q^c_{~b}) \ti_c\Big) ~.
\end{align}
Using the fact on account of the dual formulation of the Frobenius theorem i.e $(\p_a l_b - \p_b l_a) = (\p_a \rho) l_b - (\p_b \rho) l_a$, we have,
\begin{align}
\thht_{ba}- \thht_{ab} = q^c_{~a} q^d_{~b} T_{fdc}~l^f + (k_b q^c_{~a} - k_a q^c_{~b}) \ti_c ~.
\label{diffinTheta}
\end{align}
This allows us to have,
\begin{align}
\dl & = \Big(q^c_{~a} q^d_{~b} T_{fdc}~l^f + (k_b q^c_{~a} - k_a q^c_{~b}) \ti_c + T_{cab}l^c + \hw_a l_b - \hw_b l_a 
\nonumber
\\
& + (\bm{\hcvd_{k}} l_a) l_b - (\bm{\hcvd_{k}} l_b)l_a  \Big) \dxa \otimes \dxb~.
\end{align}
Using the definition of the projection tensor $q^a_{~b} = \d^a_{~b} + l^a k_b + k^a l_b$ and few lines of simple algebra it can be shown that,
\begin{equation}
q^c_{~a} q^d_{~b} T_{fdc}~l^f + (k_b q^c_{~a} - k_a q^c_{~b}) \ti_c + T_{cab}l^c  = (T_{cda}l^c k^d) l_b - (T_{cdb}l^c k^d) l_a ~.
\end{equation}
This finally results in the fact that,
\begin{align}
\dl = \Big(\hw_a l_b - \hw_b l_a + (\bm{\hcvd_{k}} l_a) l_b - (\bm{\hcvd_{k}} l_b)l_a + (T_{cda}l^c k^d) l_b - (T_{cdb}l^c k^d) l_a\Big) \dxa \otimes \dxb~.
\label{frobduall}
\end{align}
Provided we define a $1$-form $\underline{\bm{\mathfrak{T}}} \equiv (T_{cda}l^c k^d )\dxa$, we can succinctly via \eqref{frobduall} express the exterior derivative of the null normal as, 
\begin{equation}
\dl = \underline{\bm{\hw}} \wedge \underline{\bm{l}} + (\bm{\hcvd_{k} \underline{l}}) \wedge \underline{\bm{l}} + \underline{\bm{\mathfrak{T}}} \wedge \underline{\bm{l}} = \Big(\underline{\bm{\hw}} +(\bm{\hcvd_{k} \underline{l}}) + \underline{\bm{\mathfrak{T}}}\Big) \wedge \underline{\bm{l}}  ~.
\end{equation} The comparison of the above relation with $\dl = \bm{d}\rho \wedge \underline{\bm{l}}$ provides a relationship between the scalar field $\rho$ and the rotation $1$-form,
\begin{align}
	\partial_a \rho = \hw_a + \hcvd_{\bm{k}} l_a + T_{cda}l^c k^d~,
\end{align}
which is arbitrary upto a term proportional to $l_a$. 

Let us proceed to derive another expression of the rotation $1$-form. For that, notice that the exterior derivative of the auxiliary null normal $\underline{\bm{k}}$ via \eqref{auxiliarynullnormal} can be simply expressed as,
\begin{equation}
	\bm{d} \underline{\bm{k}} = \frac{1}{2 N^2} ~\bm{d} \big(\ln (\frac{N}{M})\Big) \wedge \underline{\bm{l}} ~.
	\label{exteriorderofk}
\end{equation}
Since the auxiliary null normal $\underline{\bm{k}}$ does not satisfy the dual formulation of the Frobenius theorem it can be interpreted that the hyperplane orthogonal to the auxiliary null normal is not integrable. From the definition \eqref{rotn1form2} of the rotation $1$-form, we have any vector $\bm{v} ~ \in ~\tpm$,
\begin{align}
\hw_a v^a &= -k_a (\chi^a_{~b}v^b) =- k_a  ({^{\hech}\chi^a_{~b}} \Pi^b_{~c} v^c) = - k_a \bm{\hcvd_{\Pi(v)}} l^a = - k_a (\Pi^b_{~c}v^c) \hcvd_b l^a = l^a (\Pi^b_{~c}v^c) (\hcvd_b k_a) ~.
\label{rotn1formnewdefn1}
\end{align}
From the relation of the exterior derivative of the auxiliary null normal, we have from \eqref{exteriorderofk},
\begin{align}
&\p_a k_b - \p_b k_a = \Big(\hcvd_a k_b - \hcvd_b k_a + T^c_{~ab} k_c\Big) = \frac{1}{2 N^2} \Big(\p_a \ln(\frac{N}{M}) ~l_b - \p_b \ln(\frac{N}{M}) ~ l_a \Big) ~, \nonumber \\
&  \implies\hcvd_b k_a = \hcvd_a k_b + T^{c}_{~ab}k_c  - \frac{1}{2 N^2} \Big(\p_a \ln(\frac{N}{M}) ~l_b - \p_b \ln(\frac{N}{M}) ~ l_a \Big) ~.
\label{manip1}
\end{align}
Employing \eqref{manip1} in \eqref{rotn1formnewdefn1}, we end up having,
\begin{align}
\hw_a v^a &= l^a \Pi^b_{~c} v^c \Big[\hcvd_a k_b + T^{c}_{~ab} k_c - \frac{1}{2 N^2} \Big(\p_a \ln(\frac{N}{M}) ~l_b - \p_b \ln(\frac{N}{M}) ~ l_a \Big)\Big] \nonumber \\
& =l^a \Pi^b_{~c} v^c \hcvd_a k_b + l^a \Pi^b_{~d} v^d T^c_{ab} k_c = l^a (\delta^b_{~c} + k^b l_c) v^c \hcvd_a k_b + l^a (q^b_{~d} - l^b k_d) v^d T^c_{~ab} k_c \nonumber \\
& = (l^b \hcvd_b k_a) v^a + T_{cdb}k^c l^d q^d_{~a} v^a ~.
\end{align}
Thus we finally retrieve an alternative and useful expression of the rotation $1$-form $\underline{\bm{\omega}}$ as,
\begin{equation}
\hw_a = (l^b \hcvd_b k_a) + T_{bcd} k^b l^c q^d_{~a} ~.
\label{rotn1forndefn2}
\end{equation}
The above formula allows us very easily to verify the relations \eqref{rotn1formprops}.
\subsection{The deformation rate tensor}
The deformation rate tensor $\hchi_{ab}$ essentially quantifies the rate at which the metric of the $2$-surface $S_t$ changes as we evolve along the null generators. Following \cite{Gourgoulhon:2005ng}, the deformation rate tensor is defined as,
\begin{equation}
\hchi_{ab} = \frac{1}{2}  q^i_{~a} q^j_{~b}\pounds_{\bm{l}} q_{ij} ~.
\label{deformationrate}
\end{equation}
Using the definition \eqref{Lie} of the Lie derivative of the metric tensor $g_{ab}$ along the null generators $\bm{l}$,
it is quite easy to notice that,
\begin{equation}
\hchi_{ab} = \frac{1}{2} q^c_{~a} q^d_{~b} \Big(\hcvd_c l_d + \hcvd_d l_c + (T_{cfd} + T_{dfc}) l^f\Big) ~.
\label{manip36}
\end{equation}
Then using \eqref{extsecondfundform}, the deformation rate tensor can be also be expressed as,
\begin{equation}
\hchi_{ab} = \frac{1}{2}\Big(\thht_{ab} + \thht_{ba}\Big) + \frac{1}{2} \Big(q^c_{~a} k_b + q^c_{~b} k_a\Big) \ti_c + \frac{1}{2} q^c_{~a} q^d_{~b}\Big(T_{cfd} + T_{dfc}\Big) l^f ~.
\label{deformation2ndfundformrel}
\end{equation}
A point worthwhile to notice is that in the presence of torsion in $\sptm$, the extended second fundamental form $\thht_{ab}$ and the deformation rate tensor $\hchi_{ab}$ are not equivalent. Contrary to the extended second fundamental form $\thht_{ab}$, in general the deformation rate tensor is by definition both symmetric and orthogonally transverse to the space spanned by $\bm{l}$ and $\bm{k}$, 
\begin{equation}
\hchi_{ab}l^a ~=~ 0~=~\hchi_{ab}l^b  ~~ \text{and}~~ \hchi_{ab} k^a ~=~ 0~=~ \hchi_{ab}k^b ~.
\label{deformationratetensorprops1}
\end{equation}
We can infact perform an irreducible decomposition of the deformation rate tensor in terms a symmetric trace part and a traceless symmetric part,
\begin{equation}
\hchi_{ab} = \frac{1}{2} q_{ab} ~\overset{(d)}{\hat{\theta_{\bm{l}}}} + {^{(\bm{l},d)}{\sigma_{ab}}} ~,
\label{irredecompdeform}
\end{equation}
where $\thdl$ is the outgoing expansion scalar and $\shdl$ is the traceless shear tensor corresponding to the null congruence $\bm{l}$. The reason as to why $\thdl$ is called the expansion scalar is because it represents the fractional rate of change of the area element of the transverse spacelike $2$-surface $S_t$ as we move along the null generators $\bm{l}$. We will prove this result shortly in Sec. \ref{section6} (see Eq. \eqref{thdl1}). The trace of the deformation rate tensor is given by,
\begin{align}
	\thdl &= g^{ab} \hchi_{ab} = \frac{1}{2} q^{cd} \lil q_{cd} = \frac{1}{2} q^{cd} \lil g_{cd} \nonumber \\
    & = \frac{1}{2}q^{cd} \Big(\hcvd_c l_d + \hcvd_d l_c + (T_{cfd} + T_{dfc})l^f\Big) \nonumber \\
    &= \frac{1}{2}(g^{cd} +l^c k^d + k^c l^d)	\Big(\hcvd_c l_d + \hcvd_d l_c + (T_{cfd} + T_{dfc})l^f\Big) \nonumber \\
    & = \hcvd_a l^a + T_a l^a -\kappa =\nabla_a l^a -\kappa ~.
    \label{expansionscalervalue}
    \end{align}
    In the above, we have used \eqref{deformationrate} and \eqref{manip36}.
In the spirit of \eqref{nablaalbexpansion} we will find it beneficial to expand the spacetime covariant derivative of the null normal $l_a$ in terms of the deformation rate tensor. To that extent, using \eqref{deformation2ndfundformrel} and \eqref{diffinTheta}, we can provide a relationship between the deformation rate tensor and the extended second fundamental form,
     \begin{equation}
     \hchi_{ab} = \thht_{ba} + k_a q^c_{~b} \ti_c + q^c_{~a} q^d_{~b} K_{fcd}l^f ~.
     \label{relnchintheta}
     \end{equation} 
     Finally, upon using \eqref{nablaalbexpansion}, we get our desired expression relating the covariant derivative of the null normal $l_a$ with the deformation rate tensor,
     \begin{equation}
     \hcvd_a l_b = \hchi_{ab} + \hw_a l_b - l_a (k^i \hcvd_i l_b) - k_a q^c_{~b} \ti_c - q^c_{~a} q^d_{~b} K_{fcd} l^f ~.
     \label{relnnablaalbnchi}
     \end{equation}
\subsection{Transversal deformation rate tensor}
Much like the deformation rate tensor we can also look for the projection (onto the transverse spacelike surface $S_t$) of the Lie derivative of the metric $q_{ab}$, however now in a direction transverse to $\hech$. This transverse direction is provided by the auxiliary null vector field $\bm{k}$. We define the transversal deformation rate tensor as, 
\begin{equation}
\hat{\Xi}_{ab} \equiv \fr q^c_{~a} q^d_{~b}~\lik q_{cd} ~.
\label{trandeform}
\end{equation}
Using the properties of the Lie derivative of the metric $g_{ab}$ and the fact that $q_{ab}l^a = q_{ab}k^b = 0$, the transversal deformation rate tensor can be expressed as,
\begin{align}
\xih_{ab} =\fr q^c_{~a} q^d_{~b} \lik g_{cd}  = \fr q^c_{~a} q^d_{~b} \Big(\hcvd_c k_d + \hcvd_d k_c + (T_{cfd} + T_{dfc}) k^f\Big) ~.
\label{transdeformanother}
\end{align}
From \eqref{exteriorderofk}, in the index notation, we have,
\begin{align}
(\p_a k_b - \p_b k_a) =(\hcvd_a k_b - \hcvd_b k_a + T_{cab}k^c) &= \frac{1}{2 N^2} \p_a(\ln \frac{N}{M}) l_b - \frac{1}{2 N^2} \p_b(\ln \frac{N}{M}) l_a ~.
\label{manip2}
\end{align}
Using \eqref{manip2} for the covariant derivative of the auxiliary field in \eqref{transdeformanother}, we have,
\begin{align}
\xih_{ab} &= \fr q^c_{~a} q^c_{~b} \Big[2 \hcvd_d k_c + \frac{1}{2 N^2} \p_a(\ln \frac{N}{M}) l_b - \frac{1}{2 N^2} \p_b(\ln \frac{N}{M}) l_a - T_{fcd}k^f + T_{cfd}k^f + T_{dfc}k^f \Big] ~, \nonumber \\
 &= q^c_{~a} q^d_{~b} \Big( \hcvd_d k_c + K_{fdc} k^f \Big) ~. 
\label{transdeformnew2}
\end{align}
As evident from its definition \eqref{trandeform}, the transversal deformation rate tensor is symmetric and orthogonal to the space spanned by the vectors $\bm{l}$ and $\bm{k}$ i.e.
\begin{equation}
	\xih_{ab} l^a ~ = ~0 ~=~\xih_{ab} l^b ~~ \text{and} ~~\xih_{ab} k^a ~ = ~0 ~=~\xih_{ab} k^b~. \label{transdeformprops2}
\end{equation} 

Just like for the case of the null generators, it would be of great practical interest for us to calculate the spacetime covariant derivative of the auxiliary null vector field $\bm{k}$. This quantity is again a well defined quantity, thanks to the null foliation of the spacetime $\sptm$ in the neighbourhood of $\hech$. For this, we will just have to manipulate the part $q^c_{~a} q^b_{~d} \hcvd_d k_c$ in the expression \eqref{transdeformnew2} for $\xih_{ab}$ :
\begin{align}
q^c_{~a} q^d_{~b} \hcvd_d k_c &= (\delta^c_{~a} + l^c k_a + k^c l_a)(\d^d_{~b}+ l^dk_b + k^d l_b) \hcvd_d k_c \nonumber \\
& =(\d^c_{~a} + l^ck_a) \Big(\hcvd_b k_c + k_b (l^d \hcvd_d k_c) + l_b (\k^d \hcvd_d k_c)\Big) \nonumber \\
& = \hcvd_b k_a + k_b (l^d \hcvd_d k_a) + l_b (k^d \hcvd_d k_a) + k_a (l^c \hcvd_b k_c) + k_a k_b l^c (l^d \hcvd_d k_c) + k_a l_b l^c (k^d \hcvd_d k_c) \nonumber \\
&=\hcvd_b k_a + k_b \Big[\hw_a - T_{cdf}k^c l^d q^f_{~a}\Big] + l_b (k^d \hcvd_d k_a) - k_a k^c (\hcvd_b l_c) \nonumber \\
&~~~~~~~~~~~~~~~~~~~~~-k_a k_b k^c(l^d \hcvd_d l_c) + k_a l_b l^c (k^d \hcvd_d k_c) \nonumber \\
& = \hcvd_b k_a + k_b \hw_a + l_b (\bm{\hcvd_{k}} k_a) - k_a k^c \Big(\thht_{cb} + \hw_b l_c - l_b \bm{\hcvd_{k}} l_c\Big) - k_a k_b k^c (\kappa l_c + \ti_c) \nonumber \\
&~~~~~~~~~~~~~~~~~~~~~-k_a l_b k^c(k^d \hcvd_d l_c) - k_b T_{cdf} k^c l^d q^f_{~a} \nonumber \\
& = \hcvd_b k_a + k_b \hw_a + l_b (\bm{\hcvd_k} k_a) + k_a \hw_b + \kappa k_a k_b - k_a k_b k_c \ti^c- k_b T_{cdf} k^c l^d q^f_{~a} \nonumber \\
&= \hcvd_b k_a + k_b \hw_a + l_b (\bm{\hcvd_k} k_a) + k_a \hhw_b - k_b T_{cdf} k^c l^d q^f_{~a} ~.
\label{manip3}
\end{align}
For the above result, we have used \eqref{rotn1forndefn2}  in the fourth line, \eqref{nablaalbexpansion} in the fifth line and \eqref{rotnhajireln} in the seventh line. 
Putting the value of $q^c_{~a} q^b_{~d} \hcvd_d k_c$  as obtained in \eqref{manip3} into \eqref{transdeformnew2} and rearranging, we obtain our desired quantity i.e. the covariant derivative of the auxiliary null normal in terms of the transversal deformation rate tensor,
\begin{equation}
\hcvd_a k_b = \xih_{ab} - \hhw_a k_b - k_a \hw_b -l_a (k^i \hcvd_i k_b) + k_a T_{cdf}k^c l^d q^f_{~a} - q^c_{~a} q^d_{~b} K_{fcd} k^f ~.
\label{nablaakbexpansion}
\end{equation}
In the same way, the transversal deformation rate tensor can be decomposed into a symmetric trace part and a symmetric traceless part,
\begin{equation}
\xih_{ab}  = \fr q_{ab} ~\thdk + \shdk ~.
\end{equation}
The trace of the transversal deformation rate tensor is the expansion scalar corresponding to the auxiliary $\bm{k}$ congruence. Again, as we will show later that the trace is truly the expansion scalar in the sense that it quantifies the fractional rate of change of the area element of $S_t$ as we move along $\bm{k}$. The ingoing expansion scalar corresponding to the auxiliary null vector field, via \eqref{transdeformnew2} is then given by, 
\begin{align}
\thdk &= g^{ab} \xih_{ab} = q^{cd} (\hcvd_d k_c + K_{fdc}k^f) = q^{cd} \hcvd_d k_c + \fr(g^{cd} + l^c k^d + k^c l^d) \Big(T_{dfc} + T_{cfd}\Big) k^f \nonumber \\
& = (q^{cd} \hcvd_d k_c) + T_a k^a - T_{dcf}k^d l^c k^f ~.
\label{thetadefk}
\end{align}
\subsection{The projected deviation tensor} Till now, we have discussed about three kinematical second rank tensors that are of interest to us. They are the extended second fundamental form $\thht_{ab}$, the deformation rate tensor $\hchi_{ab}$ and the transversal deformation rate tensor $\xih_{ab}$. As we shall see, all these quantities will be very relevant in the analysis of providing our advertized thermodynamic interpretation as applied to $\hech$. There also exists another second rank tensor called the deviation tensor that we wont require for our thermodynamic interpretation. In Appendix \ref{NRE3+1} we provide a detailed derivation of the NRE corresponding to the null generators $\bm{l}$ without the assumption of the geodesic constraint \eqref{geodesicconstraint}. To corroborate our results with the existing literature \cite{Dey:2017fld} we would require the construction of the deviation tensor and its relation to the extended second fundamental form. In this section, we end our discussion of the kinematics of the null hypersurface by describing the deviation tensor or effectively its projected part. On the outset, let us mention that the discussion of the deviation tensor does not require a foliation of $\sptm$ by a family of null hypersurfaces and their subsequent slicing by a stack of spacelike surfaces. All we actually require is a congruence of null trajectories (not necessarily geodesic or auto-parallel) and build upon the premise that the deviation vector field $\bm{\eta}$ is Lie transported along the null congruence. The deviation vector essentially measures the deviation between two neighbouring null trajectories. The above condition means,
\begin{equation}
[\bm{\eta}, \bm{l}] = 0 ~.
\end{equation}
In index notation this translates to,
\begin{align}
& l^a \p_i \eta^a - \eta^i \p_i l^a = 0 ~,\nonumber \\
&l^i \hcvd_i \eta^a = \eta^i \Big(\hcvd_i l^a + T^a _{ji} ~l^j\Big) \equiv \eta^i \mathcal{B}^a_{~i} ~.
\end{align}
Here $\mathcal{B}^a_{~i}$  is called the deviation tensor which measures the failure of the deviation vector to be parallel transported along the null congruence. This is given by,
\begin{equation}
\mathcal{B}_{ai} = \hcvd_i l_a + T_{aji} l^j ~.
\label{deviationtensor}
\end{equation}
The auxiliary null vector field $k^a$ to such a null congruence is as usual defined via the relations $ l_a k^a = -1$ and $k^a k_a = 0$, with the projection tensor onto a $2$-dimensional spacelike cross-section of the null congruence being $q_{ab} = g_{ab} + l_a k_b + k_a l_b$. We can easily verify that the deviation tensor is not orthogonal to both $l^a$ and $k^a$ as seen from,
\begin{align}
&\mathcal{B}_{ab} l^a = -\ti_b, ~~~ \mathcal{B}_{ab} l^b = \kappa l_a + \ti_a ~, \nonumber \\
&\mathcal{B}_{ab}k^a = - l^a \hcvd_b k_a + T_{aib} k^a l^i ~~~ \text{and}~~~\mathcal{B}_{ab} k^b = k^b \hcvd_b l_a + T_{aib} l^i k^b ~. 
\end{align}

We will project the deviation tensor onto the transverse spacelike $2$-surface of the null congruence to define the projected deviation tensor $\hb_{ab}$, 
\begin{align}
\hb_{ab} \equiv q^c_{~a} q^d_{~b} \mathcal{B}_{cd} ~.
\label{projdeviationtensor}
\end{align}
Opening up the projection tensors $q^a_{~b} = (\d^a_{~b} + l^a k_b + k^a l_b)$ in \eqref{projdeviationtensor}, a few lines of simple algebra lead us to,
\begin{align}
\hb_{ab} = \mathcal{B}_{ab} + l_b k^d \mathcal{B}_{ad} + l_a k^c \mathcal{B}_{cb} + l_a l_b k^c k^d \mathcal{B}_{cd} + \ti_a k_b - \ti_b k_a + (l_a k_b - l_b k_a) \k_c \ti^c ~.
\label{manip40}
\end{align}
Next, using the relation \eqref{deviationtensor} and \eqref{rotn1forndefn2} in the above, we obtain,
\begin{align}
  \hb_{ab} = \hcvd_b l_a + T_{akb} l^k + (\ti_a k_b - \ti_b k_a) + (l_a k_b - k_a l_b) \ti^c k_c - l_a \hw_b + l_b (k^i \hcvd_i l_a)\nonumber \\
  + (T_{akj} l^k k^j) l_b + l_a T_{ikb} k^i l^k + l_a l_b (T_{ikj} k^i l^k k^j) ~.
  \label{expansionprojdev}
\end{align}
The above relation relates the spacetime covariant derivative of the null normal with the projected deviation tensor. Finally using \eqref{nablaalbexpansion} in \eqref{expansionprojdev}, we end up deriving a relationship between the projected deviation tensor and the extended second fundamental form,
\begin{align}
\thht_{ab} = \hb_{ab} - T_{akb}l^k - (\ti_a k_b - k_a \ti_b) - (l_a k_b - k_a l_b) \ti^c k_c + (T_{aij}k^i l^j) l_b - l_a T_{pqr}k^p l^q q^r_{~b} ~. 
\label{relnprojdevn2ndfundform}
\end{align} 
We can again perform an irreducible decomposition of the projected deviation tensor into a symmetric trace part, a symmetric traceless part and an antisymmetric part,
\begin{equation}
\hb_{ab} = \fr q_{ab} \thbl + \shbl + \whbl ~. 
\label{irrepprojdevtensor}
\end{equation}
Computing the trace of the projected deviation tensor we obtain,
\begin{align}
\thbl &= g^{ab} \hb_{ab} = q^{ab} \hb_{ab} \nonumber \\
 &= (g^{ab} + l^a k^b k^a l^b) \Big[\hcvd_b l_a + T_{acb} l^c\Big] = \hcvd_a l^a - \kappa + T_a l^a ~.
\end{align}
We notice that $\thbl = \thdl$ and hence \textit{the trace of the projected deviation tensor also quantifies the expansion scalar of the null congruence}.

%%%%%%%%%%%%%%%%%%%%%%%%%%%%%%%%%%%%%%%%%%%%%%%%%%%%%%%%%%%%%%%%%%%%%%%%%%%%%%%%%%%%%%%%%%%%%%%%%%%%%%%%%%%%%%%%%%%%%%
\section{Dynamics of the null hypersurface $\hech$ provided by $\hg_{ab} k^a l^b$}\label{dynamicsh}
Now that we have provided the details for the kinematics of the null surface $\hech$, we will begin discussing the dynamics of it. The null generator $\bm{l}$ provides a notion of evolution in time \cite{Gourgoulhon:2005ng} since $l^a = \frac{dx^a}{dt}$ and is hence associated with the time vector $\bm{t}$. By dynamics, we mean to investigate the Lie derivative along $\bm{l}$ of the relevant kinematical quantities introduced in the previous section. Our aim will be focused on one particlar projection component of the Einstein tensor analogue in $\sptm$ i.e. $\hg_{ab}$. This is because we will see later that different projections of $\hat{G}_{ab}$ will lead to dynamics of these kinematical quantities. Let us decompose the vector $\hg^a_{~b} l^b$ into its subsequent components along  the basis vectors $\bm{l}$, $\bm{k}$ and $\bm{e_{{A}}}$ of $\ct_{P}(\mathcal{M})$,
\begin{equation}
	\hg^a_{~b} l^b = \phi_1 l^a + \phi_2 k^a + \phi^{{A}} e^a_{~{A}} ~,
\end{equation}
with $\phi_1 = -\hg_{ab}l^a k^b$, $\phi_2 = -\hg_{ab}l^a l^b$ and $\phi_{\tilde{A}} = (\hg_{ab}l^b q^a_{~c}) e^c_{~{A}}$.
It has been shown in the literature, at least for Riemannian manifolds, that all these three projection components can be provided physical interpretations. The projection component $G_{ab}l^a l^b$ is related to the rate of change of the outgoing expansion scalar along the null congruence $\bm{l}$ and is known as the null Raychaudhuri equation (NRE) \cite{Poisson:2009pwt, Gourgoulhon:2005ng}. On the other hand, $G_{ab}l^a q^b_{~c}$ leads to the Damour-Navier-Stokes equation (for Einstein gravity) under the membrane paradigm \cite{Damour:1979wya, thorne1986black}, whereas $G_{ab}l^a k^b$ leads to a thermodynamic identity (structurally familiar to the first law of thermodynamics) established under a certain physical process involving the null surface $\hech$. It is a worthwhile exercise to extend the formalism to a spacetime $\sptm$ involving torsion and study the physical interpretations (if possible) for such projection components. The case for the NRE corresponding to the null generators $\bm{l}$ in a metric compatible, general affine spacetime $\sptm$  has already been considered in \cite{Dey:2017fld}. However, under the umbrella of this $3+1$ perspective of null foliation we also rederive the NRE for the outgoing expansion scalar $\thdl$ in Appendix \ref{NRE3+1} and compare it with previous results. Here, we will study only the projection component $\hg_{ab}k^a l^b$ and see what interpretation(s) can be alluded to it.

Let us begin with the projection component $\hg_{ab} l^a k^b$. In doing so, we will first derive the evolution equation along $\bm{l}$ of the transversal deformation rate tensor. Then we will proceed to find the evolution rate of the expansion scalar $\thdk$ along the null generators. This will then provide our starting point towards a thermodynamic interpretation that can be attributed to the relevant projection component. At the outset, let us establish the fact that the dynamics of $\hg_{ab}l^a k^b$ will be studied entirely keeping the geodesic constraint \eqref{geodesicconstraint} in mind. We will use the fact that $\ti^a = 0$ in all subsequent analysis. Under the geodesic constraint, as we have discussed at the end of section \ref{kinema1}, that the null generators of $\hech$ are both auto-parallel and geodesic. To proceed, we start with the Ricci identity established on the manifold of the transverse spacelike $2$-surface $\sptq$, where $\bm{\hsvd}$ is the spatial covariant derivative compatible with $\bm{q}$ i.e $\hsvd_a q_{bc} = 0$. So for any spatial vector $\bm{v} ~ \in \ct_P (S_t)$,
\begin{equation}
[\hsvd_a, \hsvd_b] v^a =  \twr_{ab} ~v^a - \twt^d_{~ab} \hsvd_d v^a ~,
\label{RicciidentityonSt}
\end{equation}
where, $\twr_{ab}$  and $\twt^a_{~bc}$ are the $2$-dimensional Ricci and torsion tensors established respectively on $\sptq$. Let us first take on the left-hand side (L.H.S) of \eqref{RicciidentityonSt}. We show (via a detailed calculation in Appendix \ref{appxcommute1}) that, 
\begin{align}
[\hsvd_a, \hsvd_b] v^a = \hsvd_a(\hsvd_b v^a) - \hsvd_b (\hsvd_a v^a)  &= \Big[(\thdl - q^{cd} T_{cfd}l^f)q^m_{~b} -\hchi^m_{~b} + q^c_{~b} q^{dm} K_{fcd}l^f\Big] (k_s \hcvd_m v^s) \nonumber \\
& + \Big[(\thdk - q^{cd}T_{cfd}k^f) q^m_{~b} - \xih^m_{~b} + q^c_{~b} q^{dm} K_{fcd} k^f\Big](l_s \hcvd_m v^s) \nonumber \\
&+ q^m_{~b} q^l_{~s} q^p_{~t} \hat{R}^s_{~plm} v^t + q^c_{~b} q^d_{~s} q^m_{~f} T^f_{~cd} (\hcvd_m v^s)~. 
\label{commut1}
\end{align}

Looking at the R.H.S of \eqref{RicciidentityonSt} we have on account of $\twt^a_{~bc}$ being a tensor defined on the transverse $2$-dimensional space $S_t$,
\begin{align}
&\twr_{ab} ~v^a - \twt^d_{~ab} \hsvd_d v^a = \twr_{ab} v^a + \twt^d_{~ba} q^m_{~d} q^a _{~s} (\hcvd_m v^s) \nonumber \\
& = \twr_{ab} ~v^a + q^c_{~b} q^d_{~a} \twt^f_{~cd} q^m_{~f} q^a_{~s} (\hcvd_m v^s) = \twr_{ab} ~v^a + q^c_{~b} q^d_{~s} q^m_{~f} \twt^f_{~cd} (\hcvd_m v^s) ~.
\label{2t}
\end{align} 
Incidentally, it can be proven that the complete projection of the spacetime torsion onto the $2$-dimensional transverse space $S_t$ is equivalent to the intrinsic torsion established in $\sptq$ i.e.
\begin{equation}
q^c_{~b} q^d_{~s} q^m_{~f} T^f_{~cd} = q^c_{~b} q^d_{~s} q^m_{~f} \twt^f_{cd}~.
\label{2t4t}
\end{equation}
For the detailed derivation of this, see Appendix \ref{2t4tproof}. Upon equating the L.H.S and the R.H.S of \eqref{RicciidentityonSt} via the relations \eqref{commut1} and \eqref{2t} respectively, and using \eqref{2t4t}, we end up having,
\begin{align}
&-\underbrace{\Big[(\thdl - q^{cd} T_{cfd}l^f)q^m_{~b} -\hchi^m_{~b} + q^c_{~b} q^{dm} K_{fcd}l^f\Big](v^a \hcvd_m k_a)}_\text{expression $1$} \nonumber \\
&- \underbrace{\Big[(\thdk - q^{cd}T_{cfd}k^f) q^m_{~b} - \xih^m_{~b} + q^c_{~b} q^{dm} K_{fcd} k^f\Big](v^a \hcvd_m l_a)}_\text{expression $2$} \nonumber \\
&+ q^m_{~b} q^l_{~s} q^p_{~a} \hat{R}^s_{~plm} v^a = \twr_{ab} v^a
\label{manip11}
\end{align}
Let us focus on expression $1$. Again as usual, using \eqref{nablaakbexpansion} for the covariant derivative of the auxiliary null normal and simplifying, we obtain,
\begin{align}
\text{expression $1$} &= v^a \Big[\hchi^m_{~b} ~\xih_{ma} - \Big(\thdl - q^{cd}T_{cfd}l^f\Big)\xih_{ba} - q^c_{~b} \xih^d_{~a} (K_{fcd}l^f) - \hchi^c_{~b} q^d_{~a} (K_{fcd}k^f) \nonumber \\
& + \Big(\thdl - q^{cd}T_{cfd}l^f\Big)q^c_{~b} q^d_{~a} (K_{fcd}k^f) + q^c_{~b} q^{di} q^j_{~a} (K_{fcd}l^f) (K_{hij}k^h)\Big] ~.
\label{line1}
\end{align}
Analogously, for expression $2$, we use \eqref{relnnablaalbnchi} and simplify to yield,
\begin{align}
\text{expression $2$} &= v^a \Big[\xih^m_{~b} ~\hchi_{ma} - \Big(\thdk - q^{cd}T_{cfd}k^f\Big)\hchi_{ba} - q^c_{~b} \hchi^d_{~a} (K_{fcd}k^f) - \xih^i_{~b} q^j_{~a} (K_{hij}l^h) \nonumber \\
& + \Big(\thdk - q^{cd}T_{cfd}k^f\Big)q^i_{~b} q^j_{~a} (K_{hij}l^h) + q^c_{~b} q^{di} q^j_{~a} (K_{fcd}k^f) (K_{hij}l^h)\Big] ~.
\label{line2} 
\end{align}
Adding up both the expressions \eqref{line1} and \eqref{line2} in \eqref{manip11}, we end up having the expression of the $2$-dimensional Ricci tensor in terms of the $4$-dimensional spacetime quantities (since $v^a$ is arbitrary),
\begin{align}
\twr_{ab} &= q^m_{~b} q^l_{~s} q^p_{~a} ~\hat{R}^s_{~plm} + \Big[\Big(\hchi^{m}_{~b} \xih_{ma} + \xih^{m}_{~b} \hchi_{ma}\Big) - \Big(\thdl - q^{ij} T_{ihj}l^h\Big)\Big(\xih_{ba} - q^c_{~b}q^d_{~a} K_{fcd}k^f\Big) \nonumber \\
& -\Big(\thdk - q^{ij} T_{ihj}k^h\Big)\Big(\hchi_{ba} - q^c_{~b}q^d_{~a} K_{fcd}l^f\Big) - \Big(q^c_{~b}\xih^d_{~a} + q^d_{~a} \xih^c_{~b}\Big)K_{fcd}l^f \nonumber \\
& - \Big(q^c_{~b}\hchi^d_{~a} + q^d_{~a} \hchi^c_{~b}\Big)K_{fcd}k^f + \Big(q^c_{~b} q^{di} q^j_{~a} + q^i_{~b} q^{jc} q^d_{~a}\Big) (K_{fcd}l^f)(K_{hij}k^h)\Big]~.
\label{2r4R}
\end{align}
Now, we focus on the term $q^m_{~b} q^l_{~s} q^p_{~a} ~\hat{R}^s_{~plm}$ of \eqref{2r4R}. After some routine computations (see Appendix \ref{appxqqqr}), it can be shown that, 
\begin{align}                                              &q^m_{~b} q^l_{~s} q^p_{~a} ~\hat{R}^s_{~plm} = q^m_{~b} q^p_{~a} \hq_{rmsp} k^r l^s +   q^m_{~b} q^p_{~a} \hr_{pm} + 2 \hhw_a \hhw_b  - (\hhw_a \hp_b + \hp_a \hhw_b) \nonumber \\
& -q^m_{~b} q^p_{~a} l^r \hcvd_r \Big[2\xih_{mp} - \Big(q^i_{~m} q^j_{~p} + q^i_{~p} q^j_{~m}\Big)(K_{hij}k^h)\Big]  + q^m_{~b} q^p_{~a} \Big[\Big(\hcvd_m \hw_p + \hcvd_p \hw_m\Big) - \Big(\hcvd_m \hp_p + \hcvd_p \hp_m \Big)\Big]  \nonumber \\
& - \Big(\hchi^r_{~b} \xih_{ra} + \hchi^r_{~a} \xih_{rb}\Big) + \Big(q^j_{~a} \hchi^i_{~b} + q^j_{~b} \hchi^i_{~a}\Big) (K_{hij} k^h) + \Big(q^c_{~a} \xih^d_{~b} + q^c_{~b} \xih^d_{~a}\Big)(K_{fcd}l^f)  \nonumber \\
& -\Big(\xih^c_{~a}q^d_{~b} + \xih^c_{~b} q^d_{~a}\Big)(T_{cfd}l^f) - \Big[q^c_{~b} q^j_{~a} q^{di} + q^c_{~a} q^j_{~b} q^{di}\Big] (K_{fcd}l^f)(K_{hij}k^h) \nonumber \\
&+ \Big[q^j_{~a} q^d_{~b}q^{ci} +q^j_{~b}q^d_{~a}q^{ci}\Big](T_{cfd}l^f)(K_{hij}k^h) ~.
\label{beginningfortrace}   
\end{align}
In the above, we have defined the quantity $\hp_a$ accordingly as,
\begin{equation}
\hp_a \equiv T_{bcd} k^b l^c q^d_{~a} ~.
\end{equation} 
Obviously, the quantity $\hp_a$ is orthogonal to both $l^a $ and $k^a$ and hence is defined on the $2$-surface $S_t$. Converting the spacetime covariant derivatives present in the above relation \eqref{beginningfortrace} to Lie derivatives along the null generator $\bm{l}$, we obtain from the above,
\begin{align}
&q^m_{~b} q^l_{~s} q^p_{~a} ~\hat{R}^s_{~plm} = q^m_{~b} q^p_{~a} \hq_{rmsp} k^r l^s +   q^m_{~b} q^p_{~a} \hr_{pm} + 2 \hhw_a \hhw_b  - (\hhw_a \hp_b + \hp_a \hhw_b) \nonumber \\&- q^m_{~b} q^p_{~a} \pounds_{\bm{l}} \Big[2 \xih_{mp} - \Big(q^i_{~m} q^j_{~p} + q^i_{~p} q^j_{~m}) (K_{hij} k^h) \Big] + \Big(\hchi_{ar} \xih^r_{~b} + \hchi_{br} \xih^r_{~a}\Big) \nonumber \\
& + \hsvd_a (\hhw_b -\hp_b) + \hsvd_b (\hhw_a - \hp_a) - 2 \kappa \xih_{ab} + \Big(\xih^c_{~a} q^d_{~b} + \xih^c_{~b} q^d_{~a}\Big)(T_{cfd}l^f) \nonumber \\
& - \Big(q^c_{~a} \xih^d_{~b} + q^c_{~b} \xih^d_{~a}\Big) (K_{fcd}l^f) + \Big[q^i_{~a}(\kappa q^j_{~b} - \hchi^j_{~b}) + q^i_{~b}(\kappa q^j_{~a} - \hchi^j_{~a})\Big](K_{hij}k^h) \nonumber \\
& - \Big[\Big(q^i_{~a}q^d_{~b} + q^i_{~b} q^d_{~a}\Big)q^{jc}\Big](T_{cfd}l^f)(K_{hij}k^h) + \Big[\Big(q^c_{~a} q^i_{~b} + q^c_{~b}q^i_{~a}\Big)q^{jd}\Big](K_{fcd}l^f)(K_{hij}k^h) ~.
\label{qqqR}
\end{align}
We relegate the details of the computation to Appendix \ref{appxqqqr}. 
%%%%%%%%%%%%%%%%%%%%%%%%%%%%%%%%%%%%%%%%%%%%%%%%%%%%%%%%%%%%%%%%%%%%%%%%%%%%%%%%%%%%%%%%%%%%%%%%%%%%%%%%%%%%%%%%%%%%%%
Now all that remains to do is to put \eqref{qqqR} into \eqref{2r4R} to get to the evolution equation of the transversal deformation rate tensor,
\begin{align}
&q^m_{~b} q^p_{~a} \pounds_{\bm{l}} \xih_{mp} - \fr q^m_{~b} q^p_{~a} \pounds_{\bm{l}}\Big[\Big(q^i_{m}q^j_{~p}+ q^i_{~p} q^j_{~m}\Big)(K_{hij}k^h)\Big] =  -\fr \hr_{ab} +\fr q^m_{~b} q^p_{~a} \hq_{rmsp} k^r l^s \nonumber \\
& \fr q^m_{~b} q^p_{~a} \hr_{pm} + \Big(\hchi_{ar} \xih^r_{~b} + \hchi_{br} \xih^r_{~a}\Big) + \fr \Big(\hsvd_a(\hhw_b - \hp_b) + \hsvd_b( \hhw_a - \hp_a)\Big) - \Big(\kappa + \frac{\thdl}{2}\Big) \xih_{ab} \nonumber \\
& - \frac{\thdk}{2} \hchi_{ab} + \hhw_a \hhw_b - \fr(\hhw_a \hp_b + \hhw_b \hp_a) + \fr \Big[(\xih^c_{~a} q^d_{~b} + \xih^c_{~b} q^d_{~a}) + q^{cd} \xih_{ab}\Big](T_{cfd}l^f) \nonumber \\
& +\fr \Big[\thdk~ q^c_{~b} q^d_{~a} - q^c_{~a} \xih^d_{~b} - q^c_{~b} \xih^d_{~a} - q^c_{~b} \xih^d_{~a} -q^d_{~a} \xih^c_{~b}\Big](K_{fcd}l^f) \nonumber \\
&+ \fr \Big[q^i_{~a}\Big(\kappa q^j_{~b} - \hchi^j_{~b}\Big) + q^i_{~b}\Big(\kappa q^j_{~a} - \hchi^j_{~a}\Big) + \thdl  q^i_{~b} q^j_{~a} - \hchi^i_{~b} q^j_{~a} - \hchi^j_{~a} q^i_{~b} + q^{ij} \hchi_{ab}\Big](K_{hij}k^h) \nonumber \\
&- \fr \Big[\Big(q^i_{~a} q^d_{~b} + q^i_{~b} q^d_{~a}\Big)q^{jc}+ q^{cd} q^i_{~b} q^j_{~a}\Big](T_{cfd}l^f)(K_{hij}k^h) - \fr \Big(q^{ij}q^c_{~b}q^d_{~a}\Big)(T_{ihj}k^h)(K_{fcd}l^f) \nonumber \\
& + \fr \Big[\Big(q^c_{~a} q^i_{~b} + q^c_{~b} q^i_{~a}\Big)q^{jd} + q^c_{~b} q^{di} q^j_{~a} + q^i_{~b} q^{jc} q^d_{~a}\Big](K_{fcd}l^f)(K_{hij}k^h) ~.
\label{eqntransdeformrate}
\end{align}
So we have arrived at the evolution equation for the transversal deformation rate tensor in the metric compatible general affine spacetime $\sptm$ under the geodesic constraint ($\ti^a = 0$) i.e the null geodesic generators of $\hech$ are parallel transported along themselves. In the absence of torsion in the spacetime this equation matches with Eq. $(6.43)$ of \cite{Gourgoulhon:2005ng}. 

Now, that we have derived the evolution equation of the transversal deformation rate tensor, we proceed to study the evolution of the expansion scalar $\thdk$. This would enable us to provide the thermodynamic relation established on $\hech$. To get to the evolution equation for $\thdk$, all we need to is to take the trace of the evolution equation of the transversal deformation rate tensor \eqref{eqntransdeformrate}. However this can be avoided to the benefit of a shorter route that involves taking the trace of \eqref{2r4R} and \eqref{beginningfortrace} and then combining them. The result of the trace computation leads us to,
\begin{align}
- \k \Big(\thdk - q^{ij}T_{ihj}k^h\Big) = \Big[\fr \twor + l^r \hcvd_r \bthk - \hhw_a \hhw^a + \hhw_a \hp^a - \hsvd_a (\hhw^a- \hp^a)\nonumber \\
+ \thdl \bthk - \Big(\thdk q^{cd} - \xih^{cd}\Big)(T_{cfd}l^f)- \Big(q^{dj}q^{ci} - q^{cd}q^{ij}\Big)(T_{cbd})(K_{aij}) k^a l^b\Big] \nonumber \\
- \Big[\hat{G}_{ab}+(\hcvd_a T_b- \hcvd_b T_a) + (\hcvd_i + T_i)T^i_{~ab}\Big]k^a l^b ~.
\label{tracerablakb1}
\end{align}
All the relevant steps involved in the computation for the trace leading up to \eqref{tracerablakb1} has been shown in Appendix \ref{appxralakb1}. The above geometrical relation involves the directional derivative of the ingoing expansion scalar $\thdk$ along the null generator $\bm{l}$ being related to the component $\hg_{ab}k^a l^b$. It is in this sense that the above equation can be referred to as the NRE for $\thdk$. Let us try to motivate the reason of this particular choice of arranging the terms in \eqref{tracerablakb1}. This will be clearer when we proceed to provide a thermodynamic interpretation to the above equation. Notice that all the terms in the first squared parentheses for the R.H.S of \eqref{tracerablakb1} except for $l^r \hcvd_r \bthk$ contains geometrical quantities that are defined on the transverse $2$-surface $S_t$. The terms in the second squared parentheses for the R.H.S of \eqref{tracerablakb1} involves rather quantities defined for the spacetime $\sptm$ and are not restricted to $S_t$.
%%%%%%%%%%%%%%%%%%%%%%%%%%%%%%%%%%%%%%%%%%%%%%%%%%%%%%%%%%%%%%%%%%%%%%%%%%%%%%%%%%%%%%%%%%%%%%%%%%%%%%%%%%%%%%%%%%%%%%
\section{Thermodynamic interpretation provided to the NRE \eqref{tracerablakb1} via virtual displacement $\dlk$}\label{section6}
The relation \eqref{tracerablakb1} is geometrical in the sense that the dynamics of the gravitational field equations have not made their way into the relationship. 
At this junction, we have to use the ECKS field equation  corresponding to the metric tensor. We will rather use the form given in \eqref{manip37}. Use of this in \eqref{tracerablakb1}, we have,
\begin{align}
- \k \Big(\thdk - q^{ij}T_{ihj}k^h\Big) = \Big[\fr \twor + l^r \hcvd_r \bthk - \hhw_a \hhw^a + \hhw_a \hp^a - \hsvd_a (\hhw^a- \hp^a)\nonumber \\
+ \thdl \bthk - \Big(\thdk q^{cd} - \xih^{cd}\Big)(T_{cfd}l^f) - \Big(q^{dj}q^{ci} - q^{cd}q^{ij}\Big)(T_{cbd})(K_{aij})k^a l^b\Big] \nonumber \\
- \Big[8 \pi \mt_{ab} + \fr (\hcvd_i + T_i) \Big(3 S^i_{~ab} + S_{a~b}^{~i} + S_{b~a}^{~i}\Big)\Big]k^a l^b~.
\label{thermostart1}
\end{align}
So finally we have arrived at the equation we desired. This equation has the interpretation of a dynamical equation governing the evolution of the expansion scalar $\thdk$ along the null auto-parallel (and hence geodesic) generators of an integrable hypersurface $\hech$ in the spacetime $\sptm$.

Before proceeding to interpret \eqref{thermostart1} as a thermodynamic identity established on the generic null surface $\hech$, it is necessary to convince ourselves that $\thdk$ indeed represents the expansion scalar of the auxiliary null field $\bm{k}$. Let us represent the (in general) non-affine parameter along the auxiliary null vector field to be $\lambda_{(k)}$. Hence we have $k^a = - \frac{d x^a}{ d \lk}$. The crucial negative sign is because of the fact that the auxiliary null field $\bm{k}$ is ingoing as opposed to the null gnerators $\bm{l}$ which are outgoing. On account of the determinant of the transverse metric of the $2$-surface $S_t$ being a scalar density, we have,
\begin{align}
\frac{d \sqrt{q}}{d \lk}  &= \fr \sqrt{q} q^{AB} \frac{d}{d \lk} q_{AB} = -\fr \sqrt{q} q^{AB} k^i \hcvd_i \Big(g_{ab} e^a_{~A} e^b_{~B}\Big).  \nonumber
\end{align}
Therefore one finds,
\begin{align}
-\frac{1}{\sqrt{q}} \frac{d}{d \lk} \sqrt{q} &= \fr q^{AB} g_{ab} e^{b}_{~B} (k^i \hcvd_i e^a_{~A}) + \fr q^{AB} g_{ab} e^a_{~A} (k^i \hcvd_i e^b_{~B}) = q^{AB} g_{ab} e^b_{~B} (k^i \hcvd_i e^a_{~A}) ~.
\label{manip19}
\end{align}
Under the construction of the null hypersurface $\hech$, the basis vectors $\{\bm{e_{A}}\}$ of the tangent space established on the $2$-surface $S_t$ are Lie transported along the auxiliary null field i.e. $[\bm{k}, \bm{e_{A}}] = 0$. This results in,
\begin{equation}
k^i \hcvd_i e^a_{~A} = e^i_{~A} \hcvd_i k^a + T^a_{~bc} k^b e^c_{~A} ~.
\label{manip20}
\end{equation}
Using \eqref{manip20} in \eqref{manip19}, we obtain,
\begin{align}
-\frac{1}{\sqrt{q}} \frac{d}{d \lk} \sqrt{q} &= q^{AB} g_{ab} e^b_{~B} \Big(e^i_{~A} (\hcvd_i k^a) + T^a_{~dc} k^d\ e^c_{~A}\Big) = q^{ab} \hcvd_a k_b + q^{ab} T_{acb} k^b ~.
\label{manip21}
\end{align}
In the above, we have used the fact that $q^{AB} e^a_{~A} e^b_{~B} = q^{ab}$. Upon using the relation \eqref{nablaakbexpansion} in \eqref{manip21}, we obtain,
\begin{equation}
	-\frac{1}{\sqrt{q}} \frac{d}{d \lk} \sqrt{q} = q^{ab} \xih_{ab} - q^{cd} K_{fcd} k^f + q^{ab} T_{acb}k^c = \thdk
	\label{thdk1}
\end{equation}
Hence we indeed verify that the ingoing expansion scalar $\thdk$ represents the fractional rate of change of the $2$-surface area element $\sqrt{q}$ along the auxiliary null vector field $\bm{k}$. Strictly along the lines of the previous analysis, its quite straightforward to establish that,
\begin{equation}
\frac{1}{\sqrt{q}} \frac{d}{d \ll} \sqrt{q} = q^{ab} \hchi_{ab}= \thdl ~,
\label{thdl1}
\end{equation}
where $\ll = t$ represents the non-affine parameter for the outgoing null generators $\bm{l}$.

Finally, we arrive at the point where we discuss the physical process under which a thermodynamic interpretation can be alluded. The physical process is a virtual displacement $\delta\lambda_{(k)}$  along the auxiliary null vector field. The notion of virtual displacement has been adopted from the analysis in \cite{Chakraborty:2015aja}. The virtual displacement basically shifts our null hypersurface $\hech$ along $\bm{k}$. Consider the foliation of $\sptm$ in the neighbourbood of $\hech$ by the null family $\hech_u$. Let us suppose that $\hech$ is stationed at the value of $\lambda_{(k)} = 0$ and in the null family, there exists another surface at the value of $\lambda_{(k)} = \delta \lambda_{(k)}$. Of course, both of these null surfaces are solutions of the Einstein-Cartan spacetime. The virtual displacement is the physical process that shifts us from the null surface at $\lambda_{(k)} = 0$ to $\lambda_{(k)} = \delta \lambda_{(k)}$. Let us multiply then both sides of \eqref{thermostart1} with $\dlk$ along with a multiplicative factor of $\frac{1}{8 \pi}$. We integrate the resulting equation on the $2$-dimensional spacelike cross-section $S_t$ of $\hech$.
This results in,
\begin{align}
&-\int_{S_t} d^2 x \sqrt{q} \Big[\frac{\k}{2 \pi} \Big(\frac{1}{4} \thdk -\frac{1}{4} q^{ij}T_{ihj}k^h \Big)\Big] \dlk \nonumber \\
& = \intsd \frac{1}{8 \pi} \Big[\fr {^{(2)}}\hr + l^r \hcvd_r \Big(\thdk - q^{ij}T_{ihj}k^h\Big) -\hhw_a \hhw^a + \hhw_a \hp^a - \hsvd_a \Big(\hhw^a - \hp^a\Big) \nonumber \\
& + \thdl \Big(\thdk - q^{ij}T_{ihj}k^h\Big) - \Big(\thdk q^{cd} - \xih^{cd}\Big)(T_{fd}l^f) - \Big(q^{dj}q^{ci} - q^{cd}q^{ij}\Big)(T_{cbd})(K_{aij})k^a l^b\Big] \dlk \nonumber \\
& -\intsd \frac{1}{8 \pi} \Big[8 \pi \mt_{ab} + \fr (\hcvd_i + T_i) \Big(3 S^i_{~ab} + S_{a~b}^{~i} + S_{b~a}^{~i}\Big) \Big]k^a l^b \dlk
\label{thermiden1}
\end{align}
Let us now focus on the term in the L.H.S of \eqref{thermiden1}. We can rewrite it as,
\begin{align}
&-\int_{S_t} d^2 x \sqrt{q} \Big[\frac{\k}{2 \pi} \Big(\frac{1}{4} \thdk -\frac{1}{4} q^{ij}T_{ihj}k^h \Big)\Big] \dlk \nonumber \\
&= \ints d^2 x \frac{\k}{2 \pi} \Big[\sqrt{q} \frac{1}{\sqrt{q}} \frac{d}{d \lambda_{(k)}} \Big(\frac{\sqrt{q}}{4}\Big) + \sqrt{q}\frac{1}{4} q^{ij}T_{ihj}k^h\Big] \dlk \nonumber \\
& = \ints d^2 x \frac{\k}{2 \pi} \Big[\dlk \frac{d}{d \lambda_{(k)}}\Big(\frac{\sqrt{q}}{4}\Big) + \dlk\Big(\frac{\sqrt{q}}{4} q^{ij}T_{ihj}k^h \Big)\Big]  = \ints d^2 x~ T \Big(\delta_{\lambda_{(k)}} \snu +  \dellk\stor\Big) ~.
\label{entropy}
\end{align}
Here, we identify the temperature associated with the null surface under the process of virtual displacement $\dlk$ to be $T = (\k /2 \pi)$. We postulate that the variation of the total entropy density occurs from two contributions. First is the entropy generation term of the null surface itself. The entropy density of the null surface $\hech$ is proportional to the area element $\sqrt{q}$ of the $2$-surface $S_t$ i.e. $s_{\text{null}} = \sqrt{q}/4$ {\footnote{Same identification of entropy has been done in \cite{Chakraborty:2018qew} through Noether prescription on a Killing horizon.}}. This part of the entropy generation is purely due to the variation of cross-sectional transverse area elements $S_t$ as we move in the transverse $\bm{k}$ direction under the virtual displacement. But this not the end of the story. Due to the presence of non-trivial torsion in the spacetime, there happens to be another entropy generation term $ \dellk\stor$. In order to have an understanding for the source of it, let us consider the following particular torsion current $T_{i~~j}^{~h} q^{ij}$. Obviously the quantity $q^{ij}T_{ihj}k^h$ is negative the component of this torsion current along the null generators $\bm{l}$. We here  define the entropy variation under the virtual displacement $\dlk$ due to presence of this non-trivial torsion current $T_{i~~j}^{~h} q^{ij}$ to be $\dellk \stor$:
\begin{equation}
\dellk \stor = \dlk\Big(\frac{\sqrt{q}}{4} q^{ij}T_{ihj}k^h \Big) ~.\label{entropytorsioncurrent}
\end{equation}
Thus we see that there exists  two causes of entropy generation under the virtual displacement $\dlk$. One arises primarily due to the variation of the transverse cross-sectional area element $S_t$. The other arises due to a non-trivial torsion current flowing along the null generators. We will have something more to say on this at the end of this section.

Having done this, let us now look at the first term in the R.H.S of \eqref{thermiden1}. We identify this term to be the variation of energy $\dellk E$ associated with the physical process of virtual displacement $\dlk$,
\begin{align}
\dellk E = \intsd ~ \dlk \frac{1}{8 \pi} \Big[\fr {^{(2)}}\hr + l^r \hcvd_r \Big(\thdk - q^{ij}T_{ihj}k^h\Big) -\hhw_a \hhw^a + \hhw_a \hp^a - \hsvd_a \Big(\hhw^a - \hp^a\Big) \nonumber \\
 + \thdl \Big(\thdk - q^{ij}T_{ihj}k^h\Big) - \Big(\thdk q^{cd} - \xih^{cd}\Big)(T_{cfd}l^f) - \Big(q^{dj}q^{ci} - q^{cd}q^{ij}\Big)(T_{cbd})(K_{aij})k^a l^b\Big]~.
 \label{varenergy}
\end{align}
We can in principle perform an integration over the non-affine parameter $\lambda_{(k)}$ of the auxiliary null field to provide an expression of the energy associated with the two surface $S_t$,
\begin{align}
E = \int d \lambda_{(k)} \intsd \frac{1}{8 \pi} \Big[\fr {^{(2)}}\hr + l^r \hcvd_r \Big(\thdk - q^{ij}T_{ihj}k^h\Big) -\hhw_a \hhw^a + \hhw_a \hp^a - \hsvd_a \Big(\hhw^a - \hp^a\Big) \nonumber \\
+ \thdl \Big(\thdk - q^{ij}T_{ihj}k^h\Big) - \Big(\thdk q^{cd} - \xih^{cd}\Big)(T_{cfd}l^f) - \Big(q^{dj}q^{ci} - q^{cd}q^{ij}\Big)(T_{cbd})(K_{aij})k^a l^b\Big]~.
\label{energy}
\end{align}
Let us reiterate that our aim is to provide a thermodynamic interpretation to the NRE (corresponding to the auxiliary null field $\bm{k}$) in analogy with the first law of thermodynamics. That would be complete, if we have the liberty to interpret the following expression to be the pressure term $P$,
\begin{align}
P \equiv -\frac{1}{8 \pi} \Big[8 \pi \mt_{ab} + \fr (\hcvd_i + T_i) \Big(3 S^i_{~ab} + S_{a~b}^{~i} + S_{b~a}^{~i}\Big)\Big]k^a l^b ~.
\label{pressure}
\end{align}
The force $F$ conjugate to the physical process of virtual displacement $\dlk$ is simply then the integral of the pressure term over the transverse surface $S_t$, 
\begin{align}
F = \intsd ~ P ~.\label{force}
\end{align} 
Now once this interpretation is allowed (we will try to justify this shortly), the process of virtual displacement of the null surface $\hech$ along the auxiliary null field described via \eqref{thermiden1} can be succinctly restated as, 
\begin{align}
\ints d^2 x~ T \Big(\delta_{\lambda_{(k)}} \snu +  \dellk\stor\Big) = \dellk E + F \dlk ~.
\label{therminterpret1}
\end{align}
The above interpretation is made possible only under a virtual displacement of the null hypersurface $\hech$ in the transverse $\bm{k}$ direction. For details about the process of virtual displacement see \cite{Chakraborty:2015aja}. The virtual displacement is to be thought of as a physical process that ``virtually'' shifts the position of $\hech$ from stationed at $\lambda_{(k)} = 0$ to the position at say $\lambda_{(k)} = \d \lambda_{(k)}$. The virtual work done under this process is $F \dlk$. As a result of this, an amount of energy $\dellk E$ sweeps through the null surface. The corresponding change in the heat energy is $\ints d^2 x  T (\delta_{\lambda_{(k)}} \snu +  \dellk\stor)$.

Let us now describe the motivation behind the pressure term \eqref{pressure}. The pressure term contains  the term $-\mt_{ab}k^a l^b$. In the case of Einstein and Lanczos-Lovelock gravity, this particular term has been consistently identified as the pressure under the process of virtual displacement \cite{Chakraborty:2015aja,Kothawala:2010bf, Chakraborty:2015wma}. For static spherically symmetric spacetimes, this particular term has the value $- \mt_{ab} k^a l^b = T^r_{~r}$, which has the interpretation of being the radial or the normal pressure\cite{Hayward:1997jp, Kothawala:2010bf}. However, when dealing with the spacetime $\sptm$, we see that there are necessarily extra terms in the pressure. Notice that there are quadratic terms involving the torsion [and hence the modified torsion which can then be related to the spin angular momentum tensor via the field equation \eqref{spindensity}]. For example consider the following term in the pressure,
\begin{align}
-\frac{1}{8 \pi} \fr T_i \Big(3 S^i_{~ab} + S_{a~b}^{~i} + S_{b~a}^{~i}\Big)k^a l^b = -8 \pi \frac{1}{4} g^{ac} \tau_{aci} \Big(3 \tau^i_{~ab} + \tau_{a~b}^{~i} + \tau_{b~a}^{~i}\Big)k^a l^b ~.
\label{manip35}
\end{align} 
In arriving at the above relation, we have used \eqref{spindensity} and the fact that $T_i = \fr g^{ac} S_{aci} = \fr S^a_{~ai}$. Such quadratic terms in the spin tensor actually represent spin-spin contact interaction and hence produce a correction to the matter energy-momentum tensor \cite{Poplawski:2009fb}. Our definition of the pressure involves such spin-spin interaction terms in addition to the matter energy-momentum tensor. In addition to the energy-momentun tensor and the spin-spin contact interaction terms we also have a derivative of modified torsion tensors in the pressure term i.e $- \frac{1}{8 \pi} \fr \hcvd_i (3 S^i_{~ab} + S_{a~b}^{~i}+ S_{b~a}^{~i})k^a l^b = -\fr \hcvd_i (\tau^i_{~ab} + \tau_{a~b}^{~i} + \tau_{b~a}^{~i})k^a l^b$. In \cite{Dey:2020tkj}, while analysing the thermodynamic interpretation provided to a generic null surface (under virtual displacement) in general spacetimes without any torsion, the authors described the notion of a ``gravitational pressure'' defined as $P = -\frac{1}{8 \pi} G_{ab}k^a l^b$. Hence for a generic null surface in Einstein gravity $P = -\frac{1}{8 \pi} G_{ab}k^a l^b = -\mt_{ab}k^a l^b$ i.e the field equation gives rise to the pressure term. In the same spirit, we identify the pressure as $-\frac{1}{8 \pi}[\hg_{ab}+ (\hcvd_a T_b - \hcvd_b T_a) + (\hcvd_i + T_i)T^i_{~ab}]k^a l^b$. Once the  ECKS equation \eqref{manip37} is used on this it actually reduces to the value of the pressure \eqref{pressure}. In addition to the above motivation, there lies another reason behind the (not so obvious) definition of energy term \eqref{varenergy} and the work function \eqref{pressure} under the virtual displacement. As already mentioned in the end of previous section, we have partitioned the NRE (for $\thdk$) \eqref{tracerablakb1} in such a way, that the energy contribution arises entirely from geometrical quantities defined on the transverse submanifold $S_t$ (in addition to the scalar field $l^r \hcvd_r \Big(\thdk - q^{ij}T_{ihj}k^h\Big)$). Contrary to this, the pressure term (leading to the work function) is entirely from quantities defined in the manifold $\sptm$. In fact, it has been explicitly shown \cite{Dey:2020tkj} that at least for Einstein gravity (with zero torsion), the covaraint expression of the energy \eqref{energy} reduces to expressions of energy for well known spacetimes. For example, the computation of \eqref{energy} for the Scharzschild metric gives us the mass term. It is in this spirit, that the natural generalization of energy term for a generic $\hech$ in RC spacetime follows.
%%%%%%%%%%%%%%%%%%%%%%%%%%%%%%%%%%%%%%%%%%%%%%%%%%%%%%%%%%%%%%%%%%%%%%%%%%%%%%%%%%%%%%%%%%%%%%%%%%%%%%%%%%%%%%%%%%%%%%
\subsection{Case of completely antisymmetric torsion :}
Let us now come to the important specific case of the torsion being completely antisymmetric. Applications of completely antisymmetric torsion tensor have been discussed in string and superstring theories \cite{Fabbri:2006xq}. For the case of a string inspired gravitational theory, the Kalb-Ramond field is identified with a completely antisymmetric torsion background \cite{SenGupta:2001cs}. In the case of completely antisymmetric torsion, the expressions in our thermodynamic analysis simplify significantly. Firstly, the geodesic constraint \eqref{geodesicconstraint} no longer needs to be assumed but rather is a consequence of total antisymmetry of the torsion tensor. Let us focus on the NRE corresponding to the ingoing auxiliary null vector field $\bm{k}$ i.e. \eqref{thermostart1}. The expression simplifies to, 
\begin{align}
&- \k \thdk  = \Big[\fr \twor + l^r \hcvd_r \thdk - \hhw_a \hhw^a + \hhw_a \hp^a - \hsvd_a (\hhw^a- \hp^a) + \thdl \thdk - \fr q^{dj}q^{ci}S_{cbd}S_{aij} k^a l^b\Big] \nonumber \\
&- \Big[8 \pi \mt_{ab} + \fr \hcvd_i \Big(3 S^i_{~ab} \Big)\Big]k^a l^b~.
\label{thermostart2}
\end{align}
In the above, we have used the fact that for completely antisymmetric torsion, $S_{abc} = T_{abc}$ and $K_{abc} = \fr T_{abc}$. Proceeding ahead with the process of virtual displacement, we can attest the thermodynamic interpretation to this specific case as well. The heat energy associated with the process now is, 
\begin{align}
\ints d^2 x T \dellk \snu ~,
\end{align}
where $\snu = \frac{\sqrt{q}}{4}$. Obviously, the entropy generation term $\dellk \stor$ due to the torsion current component $q^{ij}T_{ihj}k^h$ flowing along the null generators $\bm{l}$ of $\hech$ is zero owing to the total antisymmetry of torsion. Hence under the virtual displacement process the only change in the entropy density occurs via the change in the transverse area element $\sqrt{q}$ of $\hech$. The amount of energy flow along the null hypersurface under such considerations is, 
\begin{align}
\dellk E = \intsd ~ \dlk \frac{1}{8 \pi} \Big[\fr {^{(2)}}\hr + l^r \hcvd_r \thdk -\hhw_a \hhw^a + \hhw_a \hp^a - \hsvd_a \Big(\hhw^a - \hp^a\Big)+ \thdl\thdk \nonumber \\
- \fr q^{dj}q^{ci}S_{cbd}S_{aij} k^a l^b \Big]~.
\label{varenergy1}
\end{align}
The corresponding identification of the pressure term under such a process in the case of totally antisymmetric torsion tensor is,
\begin{align}
P = -\frac{1}{8 \pi} \Big[8 \pi \mt_{ab} + \fr \hcvd_i \Big(3 S^i_{~ab} \Big)\Big]k^a l^b ~.
\label{pressure1}
\end{align}%%%%%%%%%%%%%%%%%%%%%%%%%%%%%%%%%%%%%%%%%%%%%%%%%%%%%%%%%%%%%%%%%%%%%%%%%%%%%%%%%%%%%%%%%%%%%%%%%%%%%%%%%%%%%%%%%%%%%%
\subsection{Notion of \textit{equilibrium} for the null surface $\hech$ in $\sptm$}
Now, let us discuss the case of an equilibrium null hypersurface $\hech_{\text{eq}}$, i.e. we want a truly stationary description of our null surface/horizon. 
First of all, we would require our theory to have non-propagating torsion. The Einstein-Cartan action $\aec$ is given by \eqref{totalaction},
\begin{align}
\aec = \frac{1}{16 \pi} \int_{\mathcal{V}} \sqrt{-g}~ \hr = \frac{1}{16 \pi} \int_{\mathcal{V}} \sqrt{-g} \Big[R + 2 \nabla_i T^i - T^a T_a + K^{imj}K_{mij}\Big] ~.
\end{align}
We see that this gravitational action does not contain second derivatives of the torsion term. Hence, in such theories, the torsion field itself does not propagate. 
However, the torsion can indirectly propagate through some other field with which it is coupled. For instance, here the torsion is carried by the propagation of $g_{ab}$.
%For instance, in the case of torsion coupling to photons via vacuum polarization has been studied in \cite{HariDass:2001dp}. 
In principle, whatever be the case, for a truly stationary description of our null hypersurface, we would require any torsion current flowing along the null surface $\hech$ to be zero. The first among such a non-trivial torsion current is $T_{abc}l^a l^c$. Setting this to zero implies our geodesic constraint \eqref{geodesicconstraint}. Infact, when considering the case of a Killing horizon (a stationary equilibrium description of the horizon), such a torsion current needs to be eliminated for removing inequivalent definitions of surface gravity \cite{Dey:2017fld}. The second among such torsion current that we need to consider for our purposes is $T_{ihj} q^{ij}$. The component of this torsion current flowing along the null surface (i.e along the null generators $\bm{l}$ is precisely $q^{ij}T_{ihj}k^h$. We should demand for this component to vanish in order to have a stationary description of the null surface/horizon. Infact, the authors of \cite{Dey:2017fld} have shown that only the geodesic constraint \eqref{geodesicconstraint} is required to prove the zeroth law of black-hole mechanics for a Killing horizon established in $\sptm$. They do not demand specifically the requirement that $q^{ij}T_{ihj}k^h$ be zero as well for the Killing horizon. In order to prove the zeroth law, the authors consider the specific case of a Killing horizon having a bifurfaction $2$-surface. However, not all stationary horizons have a bifurcation $2$-surface. Here we postulate that for a true stationary and hence equilibrium notion of a Kiling horizon, we require both the conditions $T_{abc}l^a l^c = 0$ and $q^{ij}T_{ijh}k^h = 0$ to be simultaneously implemented. These two constraints represent our equilibrium conditions. For such a Killing horizon established in the spacetime $\sptm$, the surface gravity and hence the temperature is constant over the horizon. Moreover the outgoing expansion scalar $\thdl$ of the null generators vanish by definition for a Killing horizon. Now if we perform a virtual displacement process for such a Killing horizon, then the thermodynamic interpretation becomes quite clear. The variation of the entropy due to the torsion term i.e $\dellk \stor$ is by default zero under our equilibrium conditions. Since the temperature is constant over the Killing horizon, while considering \eqref{therminterpret1}, we can take $T$ outside the integral. We then identify the total change of the entropy $S_{\text{null}}$ of the null surface (Killing horizon) to be $\dellk S_{\text{null}} = \ints d^2 x ~\dellk \snu$. We then finally have the thermodynamic interpretation established on the Killing horizon in $\sptm$ under the virtual displacement $\dlk$ to be,
\begin{align}
T \dellk S_{\text{null}} = \dellk E + F \dlk ~.
\label{Killingtherm}
\end{align}  
The variation of the energy term is, 
\begin{align}
\dellk E = \intsd ~ \dlk \frac{1}{8 \pi} \Big[\fr {^{(2)}}\hr + l^r \hcvd_r \thdk  -\hhw_a \hhw^a + \hhw_a \hp^a - \hsvd_a \Big(\hhw^a - \hp^a\Big) \nonumber \\
 - \Big(\thdk q^{cd} - \xih^{cd}\Big)(T_{cfd}l^f) -q^{dj}q^{ci}(T_{cbd})(K_{aij})k^a l^b\Big]~.
\end{align}
Similarly the pressure term for the virtual displacement of the Killing horizon is,
\begin{align}
P \equiv -\frac{1}{8 \pi} \Big[8 \pi \mt_{ab} + \fr (\hcvd_i + T_i) \Big(3 S^i_{~ab} + S_{a~b}^{~i} + S_{b~a}^{~i}\Big)\Big]k^a l^b ~.
\end{align}

Now, having discussed the physical interpretation of the thermodynamic identity as applied to a generic hypersurface-orthogonal null surface (satisfying the geodesic constraint) in the spacetime $\sptm$ and its relevant specifications to the case of completely antisymmetric torsion and the equilibrium case, we delve a little bit more into the possible origins of the total entropy variation term. In this regard it helps to compare our results with the interpretation provided in \cite{Dey:2017fld}. In this paper, the authors in the context of a local causal horizon established in the Riemann-Cartan spacetime $\sptm$ assume an area-entropy law, where they propose that the variation of the entropy is proportional to the variation of the horizon cross-section [see Eq. $(69)$ of \cite{Dey:2017fld}]. However, as we have seen in our case, w.r.t \eqref{entropy}, that the total variation of the entropy density is due to the sum of two contributions. One is due to the variation of the null surface/horizon cross section under the virtual displacement $\dlk$. The other is the entropy generation term $\eqref{entropytorsioncurrent}$ due to the non-zero torsion current $q^{ij}T_{ihj}$. One can surely think as to why there is no such entropy generation term due to the torsion current in the process involving the local causal horizons described in \cite{Dey:2017fld}. To our understanding, this stems from the difference in the processes involved. Even though there is no mention of a virtual displacement process in \cite{Dey:2017fld}, the entropy variation in \cite{Dey:2017fld} as applied to local causal horizons is surely due to some physical process (here that represents a local constitutive relation of entropy balance law on the local causal horizon). This physical process (involving matter fluxes across the horizon) clearly shifts the local causal horizon along its null generators $\bm{l}$. As a result the NRE corresponding to the outgoing expansion scalar $\thdl$ has been used to compute the variation of the horizon cross-section. 
Clearly, whatever the case may be, under the process of varying the local causal horizons along their null generators, there does not arise the need for a torsion current of the type $q^{ij}T_{ihj}$. However, the process that we are considering virtually shifts our null surface along the transverse auxiliary null field $\bm{k}$. The physical processes involved in both of these considerations are very different. For our case, as the NRE corresponding to the ingoing expansion scalar $\thdk$ suggests, we have to very well take into consideration the entropy generation term due to the torsion current $q^{ij}T_{ihj}$. Setting the component of this torsion current along the null generators to zero (along with the geodesic constraint), which we have seen, represents  our notion of an equilibrium horizon.

%%%%%%%%%%%%%%%%%%%%%%%%%%%%%%%%%%%%%%%%%%%%%%%%%%%%%%%%%%%%%
\section{Discussion and conclusion}\label{conclusion}
The main aim of the present analysis was to investigate whether the thermodynamic interpretation of gravitational dynamics is possible in the presence of torsion in the spacetime. We found that a particular projection of the field equation of EC theory of gravity on a generic null surface indeed provides a thermodynamic structure. The idea was originally introduced in \cite{padmanabhan2002classical, Chakraborty:2015aja} based on an infinitesimal virtual displacement along the auxiliary null vector field.  Since the original analysis was a non-covariant one, we here followed the spirit of our earlier work \cite{Dey:2020tkj} in order to provide a covariant formalism for a torsion-full spacetime.

In order to achieve this goal, we first needed to visit the problem of defining in a concrete sense both of the kinematics and dynamics of the null surface $\hech$. Since we did not find this topic which includes torsion (as far as we are aware of), the same has been constructed first in this paper. Here this had been dealt with in details following the constructions done for the ambient torsionless spacetime \cite{Gourgoulhon:2005ng}. The case of the null hypersurface in the EC spacetime is indeed important. The few salient features that separate a generic hypersurface-orthogonal null surface $\hech$ in EC gravity from the one in say Einstein (torsion-free) gravity, we encountered here, are the following. 
\begin{itemize}
\item Since the ambient connection ($\bm{\hcvd}$) is not torsion-free, we see that the connection $\bm{\hsvd}$ compatible with the spatial $2$-metric $\bm{q}$ of the null surface is also non-unique  and not torsion-free. 
\item Even inspite of the assumption of hypersurface-orthogonality of $\hech$ in the RC spacetime, we found that the twist vector does not vanish. 
\item{The relevant kinematical quantities are modified due to the presence of torsion. In particular, the extended second fundamental form (not symmetric) has only $\bm{k}$ as the degeneracy direction and not $\bm{l}$. The extended second fundamental form and the deformation rate tensor (symmetric) obviously are not equivalent.}
\end{itemize}
We studied the extrinsic geometry and the kinematics of $\hech$ in the RC spacetime in full generality. We obtained important generalizations of the geometrical quantities in the RC spacetime, viz. (extended) second fundamental form, the Weingarten map, the rotation and Hajicek one-forms, the deformation and transversal deformation rate tensors and the projected deviation tensor. We then imposed the {\it geodesic constraint} on the structure of $\hech$ while studying the dynamical evolution laws of the transversal deformation rate tensor and the ingoing expansion scalar. The geodesic constraint forces the null generators of $\hech$ to be both parallel transported along themselves as well as extremal geodesic congruences. We precisely saw that the evolution rate of the ingoing expansion scalar along the null generators are indeed related to the projection component $\hg_{ab}k^a l^b$ that we are interested in. 

Having extended this framework to the EC theory, we then incorporated the dynamics of metric tensor through equation of motion of $g_{ab}$. This provided one particular projection of EC equation on our generic null surface. Then following \cite{Chakraborty:2015aja} we provided the process of virtual displacement of $\hech$ in the transverse auxiliary null vector direction. This enabled us to interpret this evolution equation ``similar'' to the first law of thermodynamics in a covariant fashion. We saw the presence of a non-trivial torsion current $T_{i~~j}^{~h}q^{ij}$ leading to non-zero torsion current component $T_{ihj}q^{ij}k^h$ along the null generators. This led to an additional entropy generation term under the virtual displacement process. The amount of energy flow across the null surface under such process now contains additional terms depending on the non-trivial torsion tensor. Similarly, the pressure term is not defined only with respect to the matter energy-momentum tensor. It contains suitable spin-spin contact interaction terms as well as covariant derivatives of the torsion term. We mention that  the present thermodynamic interpretation is strictly based on the geodesic constraint as our evolution equation was derived within this condition.  All of the analysis, we saw, consequently reduces to the familiar form when we set the torsion to zero, i.e. say for the Einstein gravity \cite{Dey:2020tkj}. The special case of the torsion field being completely antisymmetric and its consequences were also discussed. Finally, we commented upon the case of null surface $\hech$ being in equilibrium in the EC gravity theory.

Let us at this point discuss our approach to the viewpoint of torsion. There have been predominantly two notions of torsion. It can be considered either as a geometric field or that of a background dynamical field. Here, in our analysis, we have leaned onto the geometrical perspective. This is quite evident in the way we factored the energy and work done term under the virtual displacement $\d \lambda_{(k)}$. In the NRE (of the ingoing expansion scalar $\thdk$) \eqref{tracerablakb1}, we had factored out the energy terms on the basis that they contained terms purely defined on the two-surface $S_t$ [along with the term $l^r \hcvd_r (\thdk - q^{ij}T_{ihj}k^h)$]. The work function contained terms defined entirely on the four dimensional manifold $\sptm$. This thermodynamic interpretation provided to $\hg_{ab}k^a l^b$ through the virtual displacement essentially takes this viewpoint from the very beginning that torsion is a geometric field. It is only at the end once the dynamics of the EC theory has been established (letting the torsion be sourced by the spin angular momentum tensor \eqref{spindensity}) that we can also interpret the work function or rather the pressure \eqref{pressure} in terms of the matter energy-momentum tensor $\mt_{ab}$ and the spin angular momentum tensor $\tau^a_{~bc}$. However we can right away begin with the viewpoint of torsion being a dynamical background field. This viewpoint lets the torsion terms be a part of an effective stress-energy tensor $T^{\text{eff}}_{ab}$. This effective stress-energy tensor is related to the Riemannian Einstein tensor (see Eq. $(2.5.10)$ of \cite{Poplawski:2009fb}), 
\begin{align}
G_{ab} = R_{ab} - \fr g_{ab} R = 8 \pi (\mt_{ab} + U_{ab}) = 8 \pi T^{\text{eff}}_{ab} ~,
\label{manip46}
\end{align}
where $U_{ab}$ contains terms quadratic in $\tau^a_{~bc}$ and hence represents spin-spin contact interaction terms. Obviously owing to the Bianchi identity, the effective stress-energy tensor is covariantly conserved with respect ot the Levi-Civita connection $\bm{\nabla}$. We in our approach did not proceed with such consideration of an effective stress-energy tensor [for the Riemannian spacetime $\sptml$] and went purely by the geometrical interpretation.
%\textcolor{red}{However given \eqref{tracerablakb1}, we could in principle have decomposed $\hg_{ab}$ into the (Riemannian) Einstein tensor $G_{ab}$ and torsion terms. This would have allowed us (via the dynamics) to replace the corresponding $G_{ab}k^a l^b$ in \eqref{tracerablakb1} by $T^{\text{eff}}_{ab}k^a l^b$. The process of virtual displacement under this interpretation would have surely resulted in different $\delta_{\lambda_{(k)}} E$ and $F \delta \lambda_{(k)}$ as opposed to \eqref{varenergy} and \eqref{force}. The total variation of heat density i.e $\ints d^2 x~ T \Big(\delta_{\lambda_{(k)}} \snu +  \dellk\stor\Big)$ would however be the same in both choices.}
We could however have started with the dynamical equation for ingoing Riemannian expansion scalar $\theta_{\bm{k}}$ \cite{Dey:2020tkj},
	\begin{align}
	-\kappa \theta_{(k)} &=  \Big(- {^2\mathcal{D}_a }\Omega^a - \Omega_a \Omega^a +  \theta_{(l)} \theta_{(k)} +l^i \nabla_i \theta_{(k)} + \frac{1}{2}{^2 R} \Big) - G_{ab}k^a l^b ~.
	\label{manip47}
	\end{align}
 This equation as usual relates the dynamical evolution of $\theta_{\bm{k}}$ with $G_{ab}k^a l^b$ and other Riemannian quantities. Using the fact that $\thdk = \theta_{\bm{k}}$, we can certainly use the form of the ECKS field equation \eqref{manip46} in \eqref{manip47}. Then the process of virtual displacement would on \eqref{manip47} have yielded for us different energy variation and work done terms. The pressure, as we can anticipate would depend on the matter energy tensor as well the spin-spin contact interaction terms. The variation of energy term would also be different from what had been obtained previously. Similarly looking on the L.H.S of \eqref{manip47}, we see that the entropy generation term is purely due to the change in the cross-sectional area of the null hypersurface under the virtual displacement. Under this interpretation, there is no identification of an entropy generation term due to a non-zero torsion current. The question then naturally arises as to which interpretation for the torsion field is correct. Is it good to consider it a geometric field or would it be better if torsion acted as a background field? This dilemma was also addressed in \cite{Dey:2017fld} where the EC field equations were derived from a generalized Clausius identity $\d Q = T(d S + d S_i)$ applied to a local Rindler horizon. In the paper \cite{Dey:2017fld}, the authors discussed that the internal entropy production term $d S_i$ followed quite naturally when torsion was considered as geometric field. However if torsion was proposed as a background dynamical field then such a term had to imposed by hand in an \textit{ad-hoc} fashion to recover the EC equations. We also believe that our analysis is more structured towards interpretation of torsion being a geometric field. However, our stand on this issue is by no means definitive and remains open to further scrutiny and interpretation.

Another obvious projection component in the context of the EC theory, that is not discussed in the analysis is $\hg_{ab}l^a q^b_{~c}$. We hope to return to this problem in a future work \cite{SDBRM}. We believe that our foray into this study of the connection between gravitational dynamics and thermodynamics in the case of EC gravity is indeed an interesting one. We hope, our present analysis will strengthen and complement the existing literature on thermodynamic interpretation of gravitational field equations and thereby bolstering the emergent nature gravity even in the presence of torsion.

An interesting question that crops up is whether our results can be reproduced in the context of Poincare gauge theory (PGT) \cite{Hehl:1976kj, de1986introduction, blagojevic2001gravitation, Blagojevic:2003cg}. In PGT, the basic dynamical variables are the vielbeins and the Lorentz spin connection, rather  than the metric and the general (metric compatible) connection. Due to the metric compatibility condition, the Lorentz spin connection is antisymmetric. Under PGT, the translation field strength is related to the torsion and the Lorentz field strength is related to the curvature tensor of the spacetime. Here, both the matter energy-momentum tensor and the spin density tensor source the gravitational field. Most importantly, PGT is formulated under the geometrical structure of the RC spacetime. Our central relationships \eqref{eqntransdeformrate} and \eqref{tracerablakb1}, defining the evolution dynamics of the transversal deformation rate tensor and the ingoing expansion scalar are completely geometrical (no use of gravitational field equations) and have been derived in the backdrop of the RC spacetime. Hence it is quite expected that these relations can also be exactly derived in the framework of PGT (using veilbeins and Lorentz spin connection as dynamical variables). However, while deriving \eqref{eqntransdeformrate}, we have used the geodesic constraint $\ti_b = 0$ [see Eq. \eqref{geodesicconstraint}]. Since the torsion tensor is related to the translation field strength in PGT, it would be quite instructive to understand the geometric, physical and thermodynamic conditions the geodesic constraint imposes on the translation field strength. Therefore the reformulation of the whole discussion, so far we have done here, in the language of PGT will be very interesting. For the moment, we leave this for future study. Finally, there exist many possibilities of the gravitational lagrangian under PGT. Out of these, the EC and the teleparallel \cite{aldrovandi2013teleparallel,blagojevic2001gravitation} theories are well known and studied. We have, here focused exclusively on the thermodynamic interpretation under the ECKS field equations. The EC theory is the simplest of such cases of possible gravitational Lagrangians under PGT. Here, torsion cannot propagate outside of matter sources i.e. in the absences of spin effects. Thereafter, we aim to look at the teleparallel theory in a future work as well.
%%%%%%%%%%%%%%%%%%%%%%%%%%%%%%%%%%%%%%%%%%%%%%%%%%%%%%%%%%%%%%%%%%%%%%%%%%%%%%%%%%%%%%%%%%%%%%%%%%%%%%%%%%%%%%%%%%%%%%
\section*{Acknowledgement:}
The research of one of the authors (B.R.M) is supported by the Science and Engineering Research Board (SERB), Department of Science $\&$ Technology (DST),
Government of India, under the scheme Core Research Grant (File No. CRG/2020/000616).
%%%%%%%%%%%%%%%%%%%%%%%%%%%%%%%%%%%%%%%%%%%%
\appendix
\section*{Appendices}
\section{The Null Raychaudhuri equation via the $3+1$ null construction}\label{NRE3+1}
We can reap the benefit of foliation of the spacetime $\sptm$ in the neighbourhood of $\hech$ by the null family of hypersurfaces to arrive at the NRE. The NRE determines the dynamics of the outgoing expansion scalar $\thdl$ along the null generator $\bm{l}$. To arrive at the NRE we start from the null Codacci identity established on $\hech$,
\begin{equation}
\hcvd_a (\hcvd_b l^a) - \hcvd_b (\hcvd_a l^a) = \hr_{ab}l^a - T^d_{~ab} (\hcvd_d l^a) ~.\label{Ricciiden1}
\end{equation}
Let us proceed to manipulate the first term on the L.H.S of \eqref{Ricciiden1} using the relation \eqref{nablaalbexpansion} and \eqref{expansionscalervalue},
\begin{align}
\hcvd_a (\hcvd_b l^a) &= \hcvd_a \thht^a_{~b} + l^a \hcvd_a \hw_b + \hw_b (\hcvd_a l^a) - (\hcvd_a l_b)(k^i \hcvd_i l^a) - l_b \hcvd_a (k^i \hcvd_i l^a) \nonumber \\
&= \hcvd_a \thht^a_{~b} + l^a \hcvd_a \hw_b + \hw_b \Big(\thdl + \kappa - T_a l^a\Big) \nonumber \\
&~~~~~- \Big(\thht_{ba} + \hw_a l_b - l_a (k^j \hcvd_j l_b)\Big)(k^i \hcvd_i l^a) - l_b \hcvd_a (k^i \hcvd_i l^a)  \nonumber \\
& = \hcvd_a \thht^a_{~b} + l^a \hcvd_a \hw_b + \hw_b \Big(\thdl + \k - T_a l^a\Big) - \thht_{ba} (k^i \hcvd_i l^a) \nonumber \\
&~~~~~- l_b \Big(\hw_a (k^i \hcvd_i l^a) - \hcvd_a(k^i \hcvd_i l^a)\Big) ~.
\label{manip22}
\end{align}
Let us now proceed with the second term in the R.H.S of \eqref{Ricciiden1}. Again using the relation \eqref{nablaalbexpansion} and \eqref{expansionscalervalue}, we can similarly show that,
\begin{align}
\hcvd_b(\hcvd_a l^a) &= \hcvd_b \Big(\thdl + \k -T_a l^a\Big) \nonumber \\
&= \hcvd_b \Big(\thdl + \k\Big) - l^a (\hcvd_b T_a) - T^a \thht_{ab} - \hw_b (T_a l^a) + l_b\Big(T_a (k^i \hcvd_i l^a)\Big) ~.
\label{manip23}
\end{align}
Similarly, using \eqref{nablaalbexpansion} for the second term in the R.H.S of \eqref{Ricciiden1}, we have,
\begin{align}
T^d_{~ab}(\hcvd_d l^a) = T_{dab} \thht^{ad} + T_{dab} \hw^d l^a - T_{dab} l^d (k^j \hcvd_j l^a) ~.\label{manip24}
\end{align}
Finally, upon using the relations                                                               \eqref{manip22}, \eqref{manip23} and \eqref{manip24} in the null Cadacci equation \eqref{Ricciiden1} and simplifying, we end up having,
\begin{align}
&\hcvd_a \thht^a_{~b} + l^a \hcvd_a \hw_b + \hw_b \Big(\thdl + \k\Big) - \thht_{ba} (k^i \hcvd_i l^a) - \hcvd_b (\thdl + \k) + l^a (\hcvd_b T_a) \nonumber \\
& +T^a \thht_{ab} - l_b \Big[\hw_a (k^i \hcvd_i l^a)- \hcvd_a (k^i \hcvd_i l^a) + T_a (k^i \hcvd_i l^a)\Big] = \hr_{ab} l^a - T_{fab}\thht^{af} \nonumber \\
& - T_{fab}\hw^f l^a + T_{fab}l^f (k^j \hcvd_j l^a) ~.
\label{Ricciiden2}
\end{align}
To this end, we simply need to contract the previous equation \eqref{Ricciiden2} with $l^b$. Upon using the following relations, $\hw_a l^a = \k - k_a \ti^a$, $\thht_{ba}l^b = 0$, $\thht_{ab}l^b = q^c_{~a} \ti_c$ and $T^a q^c_{~a} \ti_c = T^a \ti_a + (T^a l_a)(k^b \ti_b)$ and simplifying the above contracted (with $l^b$) relation, we have,
\begin{align}
&l^b \hcvd_a \thht^a_{~b} + l^b (l^a \hcvd_a \hw_b) + (\k - k_a \ti^a)(\thdl + \k) - l^b \hcvd_b(\thdl + \k) \nonumber \\
&+l^b l^a (\hcvd_b T_a) + T^a \ti_a + (T^a l_a)(k^b \ti_b) = \hr_{ab}l^a l^b - T_{fab} \thht^{af}l^b + \ti_a (k^j \hcvd_j l^a) ~.
\label{rablalb1}
\end{align}
Next, we manipulate the term $l^b \hcvd_a \thht^a_{~b}$ in the L.H.S of \eqref{rablalb1}. Using the fact that $\thht^a_{~b} l^b = q^{ca} \ti_c$, we have upon using \eqref{nablaalbexpansion},
\begin{align}
l^b \hcvd_a \thht^a_{~b} &= \hcvd_a (\thht^a_{~b}l^b) - \thht^a_{~b}(\hcvd_a l^b) \nonumber \\
& = \hcvd_a (q^{ca} \ti_c) - \thht^a_{~b} \Big(\thht^b_{~a} + \hw_a l^b - l_a (k^j \hcvd_j l^b)\Big) \nonumber \\
& = \hcvd_a \Big(\ti^a + (k^c \ti_c)l^a\Big) - \thht_{ab} \thht^{ba} - \hw_a \thht^a_{~b} l^b \nonumber \\
& = \hcvd_a \ti^a + (l^a \hcvd_a k_c) \ti^c + (l^a \hcvd_a \ti_c) k^c + (k^c \ti_c) (\thdl + \k) - (T_a l^a)(k^c \ti_c) \nonumber \\
& - \thht_{ab} \thht^{ba} - \hw_a \ti^a - (\k - k_a \ti^a)(k^c \ti_c) ~.
\label{manip25}
\end{align}
Upon using the relation of the rotation $1$-form \eqref{rotn1forndefn2} in the previous relation \eqref{manip25} and simplifying, we end up with,
\begin{align}
l^b \hcvd_a \thht^a_{~b} = (\hcvd_a \ti^a) - (T_{abc}k^a l^b) \ti^c + (l^a \hcvd_a \ti_c) k^c + (k^c \ti_c) (\thdl - T_a l^a)\nonumber \\
 + (k^a \ti_a)(k^b \ti_b) - \thht_{ab} \thht^{ba} ~.
 \label{lnabth}
\end{align}
After this, we focus on the term $l^b(l^a \hcvd_a \hw_b)$ in the L.H.S of \eqref{rablalb1} and use the fact that $l^b \hw_b = \k - k^a \ti_a$ along with the implementation of \eqref{rotn1forndefn2},
\begin{align}
l^b(l^a \hcvd_a \hw_b) &= l^a \hcvd_a (\k - k_c \ti^c) - \hw_b(l^a \hcvd_a l^b) \nonumber \\
& =l^a \hcvd_a \k - 2 \hw_a \ti^a + (T_{abc}k^a l^b) \ti^c - k^c(l^a \hcvd_a \ti_c) - \k^2 + \k (k_a \ti^a) ~.
\label{lanabom}
\end{align}
Adding \eqref{lnabth} and \eqref{lanabom} leads to upon simplification,
\begin{align}
l^b \hcvd_a \thht^a_{~b} + l^b l^a \hcvd_a \hw_b &= \hcvd_a \ti^a + (k_b  \ti^b)(\thdl - T_a l^a) + (k_a \ti^a)(k_b \ti^b) - \thht_{ab} \thht^{ba}\nonumber \\
& + l^a \hcvd_a \k - 2\hw_a \ti^a - \k^2 + \k (k_a \ti^a) ~.
\label{manip26}
\end{align}
Putting the above relation \eqref{manip26} in  \eqref{rablalb1} and  simplifying,
\begin{align}
&\hcvd_a \ti^a + (k_a \ti^a)(k_b \ti^b) - \thht_{ab} \thht^{ba} - 2 \hw_a \ti^a + \k \thdl - l^b \hcvd_b \thdl + l^b l^a (\hcvd_b T_a) + T^a \ti_a \nonumber \\ 
&= \hr_{ab} l^a l^b - T_{fab} \thht^{af} l^b + \ti_a (k^j \hcvd_j l^a) ~.
\label{rablalb2}
\end{align}
After this, we consider a further manipulation of the term $\hcvd_a \ti^a$ using the relation \eqref{nablaalbexpansion},
\begin{align}
\hcvd_a \ti^a &= \hcvd_a (T_{b~~c}^{~a} ~l^b l^c) = (\hcvd_a T_{b~~c}^{~a})l^b l^c + T_{b~~c}^{~a} (\hcvd_a l^b)l^c + T_{b~~c}^{~a}l^b (\hcvd_a l^c) \nonumber \\
& = (\hcvd_a T_{b~~c}^{~a})l^b l^c + T_{bac}l^c \thht^{ba} + T_{bac}l^b \thht^{ca} + 2 \ti^a \hw_a + \ti_c(k^j \hcvd_j l^c)~.
\label{manip27}
\end{align}
Putting the above relation \eqref{manip27} in \eqref{rablalb2} and further simplifying leads us to,
\begin{align}
(\hcvd_a T_{b~~c}^{~a}) l^b l^c + (T_{abc} + T_{cba}+ T_{bac}) \thht^{ab} l^c + (k_a \ti^a)( k_b \ti^b) - \thht_{ab} \thht^{ba} \nonumber \\
+ \k \thdl - l^b \hcvd_b \thdl + l^b l^a (\hcvd_b T_a) + T^a \ti_a = \hr_{ab}l^a l^b~.
\label{rablalb3}
\end{align}
We should be careful not to assign \eqref{rablalb3} as being interpreted as the NRE. This is because we notice the trace of the extended second fundamental form $\thht_{ab}$ is not the true outgoing expansion scalar $\thdl$. It would be better to rewrite any $\thht_{ab}$ occurring in \eqref{rablalb3} by the corresponding projected deviation tensor $\hb_{ab}$ or the deformation rate tensor $\hchi_{ab}$, since the trace of both of these tensors gives the true outgoing expansion scalar $\thdl$. This can be done by the virtue of \eqref{relnprojdevn2ndfundform} or \eqref{relnchintheta}. However here, we will progress with the projected deviation tensor in order to corroborate our results with \cite{Dey:2017fld}. Using \eqref{relnprojdevn2ndfundform} through some moderate but straightforward algebra, it can be shown that,
\begin{align}
\thht_{ab} \thht^{ba} = \hb_{ab} \hb^{ba} - 2 \hb^{ab}l^c T_{bca} + (T_{aib}l^i)(T^{bka}l_k) + (\ti_a k^a)( \ti_b k^b) - 2 \ti^a T_{aib}l^i k^b ~.
\label{manip28}
\end{align}
Using the symmetries of the torsion tensor, similar straightforward algebra shows that,
\begin{align}
l^c (T_{abc} + T_{cba} + T_{bac}) \thht^{ab} = -\hb^{ab} l^c (T_{acb} + T_{bca} + T_{cab}) + T^{akb} l_k l^c (T_{acb} + T_{bca} + T_{cab}) \nonumber \\
+2 T_{cab} l^c \ti^a k^b - 2 (\ti_a k^a)(\ti_b k^b) - 2 T_{ijb} k^i l^j \ti^b ~.
\label{manip29}
\end{align}
Combining the last two relations \eqref{manip28} and \eqref{manip29}, we have after some simplification,
\begin{align}
&l^c (T_{abc} + T_{cba} + T_{bac}) \thht^{ab} - \thht_{ab} \thht^{ba} = - \hb^{ab}(T_{acb} - T_{bca} + T_{cab}) l^c + T^{a~~b}_{~~i} l^i l^j (T_{ajb}+ T_{jab}) \nonumber \\
&- \hb^{ab} \hb_{ba} - 3 (\ti_a k^a)(\ti_b k^b) - 2 \ti^a T_{abc}k^b l^c - 2 \ti^c T_{abc}k^a l^b + 2 \ti^b T_{cba}l^c k^a ~.
\label{manip30}
\end{align}
Putting \eqref{manip30} in \eqref{rablalb3}, we obtain as a result,
\begin{align}
l^a \hcvd_a \thdl &= -\hr_{ab} l^a l^b + (\hcvd^a T_{bac}) l^b l^c + \k \thdl + l^b l^a (\hcvd_b T_a) + T^a \ti_a \nonumber \\
&- \hb^{ab} (T_{acb} - T_{bca} + T_{cab})l^c + l^i l^j T^{a~~b}_{~~i}(T_{ajb} + T_{jab}) - \hb^{ab} \hb_{ba} \nonumber \\
&+ \Big[-2 (\ti_a k^a)(\ti_b k^b) - 2 \ti^a k^b l^c(T_{abc}+ T_{bca} + T_{cba})\Big] ~.
\end{align}
As usual using \eqref{irrepprojdevtensor} we have finally the NRE corresponding to a hypersurface-orthogonal null congruence,
\begin{align}
l^a \hcvd_a \thdl &= -\hr_{ab} l^a l^b + (\hcvd^a T_{bac}) l^b l^c + \k \thdl + l^b l^a (\hcvd_b T_a) + T^a \ti_a \nonumber \\
&- \whblu (T_{acb} - T_{bca} + T_{cab})l^c + l^i l^j T^{a~~b}_{~~i}(T_{ajb} + T_{jab}) \nonumber \\
& - \fr (\thdl)^2 - \shbl \shblu + \whbl\whblu\nonumber \\
&+ \Big[-2 (\ti_a k^a)(\ti_b k^b) - 2 \ti^a k^b l^c(T_{abc}+ T_{bca} + T_{cba})\Big] ~.
\label{NRE1}
\end{align}
The above Eq. \eqref{NRE1} represents the evolution of the outgoing expansion scalar $\thdl$ along the null congruence $\bm{l}$. Notice, that as of yet we have not imposed the geodesic constraint and hence even though the congruence $\bm{l}$ is geodesic, it not auto-parallel. The present Eq. \eqref{NRE1} should be matched with Eq. ($66$) of \cite{Dey:2017fld}. Eq. ($66$) of \cite{Dey:2017fld} has been written down for an affinely parametrized null congruence ($\k = 0$). Our Eq. \eqref{NRE1} matches exactly with Eq. ($66$) of \cite{Dey:2017fld} except for the last terms in the squared parantheses (i.e. the terms containing $\ti_a$).  Even though the authors of \cite{Dey:2017fld} claim that their Eq. ($66$) represents the NRE in its full generality, we believe that they have missed the terms in the square parentheses. Notice that the NRE \eqref{NRE1} in this generality depends upon the auxiliary null vector field $\bm{k}$. This is somewhat a rather peculiar feature. This is because, for our construction the auxiliary null vector is uniquely defined. The auxiliary null field is transverse to the null generator as well as being orthogonal to the $2$-dimensional sub-space $S_t$. This necessarily implies that the evolution of the outgoing expansion scalar $\thdl$ along the null generator $\bm{l}$ actually encodes information of a direction that is transverse to the null generators and orthogonal to the spacelike submanifold of $\hech$. It is quite an instructive exercise to verify that if we decompose \eqref{NRE1} into its pure Riemannian parts and the pure torsion terms on both sides of the equation, we end up having the NRE for the outgoing expansion scalar $\theta_{\bm{l}}$ of $\hech$ established in the Riemannian spacetime $\sptml$ (provided with the Levi-Civita connection $\bm{\nabla}$). For arriving at the result, we make the following observations based on the projected deviation tensor. It can quite simply be established that,
\begin{align}
\hb_{ab} &= \hbr_{ab} + (K_{abc}- T_{abc})l^c + \Big[l_a (K_{dbc} - T_{dbc})k^d l^c + l_b (K_{adc}-T_{adc})k^d l^c\Big] + l_a l_b (K_{cdf} - T_{cdf})k^c k^d l^f \nonumber \\
&+ (\ti_a k_b - k_a \ti_b) + (l_a k_b - k_a l_b) (k_i \ti^i) ~,
\label{manip41}
\end{align}
where $\hbr_{ab} = q_a^{~i}q_b^{~j}(\nabla_j l_i)$ is the projected deviation tensor as computed for the null congruence $\bm{l}$ in the spacetime $\sptml$ endowed with the Levi-Civita connection. Taking the trace of \eqref{manip41} on both sides leads to the fact that the outgoing expansion scalars for the Riemann-Cartan and the Riemannian versions are the same i.e. 
\begin{align}
\thdl = \theta_{\bm{l}} = \hcvd_a l^a - \kappa + T_a l^a = \nabla_a l^a - \kappa ~.
\label{manip42}
\end{align}
Similarly, the shear torsions are related by,
\begin{align}
\shbl = \shblr + l_a l_b (K_{cdf} - T_{cdf})k^c k^d l^f~ ,
\label{manip43}
\end{align}
where $\shblr = \hbr_{(ab)} - \frac{1}{2}q_{ab} \theta_{\bm{l}}$. For a hypersurface orthogonal null congruence $\hech$ generated by $\bm{l}$ in $\sptml$, we have the antisymmetric part of the projected deviation tensor $\hbr_{ab}$ to be zero i.e $\hbr_{[ab]} = 0$ \cite{Poisson:2009pwt}. Hence,
\begin{align}
\whbl = \hb_{[ab]} = K_{acb}l^c + (l_a K_{dcb} - l_b K_{dca}) k^d l^c + (\ti_a k_b - k_a \ti_b) + (l_a k_b - k_a l_b) (k_i \ti^i) ~.
\label{manip44}
\end{align}
Finally we would require the decomposition of the Ricci tensor in $\sptm$ in terms of the pure Riemannian counterpart and pure torsion terms, i.e,
\begin{align}
\hr_{ab} = R_{ab} + \hcvd_i K^i_{~ba} + \hcvd_b T_a + T^i_{~jb} K^j_{~ia} + K^i_{~ja} K^j_{~bi} + T_i K^i_{~ba} ~.
\label{manip45}
\end{align}
Putting \eqref{manip42}, \eqref{manip43}, \eqref{manip44} and \eqref{manip45} in \eqref{NRE1}, leads upon simplification to the well known NRE for $\theta_{\bm{l}}$,
\begin{align}
	l^a \nabla_a \theta_{\bm{l}} = -R_{ab}l^a l^b + \kappa \theta_{\bm{l}} - \frac{1}{2} \theta_{\bm{l}}^2 - \shblr \shblru ~.
\end{align}

Finally, if we want to consider a system of hypersurface-orthogonal auto-parallel geodesic null congruence generating $\hech$, then we have to impose the geodesic constraint $\ti_a = 0$ in \eqref{NRE1}. For this particular case then, we have,
\begin{align}
l^a \hcvd_a \thdl &= -\hr_{ab} l^a l^b + \k \thdl - \fr (\thdl)^2 - \shbl \shblu + \whbl\whblu\nonumber \\
&- \whblu (T_{acb} - T_{bca} + T_{cab})l^c + l^i l^j T^{a~~b}_{~~i}(T_{ajb} + T_{jab}) \nonumber \\
&+ (\hcvd^a T_{bac}) l^b l^c + l^b l^a (\hcvd_b T_a) ~.
\label{NRE2}
\end{align}
The above equation under the geodesic constraint  determines the evolution of the outgoing expansion scalar $\thdl$ along $\bm{l}$ and contains explicitly no knowledge of the auxiliary null field $\bm{k}$.
%%%%%%%%%%%%%%%%%%%%%%%%%%%%%%%%%%%%%%%%%%%%%%%%%%%%%%%%%%%%%%%%%%%%%%%%%%%%%%%%%%%%%%%%%%%%%%%%%%%%%%%%%%%%%%%%%%%%%%
\section{Derivation of the relation \eqref{commut1}}\label{appxcommute1}

Let us manipulate the first term i.e $\hsvd_a(\hsvd_b v^a)$ of \eqref{RicciidentityonSt},
\begin{align}
\hsvd_a(\hsvd_b v^a) &= q^i_{~b} q^l_{~k} \hcvd_l (q^m_{~i} q^k_{~s} \hcvd_m v^s)  \nonumber \\
& = q^i_{~b} q^l_{~s} \hcvd_l(\d^m_{~i} + l^m k_i+ k^m l_i) (\hcvd_m v^s) + q^m_{~b} q^l_{~k} \hcvd_l (\d^k_{~s} + l^k k_s + k^k l_s) (\hcvd_m v^s) \nonumber \\
&~~~~~~~~~~~~~~~~~~~~~~~~~~~~~~~~~ + q^m_{~b} q^l_{~s} \hcvd_l \hcvd_m v^s ~. \nonumber \\
&=  q^i_{~b} q^l_{~s} l^m (\hcvd_l k_i) (\hcvd_m v^s) +  q^i_{~b} q^l_{~s} k^m (\hcvd_l l_i) (\hcvd_m v^s) + q^m_{~b}q^l_{~k} k_s (\hcvd_l l^k)(\hcvd_m v^s) \nonumber \\
&~~~~~~~~~~~+ q^m_{~b} q^l_{~k} l_s (\hcvd_l k^k)(\hcvd_m v^s) + q^m_{~b} q^l_{~s} \hcvd_l \hcvd_m v^s ~. 
\label{manip4s}
\end{align}
Upon using \eqref{nablaakbexpansion} and \eqref{relnnablaalbnchi} in \eqref{manip4s},we have,
\begin{align}
\hsvd_a(\hsvd_b v^a) &= q^i_{~b} q^l_{~s} l^m \Big[\xih_{li} - q^c_{~l} q^d_{~i} K_{fcd} k^f\Big] (\hcvd_m v^s) + q^i_{~b} q^l_{~s} k^m \Big[\hchi_{li} - q^c_{~l} q^d_{~i} K_{fcd}l^f\Big](\hcvd_m v^s) \nonumber \\
&+ q^m_{~b} q^{lk} l_s \Big[\xih_{lk} -q^c_{l}q^d_{~k} K_{fcd}k^f\Big] (\hcvd_m v^s) + q^m_{~b} q^{lk} k_s \Big[\hchi_{lk} - q^c_{~l} q^d_{~k} K_{fcd}l^f\Big] (\hcvd_m v^s)\nonumber \\
& + q^m_{~b}q^l_{~s} \hcvd_l \hcvd_m v^s ~.
\end{align}
The above expression can be very easily expressed as, 
\begin{align}
\hsvd_a(\hsvd_b v^a) &= \Big[\xih_{sb} - q^c_{~s} q^d_{~b} K_{fcd}k^f\Big] (l^m \hcvd_m v^s) + \Big[\hchi_{sb} - q^c_{~s} q^d_{~b} K_{fcd} l^f\Big] (k^m \hcvd_m v^s) \nonumber \\
&+ \Big[(\thdk - q^{cd} K_{fcd}k^f)q^m_{~b}\Big] (l_s \hcvd_m v^s) + \Big[(\thdl - q^{cd} K_{fcd}l^f)q^m_{~b}\Big] (k_s \hcvd_m v^s) \nonumber \\
& + q^m_{~b}q^l_{~s} \hcvd_l \hcvd_m v^s ~.
\label{dadbmanip1}
\end{align}
Let us now deal with the second term $\hsvd_b (\hsvd_a v^a)$ of \eqref{RicciidentityonSt},
\begin{align}
\hsvd_b (\hsvd_a v^a) = q^m_{~s} q^r_{~b} \hcvd_r (q^i_{~m} + q^s_{~j} \hcvd_i v^j) ~.
\end{align}
Using an exactly similar analysis as was done for $\hsvd_a (\hsvd_b v^a)$, it can be verified that,
\begin{align}
\hsvd_b (\hsvd_a v^a) &= \Big[\xih_{bs} - q^c_{~b} q^d_{~s} K_{fcd}k^f\Big](l^m \hcvd_m v^s) + \Big[\hchi_{bs} - q^c_{~b} q^d_{~s} K_{fcd}l^f\Big](k^m \hcvd_m v^s) \nonumber \\
& + \Big[\hchi^m_{~b} - q^c_{~b} q^{dm} K_{fcd}l^f\Big](k_s \hcvd_m v^s) + \Big[\xih^m_{~b} - q^c_{~b} q^{dm} K_{fcd}k^f\Big](l_s \hcvd_m v^s) \nonumber \\
& + q^m_{~b} q^l_{~s} \hcvd_m \hcvd_l v^s ~.
\label{dadbmanip2}
\end{align}
Before proceeding forward let us list a few results obtained from the properties of the torsion and contorsion tensor, 
\begin{align}
&(q^c_{~b} q^d_{~s} K_{fcd} - q^d_{~s} q^c_{~b} K_{fdc})k^f = q^c_{~b} q^d_{~s} (K_{fcd}- K_{fdc}) k^f  = q^c_{~b} q^d_{~s} (T_{fcd} k^f)  ~. \label{manip4}\\
&(q^c_{~b} q^d_{~s} K_{fcd} - q^d_{~s} q^c_{~b} K_{fdc})l^f = q^c_{~b} q^d_{~s} (T_{fcd} l^f) ~. \label{manip5}
\end{align}
Finally, subtracting \eqref{dadbmanip2} from \eqref{dadbmanip1} and utilizing the symmetry of the deformation rate and transversal deformation rate tensors along with, \eqref{manip4} and \eqref{manip5} we have our desired result \eqref{commut1}. 

%%%%%%%%%%%%%%%%%%%%%%%%%%%%%%%%%%%%%%%%%%%%%%%%%%%%%%%%%%%%%%%%%%%%%%%%%%%%%%%%%%%%%%%%%%%%%%%%%%%%%%%%%%%%%%%%%%%%%%
\section{Proof of Eqn. \eqref{2t4t}}\label{2t4tproof}
Due to the presence of torsion in the ambient spacetime $\sptm$, the submanifold $\sptq$ is not torsion free with an intrinsic $\twt^{a}_{~bc}$ present in it. For any two vectors $(\bmx,\bmy) ~\in~ \ct_P(S_t) \otimes \ct_{P}(S_t) $, we have from the definition of torsion as,
\begin{equation}
\bmt (\bmx, \bmy) = \hsvd_{\bmx} \bmy - \hsvd_{\bmy} \bmx - {^{(2)}[\bmx,\bmy]} ~,
\label{torsionbasicdef}
\end{equation}
where ${^{(2)}[\bmx,\bmy]} = {^{(2)}\bm{\pounds_{\bmx} \bmy}}$ is the intrinsic Lie bracket defined for the manifold $\sptq$. In index notation, this translates to, 
\begin{equation}
{^{(2)}T^a_{~bc}} X^b Y^c = X^i \hsvd_i Y^a - Y^i \hsvd_i X^a- {^{(2)}\pounds_{\bmx}} Y^a ~.
\label{indextorsiondefn}
\end{equation}
Now, since the vectors $\bmx$ and $\bmy$ lie in the tangent space established on $S_t$, the Lie bracket of these two vectors also belongs to $\ct_P(S_t)$. Following from the Frobenius theorem \cite{Boersma:1994pc}, we have,
\begin{equation}
{^{(2)}[\bmx,\bmy]}^a = q^a_{~b}[\bmx,\bmy]^b~.\label{frobeunresolved}
\end{equation}
Expanding the above relation \eqref{frobeunresolved}, we have,
\begin{align}
X^b \hsvd_b Y^a - Y^b \hsvd_b X^a - {^{(2)}}T^a_{~bc} X^b Y^c = q^a_{~b}\Big[X^c \hcvd_c Y^b - Y^c \hcvd_c X^b - T^b_{~cd}X^c Y^d\Big] ~.
\label{manip39}
\end{align}
Using the fact that $\bm{X}$ and $\bm{Y}$ are spatial tangent vectors on $S_t$, we have as consequence, $q^a_{~b}X^c \hcvd_c Y^b = X^c \hsvd_c Y^a$ and $q^a_{~b}Y^c \hcvd_c X^b = \Y^c \hsvd_c X^a$. Using these relations in \eqref{manip39} and simplifying, we obtain the following,
\begin{align}
{^{(2)}T^{a}_{~bc}} X^b Y^c = T^{a}_{~bc} X^b Y^c + l^a k^d T_{dbc}X^b Y^c + k^a l^d T_{dbc} X^b Y^c ~.
\end{align} 
From this, it follows that,
\begin{align}
{^{(2)}T^{a}_{~bc}}q^b_{~i} q^c_{~j}  X^i Y^j = T^{a}_{~bc}q^b_{~i} q^c_{~j}  X^i Y^j + l^a k^d T_{dbc}q^b_{~i} q^c_{~j} X^i Y^j + k^a l^d T_{dbc}q^b_{~i} q^c_{~j}  X^i Y^j ~.
\label{manip9.5}
\end{align}
From the above we can have, 
\begin{equation}
q^f_{~a}{^{(2)}T^{a}_{~bc}}q^b_{~i} q^c_{~j}  X^i Y^j = q^f_{~a} T^{a}_{~bc}q^b_{~i} q^c_{~j}  X^i Y^j ~.
\label{manip10}
\end{equation}
Since all the indices of the $4$-dimension torsion tensor have been projected onto the surface $S_t$ we have finally the result,
\begin{equation}
q^c_{~b} q^d_{~s} q^m_{~f} T^f_{~cd} = q^c_{~b} q^d_{~s} q^m_{~f} \twt^f_{cd} = \twt^m_{~bs}~.
\label{manip10.5}
\end{equation} 
%%%%%%%%%%%%%%%%%%%%%%%%%%%%%%%%%%%%%%%%%%%%%%%%%%%%%%%%%%%%%%%%%%%%%%%%%%%%%%%%%%%%%%%%%%%%%%%%%%%%%%%%%%%%%%%%%%%%%%
\section{Derivation of the relations \eqref{beginningfortrace} and \eqref{qqqR}}\label{appxqqqr}
We begin by noticing that,
\begin{align}
q^m_{~b} q^l_{~s} q^p_{~a} ~\hat{R}^s_{~plm} = q^m_{~b} q^p_{~a} \Big(\hr_{pm} + \hr_{sprm} l^r k^s + \hr_{sprm} k^r l^s\Big) ~.
\label{manip12}
\end{align}
Let us list a few relevant properties involving curvature tensors in metric affine spacetime $\sptm$. Even though the Riemann curvature tensor is antisymmetric in the first two and  the last two indices, it is not symmetric under pairwise exchange, 
\begin{align}
\hr_{cdab} = \hr_{abcd} + \hq_{abcd} ~,
\text{where~~} \hq_{abcd} &= -\frac{3}{2}\Big(\hcvd_{[b} T_{|a| cd]} - \hcvd_{[a}T_{|b|cd]} - \hcvd_{[d}T_{|c|ab]} + \hcvd_{[c}T_{|d|ab]} \nonumber \\
&+T_{ae[b}T^e_{~cd]} - T_{be[a}T^{e}_{~cd]} - T_{ce[d}T^e_{~ab]} + T_{de[c}T^e_{~ab]}\Big) ~,
\label{riemann1}
\end{align}
where $||$ indicates the enclosed index barred from antisymmetrization. Using the above property \eqref{riemann1} in \eqref{manip12}, we obtain,
\begin{align}
q^m_{~b} q^l_{~s} q^p_{~a} ~\hat{R}^s_{~plm} = q^m_{~b} q^p_{~a} \Big(\hr_{pm} + l^r (\hr_{sprm} k^s) + l^r (\hr_{smrp} k^s) + \hq_{rmsp}k^r l^s\Big) ~.\label{manip13}
\end{align}
Using the Ricci identity \eqref{ricciidenx} for the auxiliary vector field $\bf{k}$, we have, 
\begin{align}
q^m_{~b} q^l_{~s} q^p_{~a} ~\hat{R}^s_{~plm} &= q^m_{~b} q^p_{~a} \Big(\hr_{pm} - l^r \hcvd_r \hcvd_m k_p + l^r \hcvd_m \hcvd_r k_p-l^r T^{d}_{~rm }(\hcvd_d k_p) \nonumber \\
& -l^r \hcvd_r \hcvd_p k_m + l^r \hcvd_p \hcvd_r k_m - l^r T^d_{~rp}(\hcvd_d k_m)\Big) + q^m_{~b} q^p_{~a} \hq_{rmsp} k^r l^s  \nonumber \\
&= q^m_{~b} q^p_{~a} \Big[\hr_{pm} ~~\underbrace{-l^r\hcvd_r(\hcvd_m k_p) - l^r \hcvd_r (\hcvd_p k_m)}_{A_1}  ~~\underbrace{+\hcvd_m (l^r \hcvd_r k_p) + \hcvd_p (l^r \cvd_r k_m)}_{A_2} \nonumber \\
& \underbrace{-(\hcvd_m l^r)(\hcvd_r k_p) -(\hcvd_p l^r)(\hcvd_r k_m)}_{A_3}~~\underbrace{-l^r T^{d}_{~rm}(\hcvd_d k_p) - l^r T^d_{~rp} (\hcvd_d k_m)}_{A_4}\Big] \nonumber \\
&+ q^m_{~b} q^p_{~a} \hq_{rmsp} k^r l^s ~.
\label{manip14}
\end{align}
Let us build on this calculation step by step. We first evaluate the projection of the quantity $A_1$. Using \eqref{nablaakbexpansion} and the symmetry of the transversal deformation rate tensor we obtain, 
\begin{align}
& q^m_{~b} q^p_{~a}  \Big(-l^r\hcvd_r(\hcvd_m k_p) - l^r \hcvd_r (\hcvd_p k_m)\Big) = -q^m_{~b} q^p_{~a}  l^r \Big[2 \hcvd_r \xih_{mp} - \hhw_m (\hcvd_r k_p) - \hhw_p(\hcvd_r k_m) \nonumber \\
& - (\hcvd_r k_m) \hw_p - (\hcvd_r k_p) \hw_m - (\hcvd_r l_m)(k^i \hcvd_i k_p)- (\hcvd_r l_p)(k^i \hcvd_i k_m) + (\hcvd_r k_m)(T_{dep}k^d l^e)  \nonumber \\
&+ (\hcvd_r k_p)(T_{dem}k^d l^e) - \hcvd_r\Big((q^c_{~m}q^d_{~p} + q^c_{~p} q^d_{~m})K_{fcd}k^f\Big)\Big] 
\label{manip15}
\end{align}
Upon this present relation \eqref{manip15}, we will use the auto-parallel equation under the geodesic constraint i.e. $l^b \hcvd_b l^a = \kappa l^a$ as well as \eqref{rotn1forndefn2}. Let us further introduce a notation to reduce the clutter of indices, 
\begin{equation}
\hp_a \equiv T_{bcd} k^b l^c q^d_{~a} ~.
\end{equation}
Finally simplifying \eqref{manip15}, we have,
\begin{align}
q^m_{~b} q^p_{~a}  \Big(-l^r\hcvd_r(\hcvd_m k_p) - l^r \hcvd_r (\hcvd_p k_m)\Big) &= - q^m_{~b} q^p_{~a} l^r \hcvd_r \Big[2 \xih_{mp} - (q^c_{~m} q^d_{~p} + q^c_{~p}q^d_{~m}) (K_{fcd}k^f)\Big] \nonumber \\
& + 4 \hhw_a \hhw_b - 3 \hp_a \hhw_b - 3 \hhw_a \hp_b + 2 \hp_a \hp_b ~.
\label{addmanip1}
\end{align}
Proceeding to the next term in \eqref{manip14}, and using $l^r \hcvd_r k_p = \hw_p - \hp_p$, we have, 
\begin{align}
q^m_{~b} q^p_{~a} \Big(\hcvd_m(l^r \hcvd_r k_p) + \hcvd_p (l^r \hcvd_r k_m)\Big) = q^m_{~b} q^p_{~a} \Big((\hcvd_m \hw_p + \hcvd_p \hw_m) - (\hcvd_m \hp_p +\hcvd_p \hp_m)\Big) ~.\label{addmanip2}
\end{align} 
Proceeding on to the spatial projection of the term $A_3$, we again as usual use the relations \eqref{nablaakbexpansion} and \eqref{relnnablaalbnchi} and simplify to have,
\begin{align}
&q^m_{~b} q^p_{~a} \Big(-(\hcvd_m l^r)(\hcvd_r k_p) -(\hcvd_p l^r)(\hcvd_r k_m)\Big) = -\hchi^r_{~b} \xih_{ra} - \hchi^r_{~a} \xih_{rb} - 2 \hhw_a \hhw_b + (\hhw_a \hp_b + \hhw_b \hp_a) \nonumber \\
&~~~~~~~~~~~~~ + \Big[q^j_{~a} \hchi^i_{~b} + q^j_{~b} \hchi^i_{~a}\Big](K_{hij}k^h) + \Big[q^c_{~a} \xih^d_{~b} + q^c_{~b} \xih^d_{~a}\Big](K_{fcd}l^f) \nonumber \\
& ~~~~~~~~~~~~- \Big[q^c_{~b} q^{di} q^j_{~a} + q^c_{~a}q^{di} q^j_{~b}\Big](K_{fcd} l^f) (K_{hij}k^h) ~. 
\label{addmanip3}
\end{align}
Similar analysis on the spatial projection of term $A_4$ leads us to, 
\begin{align}
&q^m_{~b} q^p_{~a}\Big(-l^r T^{d}_{~rm}(\hcvd_d k_p) - l^r T^d_{~rp} (\hcvd_d k_m)\Big) = -\Big(\xih^c_{~a} q^d_{~b} +\xih^c_{~b} q^d_{~a}\Big)(T_{cfd}l^f) -2 \hp_a \hp_b \nonumber \\
& +(\hhw_a \hp_b + \hhw_b \hp_a) + \Big(q^d_{~b} q^j_{~a} q^{ci} + q^d_{~a} q^j_{~b} q^{ci}\Big)(T_{cdf}l^f)(K_{hij}k^h) ~.
\label{addmanip4}
\end{align}
Adding up \eqref{addmanip1}, \eqref{addmanip2} , \eqref{addmanip4} and \eqref{addmanip4} in \eqref{manip14} and proceeding to simplify, we have our result \eqref{beginningfortrace}.

Notice that in \eqref{beginningfortrace}, there exists the term  $q^m_{~b} q^p_{~a} l^r \hcvd_r (2\xih_{mp} ) $. We will convert the covariant derivative into a Lie derivative term along the null generator to go ahead towards our construction of the evolution equation of the transversal deformation rate tensor. It is quite easy to show that, 
\begin{align}
\pounds_{\bm{l}} \xih_{mp} &= l^i \p_i \xih_{mp} + \xih_{mi} \p_p l^i  + \xih_{ip} \p_m l^i \nonumber \\
& = l^r \hcvd_r \xih_{mp} + \xih^i_{~m}(\hcvd_p l_i)+ \xih^i_{~p}(\hcvd_m l_i) + \xih^s_{~p} (T_{sim}l^i) + \xih^s_{~m}(T_{sip}l^i) ~.
\label{liecov}
\end{align}  
Projecting the Lie derivative along the null generator $\bm{l}$ of the transversal deviation rate tensor $\pounds_{\bm{l}} \xih_{mp}$ on to the spatial cross section $S_t$ , we the use \eqref{relnnablaalbnchi} and simplify to have,
\begin{align}
-2 q^m_{~b} q^p_{~a} l^r \hcvd_r (2\xih_{mp} )  &= -2q^m_{~b} q^p_{~a} \pounds_{\bm{l}} \xih_{mp} + 2 \Big[\xih^c_{~a} q^d_{~b} + \xih^c_{~b} q^d_{~a}\Big](T_{cfd}l^f) \nonumber \\
& 2 \Big(\hchi_{ai} \xih^i_{~b} + \hchi_{bi} \xih^i_{~a}\Big) - 2 \Big[q^c_{~a} \xih^d_{~b} + q^c_{~b} \xih^d_{~a}\Big](K_{fcd}l^f) ~.
\label{liecovtransversal}
\end{align}
There also exists the term $q^m_{~b} q^p_{~a} l^r \Big[\hcvd_r \Big(q^i_{~m} q^j_{~p} + q^i_{~p} q^j_{~m}\Big)(K_{hij}k^h)\Big]$ in \eqref{beginningfortrace}. In a similar fashion, following \eqref{liecov} and \eqref{liecovtransversal}, we want to convert the covariant derivative into a Lie derivative term. After a few lines of simple algebra, we have,
\begin{align}
&q^m_{~b} q^p_{~a} l^r \Big[\hcvd_r \Big(q^i_{~m} q^j_{~p} + q^i_{~p} q^j_{~m}\Big)(K_{hij}k^h)\Big] = q^m_{~b} q^p_{~a} \pounds_{\bm{l}} \Big[ \Big(q^i_{~m} q^j_{~p} + q^i_{~p} q^j_{~m}\Big)(K_{hij}k^h)\Big] \nonumber \\
& - \Big[\Big(q^j_{~a} q^d_{~b} + q^j_{~b} q^d_{~a}\Big)q^{ic} + \Big(q^i_{~a} q^d_{~b} + q^i_{~b} q^d_{~a}\Big)q^{jc}\Big](T_{cfd}l^f)(K_{hij}k^h) \nonumber \\
&-\Big[\hchi^i_{~a} q^j_{~b} + \hchi^i_{~b} q^j_{~a} + \hchi^j_{~a} q^i_{~b} + \hchi^j_{~b} q^i_{~a}\Big](K_{hij}k^h) \nonumber \\
&+ \Big[\Big(q^c_{~a} q^j_{~b}+ q^c_{~b} q^j_{~a}\Big)q^{id} + \Big(q^c_{~a} q^i_{~b} + q^c_{~b}q^i_{~a}\Big)q^{jd}\Big](K_{fcd}l^f)(K_{hij}k^h) ~.
\label{liecovtidelphi}
\end{align}
We have one more transformation to do. We consider the sixth term of the R.H.S. of \eqref{beginningfortrace} i.e. $q^m_{~b} q^p_{~a} \Big[\Big(\hcvd_m \hw_p + \hcvd_p \hw_m\Big) - \Big(\hcvd_m \hp_p + \hcvd_p \hp_m \Big)\Big] $. We aim to convert the spacetime covariant derivatives into spatial derivatives. Notice that $\hp_a$ acts on the tangent space of $S_t$ and hence is spatial $1$-form. Hence upon using the relation $\hw_a = \hhw_a - \kappa l_a$ under the geodesic constraint, we have,
\begin{align}
q^m_{~b} q^p_{~a} \Big[\Big(\hcvd_m \hw_p + \hcvd_p \hw_m\Big) - \Big(\hcvd_m \hp_p + \hcvd_p \hp_m \Big)\Big]  &= q^m_{~b} q^p_{~a} \Big[(\hcvd_m \hhw_p + \hcvd_p \hhw_m) - \kappa(\hcvd_m k_p) - \kappa(\hcvd_p k_m)\Big] \nonumber \\
& - (\hsvd_a \hp_b + \hsvd_b \hp_a) ~.
\label{manip16}
\end{align}
Upon using \eqref{nablaakbexpansion} in the above Eq. \eqref{manip16}, we have,
\begin{align}
q^m_{~b} q^p_{~a} \Big[\Big(\hcvd_m \hw_p + \hcvd_p \hw_m\Big) - \Big(\hcvd_m \hp_p + \hcvd_p \hp_m \Big)\Big] & = \hsvd_a (\hhw_b - \hp_b) + \hsvd_b(\hhw_a - \hp_a) - 2 \kappa \xih_{ab} \nonumber \\
& + \kappa \Big(q^i_{~b} q^j_{~a} + q^i_{~a} q^j_{~b}\Big)(K_{hij}k^h) ~.
\label{spatialomega}
\end{align}
At the end of this, finally using \eqref{liecovtransversal}, \eqref{liecovtidelphi} and \eqref{spatialomega} in \eqref{beginningfortrace}, we obtain, after some simplification, our desired result \eqref{qqqR}.
%%%%%%%%%%%%%%%%%%%%%%%%%%%%%%%%%%%%%%%%%%%%%%%%%%%%%%%%%%%%%%%%%%%%%%%%%%%%%%%%%%%%%%%%%%%%%%%%%%%%%%%%%%%%%%%%%%%%%%
\section{Derivation of the result \eqref{tracerablakb1}} \label{appxralakb1}
Let us for the benefit of the reader list the individual traces of the terms in the R.H.S of \eqref{beginningfortrace}. 
\begin{enumerate}
	\item {$g^{ab}	q^m_{~b} q^p_{~a} \hq_{rmsp} k^r l^s = q^{bd} \hq_{abcd} k^a l^c ~.$} 
	\item{$ g^{ab} q^m_{~b} q^p_{~a} \hr_{pm} = \hat{R} + \Big(\hr_{ab}l^a k^b+ \hr_{ab} k^a l^b\Big)~.$}
	\item{$-g^{ab}q^m_{~b} q^p_{~a} l^r \hcvd_r \Big[2\xih_{mp} - \Big(q^i_{~m} q^j_{~p} + q^i_{~p} q^j_{~m}\Big)(K_{hij}k^h)\Big] = -2 l^r \hcvd_r \Big(\thdk - q^{ij}T_{ihj}k^h\Big)~.$}
	\item{$g^{ab}(2 \hhw_a \hhw^a) = 2 \hhw^a \hhw_a~.$}
	\item{$-g^{ab}(\hhw_a \hp_b + \hp_a \hhw_b) = -2 \hhw_a \hp^a~.$}
	\item{$g^{ab}q^m_{~b} q^p_{~a} \Big[\Big(\hcvd_m \hw_p + \hcvd_p \hw_m\Big) - \Big(\hcvd_m \hp_p + \hcvd_p \hp_m \Big)\Big] = 2 \hsvd_a (\hhw^a - \hp^a) - 2 \kappa \Big(\thdk - q^{ij}T_{ihj}k^h\Big)~.$}
	\item{$-g^{ab} \Big(\hchi^r_{~b} \xih_{ra} + \hchi^r_{~a} \xih_{rb}\Big) = -2 \xih_{ab} \hchi^{ab}~.$}
	\item{$g^{ab} \Big(q^j_{~a} \hchi^i_{~b} + q^j_{~b} \hchi^i_{~a}\Big) (K_{hij} k^h) = 2 \hchi^{ij} T_{ihj}k^h~.$}
	\item{$g^{ab}\Big(q^c_{~a} \xih^d_{~b} + q^c_{~b} \xih^d_{~a}\Big)(K_{fcd}l^f)  = 2 \xih^{cd} T_{cfd}l^f~.$}
	\item{$-g^{ab} \Big(\xih^c_{~a}q^d_{~b} + \xih^c_{~b} q^d_{~a}\Big)(T_{cfd}l^f) = -2 \xih^{cd} T_{cfd}l^f~.$}
	\item{$-g^{ab} \Big[q^c_{~b} q^j_{~a} q^{di} + q^c_{~a} q^j_{~b} q^{di}\Big] (K_{fcd}l^f)(K_{hij}k^h) = -2 q^{cj} q^{di} (K_{fcd} l^f)(K_{hij}k^h) ~.$}
	\item{$g^{ab}\Big[q^j_{~a} q^d_{~b}q^{ci} +q^j_{~b}q^d_{~a}q^{ci}\Big](T_{cfd}l^f)(K_{hij}k^h)=2 q^{dj}q^{ci}(T_{cfd}l^f)(K_{hij}k^h)~. $}
\end{enumerate}
Adding up the traces we have,
\begin{align}
&g^{ab} (q^{m}_{~b} q^l_{~s} q^p_{~a} \hat{R}^s_{~plm}) = q^{bd} \hq_{abcd} k^a l^c + \hat{R} + \Big(\hr_{ab}l^a k^b+ \hr_{ab} k^a l^b\Big) -2 l^r \hcvd_r \Big(\thdk - q^{ij}T_{ihj}k^h\Big)\nonumber \\
& + 2 \hhw^a \hhw_a -2 \hhw_a \hp^a + 2 \hsvd_a (\hhw^a - \hp^a) - 2 \kappa \Big(\thdk - q^{ij}T_{ihj}k^h\Big) -2 \xih_{ab} \hchi^{ab} + 2 \hchi^{ij} T_{ihj}k^h \nonumber \\
& -2 q^{cj} q^{di} (K_{fcd} l^f)(K_{hij}k^h) + 2 q^{dj}q^{ci}(T_{cfd}l^f)(K_{hij}k^h)~.
\label{manip17}
\end{align}
Now we have to take the trace of \eqref{2r4R} and put the value of $g^{ab} (q^{m}_{~b} q^l_{~s} q^p_{~a} \hat{R}^s_{~plm})$ from \eqref{manip17}. That leads us to, upon simplification,
\begin{align}
&-\kappa \Big(\thdk - q^{ij}T_{ihj}k^h\Big) = \fr  {^{(2)} \hr}  -\fr  q^{bd} \hq_{abcd} k^a l^c -\fr \hr -\fr \Big(\hr_{ab} l^a k^b + \hr_{ab} k^a l^b\Big) - \hhw_a \hhw^a + \hhw_a \hp^a \nonumber \\
& +  l^r \hcvd_r \Big(\thdk - q^{ij} T_{ihj}k^h\Big) - \hsvd_a (\hhw^a - \hp^a) + \thdl \Big(\thdk - q^{ij} T_{ihj}k^h\Big) - \Big[\thdk q^{cd} - \xih^{cd}\Big](T_{cfd} l^f) \nonumber \\
& - \Big[q^{dj}q^{ci} - q^{cd} q^{ij}\Big](T_{cfd}l^f)(K_{hij}k^h) ~.
\label{tracerablakb}
\end{align}
Onwards, using the symmetries of the tensor $\hq_{abcd}$, we go on to compute the quantity $q^{bd}\hq_{abcd} k^a l^c$. The tensor $\hq_{abcd}$ like  the curvature tensor is antisymmteric in the first and the second pair of indices. We have hence upon using the symmetries of $\hq_{abcd}$,
\begin{equation}
q^{bd}\hq_{abcd} k^a l^c = g^{bd} \hq_{abcd} k^a l^c - \hq_{abcd}k^a k^c l^b l^d ~.
\end{equation} 
From \eqref{riemann1}, we see that all the individual terms inside the expression for $\hq_{abcd}$ is antisymmetric in either $a$ and $c$ or $b$ and $d$. Hence $\hq_{abcd}k^a k^c l^b l^d$ vanishes. Hence,
\begin{align}
q^{bd}\hq_{abcd} k^a l^c &= - \frac{3}{2} g^{bd} k^a l^c\Big(\hcvd_{[b} T_{|a| cd]} - \hcvd_{[a}T_{|b|cd]} - \hcvd_{[d}T_{|c|ab]} + \hcvd_{[c}T_{|d|ab]} \nonumber \\
&+T_{ae[b}T^e_{~cd]} - T_{be[a}T^{e}_{~cd]} - T_{ce[d}T^e_{~ab]} + T_{de[c}T^e_{~ab]}\Big) ~ \nonumber \\
& = - \frac{3}{2}g^{bd} k^a l^c \Big(- \hcvd_{[a}T_{|b|cd]} + \hcvd_{[c}T_{|d|ab]} - T_{be[a}T^{e}_{~cd]} + T_{de[c}T^e_{~ab]} \Big) ~.
\label{manip18}
\end{align}
Upon expanding the antisymmetric parts, we have,
\begin{align}
&- \frac{3}{2}g^{bd} k^a l^c \Big(- \hcvd_{[a}T_{|b|cd]} + \hcvd_{[c}T_{|d|ab]}\Big) = k^a l^c \Big(\hcvd_a T_c - \hcvd_c T_a + \hcvd_d T^d_{ac}\Big) \nonumber \\
&- \frac{3}{2}g^{bd} k^a l^c \Big(- T_{be[a}T^{e}_{~cd]} + T_{de[c}T^e_{~ab]}\Big) = k^a l^c \Big(g^{bd}(T_{bea}T^e_{~cd} - T_{bec}T^e_{~ad}) + T_eT^e_{~ac}\Big)~.
\end{align}
Hence we have as a result,
\begin{align}
q^{bd} \hq_{abcd} k^a l^c = \Big[\Big(\hcvd_a T_b - \hcvd_b T_a\Big) + \hcvd_i T^i_{~ab} + \Big(T^i_{~ea} T^e_{~bi} - T^i_{~eb} T^e_{~ai}\Big) + T_i T^i_{~ab}\Big]k^a l^b ~.
\label{qabcdpart1}
\end{align}
However, owing to the symmetry property of the torsion tensor, it is quite easy to see that,
\begin{align}
T^i_{~ea} T^e_{~bi} - T^i_{~eb} T^e_{~ai}= 0 ~.
\end{align}
Hence we have,
\begin{align}
q^{bd} \hq_{abcd} k^a l^c = \Big[\Big(\hcvd_a T_b - \hcvd_b T_a\Big) + (\hcvd_i + T_i) ~T^i_{~ab}\Big]k^a l^b ~.
\label{qabcdpart}
\end{align}
Next, we will deal with the term $(\hr_{ab}l^a k^b + \hr_{ab}k^a l^b) = (\hr_{ab}+ \hr_{ba}) k^a l^b$ in \eqref{tracerablakb}. Following simply the definition of the Riemann-curvature tensor in the spacetime $\sptm$, we have for the Ricci tensor,
\begin{align}
\hr_{ba} = \hr_{ab} + \Big(\hcvd_a T_b - \hcvd_b T_a\Big) + \Big(\hcvd_i + T_i\Big)T^i_{~ab} ~.
\label{Riccitensymmetry}
\end{align}
Using the above relation \eqref{Riccitensymmetry} and the fact that $\hr = -g_{ab}l^a k^b \hr$, we obtain,
\begin{align}
-\fr \Big(\hr_{ab}l^a k^b+ \hr_{ab}k^a l^b\Big) - \fr \hr = - \Big[\hat{G}_{ab} + \fr (\hcvd_a T_b - \hcvd_b T_a) + \fr(\hcvd_i + T_i)T^i_{~ab}\Big]k^a l^b ~,
\label{manip31}
\end{align}
where as usual we have $\hat{G}_{ab} = \hr_{ab} - \fr g_{ab} \hr$.
Using the relations \eqref{manip31} and \eqref{qabcdpart}, we have,
\begin{align}
-\fr q^{bd} \hq_{abcd} k^al^c -\fr \Big(\hr_{ab}l^a k^b+ \hr_{ab}k^a l^b\Big) - \fr \hr = -\hat{G}_{ab}k^a l^b \nonumber \\
- \Big[(\hcvd_a T_b- \hcvd_b T_a) + (\hcvd_i + T_i)T^i_{~ab}\Big]k^a l^b
\label{manip32}
\end{align}
Let us then rewrite \eqref{tracerablakb} using \eqref{manip32} and hence end up with the relation \eqref{tracerablakb1}.
%%%%%%%%%%%%%%%%%%%%%%%%%%%%%%%%%%%%%%%%%%%%%%%%%%%%%%%%%%%%%%%%%%%%%%%%%%%%%%%%%%%%%%%%%%%%%%%%%%%%%%%%%%%%%%%%%%%%%%
\bibliographystyle{elsarticle-num}
\bibliography{torsion_references1}

\begin{thebibliography}{10}
\expandafter\ifx\csname url\endcsname\relax
  \def\url#1{\texttt{#1}}\fi
\expandafter\ifx\csname urlprefix\endcsname\relax\def\urlprefix{URL }\fi
\expandafter\ifx\csname href\endcsname\relax
  \def\href#1#2{#2} \def\path#1{#1}\fi

\bibitem{Bekenstein:1973ur}
J.~D. Bekenstein, {Black holes and entropy}, Phys. Rev. D 7 (1973) 2333--2346.
\newblock \href {https://doi.org/10.1103/PhysRevD.7.2333}
  {\path{doi:10.1103/PhysRevD.7.2333}}.

\bibitem{Bardeen:1973gs}
J.~M. Bardeen, B.~Carter, S.~Hawking, {The Four laws of black hole mechanics},
  Commun. Math. Phys. 31 (1973) 161--170.
\newblock \href {https://doi.org/10.1007/BF01645742}
  {\path{doi:10.1007/BF01645742}}.

\bibitem{Hawking:1971vc}
S.~Hawking, {Black holes in general relativity}, Commun. Math. Phys. 25 (1972)
  152--166.
\newblock \href {https://doi.org/10.1007/BF01877517}
  {\path{doi:10.1007/BF01877517}}.

\bibitem{Hawking:1976de}
S.~Hawking, {Black Holes and Thermodynamics}, Phys. Rev. D 13 (1976) 191--197.
\newblock \href {https://doi.org/10.1103/PhysRevD.13.191}
  {\path{doi:10.1103/PhysRevD.13.191}}.

\bibitem{Davies:1978mf}
P.~Davies, {Thermodynamics of Black Holes}, Proc. Roy. Soc. Lond. A A353 (1977)
  499--521.
\newblock \href {https://doi.org/10.1098/rspa.1977.0047}
  {\path{doi:10.1098/rspa.1977.0047}}.

\bibitem{Wald:1999vt}
R.~M. Wald, {The thermodynamics of black holes}, Living Rev. Rel. 4 (2001) 6.
\newblock \href {http://arxiv.org/abs/gr-qc/9912119}
  {\path{arXiv:gr-qc/9912119}}, \href {https://doi.org/10.12942/lrr-2001-6}
  {\path{doi:10.12942/lrr-2001-6}}.

\bibitem{Carlip:2014pma}
S.~Carlip, {Black Hole Thermodynamics}, Int. J. Mod. Phys. D 23 (2014) 1430023.
\newblock \href {http://arxiv.org/abs/1410.1486} {\path{arXiv:1410.1486}},
  \href {https://doi.org/10.1142/S0218271814300237}
  {\path{doi:10.1142/S0218271814300237}}.

\bibitem{Wall:2018ydq}
A.~C. Wall, {A Survey of Black Hole Thermodynamics.~~} (4 2018).
\newblock \href {http://arxiv.org/abs/1804.10610} {\path{arXiv:1804.10610}}.

\bibitem{Kolekar:2011bb}
S.~Kolekar, D.~Kothawala, T.~Padmanabhan, {Two Aspects of Black Hole Entropy in
  Lanczos-Lovelock Models of Gravity}, Phys. Rev. D 85 (2012) 064031.
\newblock \href {http://arxiv.org/abs/1111.0973} {\path{arXiv:1111.0973}},
  \href {https://doi.org/10.1103/PhysRevD.85.064031}
  {\path{doi:10.1103/PhysRevD.85.064031}}.

\bibitem{Kolekar:2012tq}
S.~Kolekar, T.~Padmanabhan, S.~Sarkar, {Entropy Increase during Physical
  Processes for Black Holes in Lanczos-Lovelock Gravity}, Phys. Rev. D 86
  (2012) 021501.
\newblock \href {http://arxiv.org/abs/1201.2947} {\path{arXiv:1201.2947}},
  \href {https://doi.org/10.1103/PhysRevD.86.021501}
  {\path{doi:10.1103/PhysRevD.86.021501}}.

\bibitem{Mureika:2015sda}
J.~R. Mureika, J.~W. Moffat, M.~Faizal, {Black hole thermodynamics in MOdified
  Gravity (MOG)}, Phys. Lett. B 757 (2016) 528--536.
\newblock \href {http://arxiv.org/abs/1504.08226} {\path{arXiv:1504.08226}},
  \href {https://doi.org/10.1016/j.physletb.2016.04.041}
  {\path{doi:10.1016/j.physletb.2016.04.041}}.

\bibitem{Soroushfar:2016nbu}
S.~Soroushfar, R.~Saffari, N.~Kamvar, {Thermodynamic geometry of black holes in
  f(R) gravity}, Eur. Phys. J. C 76~(9) (2016) 476.
\newblock \href {http://arxiv.org/abs/1605.00767} {\path{arXiv:1605.00767}},
  \href {https://doi.org/10.1140/epjc/s10052-016-4311-6}
  {\path{doi:10.1140/epjc/s10052-016-4311-6}}.

\bibitem{Aros:2019quj}
R.~Aros, M.~Estrada, {Regular black holes and its thermodynamics in Lovelock
  gravity}, Eur. Phys. J. C 79~(3) (2019) 259.
\newblock \href {http://arxiv.org/abs/1901.08724} {\path{arXiv:1901.08724}},
  \href {https://doi.org/10.1140/epjc/s10052-019-6783-7}
  {\path{doi:10.1140/epjc/s10052-019-6783-7}}.

\bibitem{Sarkar:2019xfd}
S.~Sarkar, {Black Hole Thermodynamics: General Relativity and Beyond}, Gen.
  Rel. Grav. 51~(5) (2019) 63.
\newblock \href {http://arxiv.org/abs/1905.04466} {\path{arXiv:1905.04466}},
  \href {https://doi.org/10.1007/s10714-019-2545-y}
  {\path{doi:10.1007/s10714-019-2545-y}}.

\bibitem{Hayward:1997jp}
S.~A. Hayward, {Unified first law of black hole dynamics and relativistic
  thermodynamics}, Class. Quant. Grav. 15 (1998) 3147--3162.
\newblock \href {http://arxiv.org/abs/gr-qc/9710089}
  {\path{arXiv:gr-qc/9710089}}, \href
  {https://doi.org/10.1088/0264-9381/15/10/017}
  {\path{doi:10.1088/0264-9381/15/10/017}}.

\bibitem{Hayward:1998ee}
S.~A. Hayward, S.~Mukohyama, M.~C. Ashworth, {Dynamic black hole entropy},
  Phys. Lett. A 256 (1999) 347--350.
\newblock \href {http://arxiv.org/abs/gr-qc/9810006}
  {\path{arXiv:gr-qc/9810006}}, \href
  {https://doi.org/10.1016/S0375-9601(99)00225-X}
  {\path{doi:10.1016/S0375-9601(99)00225-X}}.

\bibitem{Ashtekar:2003hk}
A.~Ashtekar, B.~Krishnan, {Dynamical horizons and their properties}, Phys. Rev.
  D 68 (2003) 104030.
\newblock \href {http://arxiv.org/abs/gr-qc/0308033}
  {\path{arXiv:gr-qc/0308033}}, \href
  {https://doi.org/10.1103/PhysRevD.68.104030}
  {\path{doi:10.1103/PhysRevD.68.104030}}.

\bibitem{Majhi:2014hpa}
B.~R. Majhi, {Thermodynamics of Sultana-Dyer Black Hole}, JCAP 05 (2014) 014.
\newblock \href {http://arxiv.org/abs/1403.4058} {\path{arXiv:1403.4058}},
  \href {https://doi.org/10.1088/1475-7516/2014/05/014}
  {\path{doi:10.1088/1475-7516/2014/05/014}}.

\bibitem{Majhi:2014lka}
B.~R. Majhi, {Conformal Transformation, Near Horizon Symmetry, Virasoro Algebra
  and Entropy}, Phys. Rev. D 90~(4) (2014) 044020.
\newblock \href {http://arxiv.org/abs/1404.6930} {\path{arXiv:1404.6930}},
  \href {https://doi.org/10.1103/PhysRevD.90.044020}
  {\path{doi:10.1103/PhysRevD.90.044020}}.

\bibitem{Majhi:2015tpa}
B.~R. Majhi, {Near horizon hidden symmetry and entropy of Sultana-Dyer black
  hole: A time dependent case}, Phys. Rev. D 92~(6) (2015) 064026.
\newblock \href {http://arxiv.org/abs/1505.03310} {\path{arXiv:1505.03310}},
  \href {https://doi.org/10.1103/PhysRevD.92.064026}
  {\path{doi:10.1103/PhysRevD.92.064026}}.

\bibitem{Bhattacharya:2016kbm}
K.~Bhattacharya, B.~R. Majhi, {Temperature and thermodynamic structure of
  Einstein\textquoteright{}s equations for a cosmological black hole}, Phys.
  Rev. D 94~(2) (2016) 024033.
\newblock \href {http://arxiv.org/abs/1602.07879} {\path{arXiv:1602.07879}},
  \href {https://doi.org/10.1103/PhysRevD.94.024033}
  {\path{doi:10.1103/PhysRevD.94.024033}}.

\bibitem{Nielsen:2008cr}
A.~B. Nielsen, {Black holes and black hole thermodynamics without event
  horizons}, Gen. Rel. Grav. 41 (2009) 1539--1584.
\newblock \href {http://arxiv.org/abs/0809.3850} {\path{arXiv:0809.3850}},
  \href {https://doi.org/10.1007/s10714-008-0739-9}
  {\path{doi:10.1007/s10714-008-0739-9}}.

\bibitem{Cai:2008mh}
R.-G. Cai, L.-M. Cao, Y.-P. Hu, S.~P. Kim, {Generalized Vaidya Spacetime in
  Lovelock Gravity and Thermodynamics on Apparent Horizon}, Phys. Rev. D 78
  (2008) 124012.
\newblock \href {http://arxiv.org/abs/0810.2610} {\path{arXiv:0810.2610}},
  \href {https://doi.org/10.1103/PhysRevD.78.124012}
  {\path{doi:10.1103/PhysRevD.78.124012}}.

\bibitem{Jacobson:1995ab}
T.~Jacobson, {Thermodynamics of space-time: The Einstein equation of state},
  Phys. Rev. Lett. 75 (1995) 1260--1263.
\newblock \href {http://arxiv.org/abs/gr-qc/9504004}
  {\path{arXiv:gr-qc/9504004}}, \href
  {https://doi.org/10.1103/PhysRevLett.75.1260}
  {\path{doi:10.1103/PhysRevLett.75.1260}}.

\bibitem{Padmanabhan:2002sha}
T.~Padmanabhan, {Classical and quantum thermodynamics of horizons in
  spherically symmetric space-times}, Class. Quant. Grav. 19 (2002) 5387--5408.
\newblock \href {http://arxiv.org/abs/gr-qc/0204019}
  {\path{arXiv:gr-qc/0204019}}, \href
  {https://doi.org/10.1088/0264-9381/19/21/306}
  {\path{doi:10.1088/0264-9381/19/21/306}}.

\bibitem{Padmanabhan:2003gd}
T.~Padmanabhan, {Gravity and the thermodynamics of horizons}, Phys. Rept. 406
  (2005) 49--125.
\newblock \href {http://arxiv.org/abs/gr-qc/0311036}
  {\path{arXiv:gr-qc/0311036}}, \href
  {https://doi.org/10.1016/j.physrep.2004.10.003}
  {\path{doi:10.1016/j.physrep.2004.10.003}}.

\bibitem{Padmanabhan:2009vy}
T.~Padmanabhan, {Thermodynamical Aspects of Gravity: New insights}, Rept. Prog.
  Phys. 73 (2010) 046901.
\newblock \href {http://arxiv.org/abs/0911.5004} {\path{arXiv:0911.5004}},
  \href {https://doi.org/10.1088/0034-4885/73/4/046901}
  {\path{doi:10.1088/0034-4885/73/4/046901}}.

\bibitem{Kothawala:2007em}
D.~Kothawala, S.~Sarkar, T.~Padmanabhan, {Einstein's equations as a
  thermodynamic identity: The Cases of stationary axisymmetric horizons and
  evolving spherically symmetric horizons}, Phys. Lett. B 652 (2007) 338--342.
\newblock \href {http://arxiv.org/abs/gr-qc/0701002}
  {\path{arXiv:gr-qc/0701002}}, \href
  {https://doi.org/10.1016/j.physletb.2007.07.021}
  {\path{doi:10.1016/j.physletb.2007.07.021}}.

\bibitem{Paranjape:2006ca}
A.~Paranjape, S.~Sarkar, T.~Padmanabhan, {Thermodynamic route to field
  equations in Lancos-Lovelock gravity}, Phys. Rev. D 74 (2006) 104015.
\newblock \href {http://arxiv.org/abs/hep-th/0607240}
  {\path{arXiv:hep-th/0607240}}, \href
  {https://doi.org/10.1103/PhysRevD.74.104015}
  {\path{doi:10.1103/PhysRevD.74.104015}}.

\bibitem{Chakraborty:2015aja}
S.~Chakraborty, K.~Parattu, T.~Padmanabhan, {Gravitational field equations near
  an arbitrary null surface expressed as a thermodynamic identity}, JHEP 10
  (2015) 097.
\newblock \href {http://arxiv.org/abs/1505.05297} {\path{arXiv:1505.05297}},
  \href {https://doi.org/10.1007/JHEP10(2015)097}
  {\path{doi:10.1007/JHEP10(2015)097}}.

\bibitem{unruh1976notes}
W.~G. Unruh, Notes on black-hole evaporation, Physical Review D 14~(4) (1976)
  870.

\bibitem{unruh1984happens}
W.~G. Unruh, R.~M. Wald, What happens when an accelerating observer detects a
  rindler particle, Physical Review D 29~(6) (1984) 1047.

\bibitem{Parikh:2018anm}
M.~Parikh, S.~Sarkar, A.~Svesko, {Local first law of gravity}, Phys. Rev. D
  101~(10) (2020) 104043.
\newblock \href {http://arxiv.org/abs/1801.07306} {\path{arXiv:1801.07306}},
  \href {https://doi.org/10.1103/PhysRevD.101.104043}
  {\path{doi:10.1103/PhysRevD.101.104043}}.

\bibitem{Adami:2021kvx}
H.~Adami, M.~M. Sheikh-Jabbari, V.~Taghiloo, H.~Yavartanoo, {Null Surface
  Thermodynamics} (10 2021).
\newblock \href {http://arxiv.org/abs/2110.04224} {\path{arXiv:2110.04224}}.

\bibitem{Padmanabhan:2007en}
T.~Padmanabhan, A.~Paranjape, {Entropy of null surfaces and dynamics of
  spacetime}, Phys. Rev. D 75 (2007) 064004.
\newblock \href {http://arxiv.org/abs/gr-qc/0701003}
  {\path{arXiv:gr-qc/0701003}}, \href
  {https://doi.org/10.1103/PhysRevD.75.064004}
  {\path{doi:10.1103/PhysRevD.75.064004}}.

\bibitem{Padmanabhan:2007xy}
T.~Padmanabhan, {Dark energy and gravity}, Gen. Rel. Grav. 40 (2008) 529--564.
\newblock \href {http://arxiv.org/abs/0705.2533} {\path{arXiv:0705.2533}},
  \href {https://doi.org/10.1007/s10714-007-0555-7}
  {\path{doi:10.1007/s10714-007-0555-7}}.

\bibitem{Damour:1979wya}
T.~Damour, {Quelques proprietes mecaniques, electromagnet iques,
  thermodynamiques et quantiques des trous noir}, Ph.D. thesis, Paris U.,
  VI-VII (1979).

\bibitem{damour1982surface}
T.~Damour,
  \href{https://www.ihes.fr/~damour/Articles/surfaceeffects.pdf}{Surface
  effects in black hole physics}, 1982.
\newline\urlprefix\url{https://www.ihes.fr/~damour/Articles/surfaceeffects.pdf}

\bibitem{Padmanabhan:2010rp}
T.~Padmanabhan, {Entropy density of spacetime and the Navier-Stokes fluid
  dynamics of null surfaces}, Phys. Rev. D 83 (2011) 044048.
\newblock \href {http://arxiv.org/abs/1012.0119} {\path{arXiv:1012.0119}},
  \href {https://doi.org/10.1103/PhysRevD.83.044048}
  {\path{doi:10.1103/PhysRevD.83.044048}}.

\bibitem{Kolekar:2011gw}
S.~Kolekar, T.~Padmanabhan, {Action Principle for the Fluid-Gravity
  Correspondence and Emergent Gravity}, Phys. Rev. D 85 (2012) 024004.
\newblock \href {http://arxiv.org/abs/1109.5353} {\path{arXiv:1109.5353}},
  \href {https://doi.org/10.1103/PhysRevD.85.024004}
  {\path{doi:10.1103/PhysRevD.85.024004}}.

\bibitem{Poisson:2009pwt}
E.~Poisson, {A Relativist's Toolkit: The Mathematics of Black-Hole Mechanics},
  Cambridge University Press, Cambridge, U.K., 2009.
\newblock \href {https://doi.org/10.1017/CBO9780511606601}
  {\path{doi:10.1017/CBO9780511606601}}.

\bibitem{Mohd:2013jva}
A.~Mohd, S.~Sarkar, {Thermodynamics of Local Causal Horizons}, Phys. Rev. D
  88~(2) (2013) 024026.
\newblock \href {http://arxiv.org/abs/1304.2008} {\path{arXiv:1304.2008}},
  \href {https://doi.org/10.1103/PhysRevD.88.024026}
  {\path{doi:10.1103/PhysRevD.88.024026}}.

\bibitem{Parattu:2017cjd}
K.~Parattu, {Einstein Equations from/as Thermodynamics of Spacetime}, Fundam.
  Theor. Phys. 187 (2017) 339--352.
\newblock \href {https://doi.org/10.1007/978-3-319-51700-1_20}
  {\path{doi:10.1007/978-3-319-51700-1_20}}.

\bibitem{Eling:2006aw}
C.~Eling, R.~Guedens, T.~Jacobson, {Non-equilibrium thermodynamics of
  spacetime}, Phys. Rev. Lett. 96 (2006) 121301.
\newblock \href {http://arxiv.org/abs/gr-qc/0602001}
  {\path{arXiv:gr-qc/0602001}}, \href
  {https://doi.org/10.1103/PhysRevLett.96.121301}
  {\path{doi:10.1103/PhysRevLett.96.121301}}.

\bibitem{Chirco:2009dc}
G.~Chirco, S.~Liberati, {Non-equilibrium Thermodynamics of Spacetime: The Role
  of Gravitational Dissipation}, Phys. Rev. D 81 (2010) 024016.
\newblock \href {http://arxiv.org/abs/0909.4194} {\path{arXiv:0909.4194}},
  \href {https://doi.org/10.1103/PhysRevD.81.024016}
  {\path{doi:10.1103/PhysRevD.81.024016}}.

\bibitem{Dey:2017fld}
R.~Dey, S.~Liberati, D.~Pranzetti, {Spacetime thermodynamics in the presence of
  torsion}, Phys. Rev. D 96~(12) (2017) 124032.
\newblock \href {http://arxiv.org/abs/1709.04031} {\path{arXiv:1709.04031}},
  \href {https://doi.org/10.1103/PhysRevD.96.124032}
  {\path{doi:10.1103/PhysRevD.96.124032}}.

\bibitem{Kothawala:2009kc}
D.~Kothawala, T.~Padmanabhan, {Thermodynamic structure of Lanczos-Lovelock
  field equations from near-horizon symmetries}, Phys. Rev. D 79 (2009) 104020.
\newblock \href {http://arxiv.org/abs/0904.0215} {\path{arXiv:0904.0215}},
  \href {https://doi.org/10.1103/PhysRevD.79.104020}
  {\path{doi:10.1103/PhysRevD.79.104020}}.

\bibitem{Chakraborty:2015wma}
S.~Chakraborty, {Lanczos-Lovelock gravity from a thermodynamic perspective},
  JHEP 08 (2015) 029.
\newblock \href {http://arxiv.org/abs/1505.07272} {\path{arXiv:1505.07272}},
  \href {https://doi.org/10.1007/JHEP08(2015)029}
  {\path{doi:10.1007/JHEP08(2015)029}}.

\bibitem{Dalui:2021sme}
S.~Dalui, B.~R. Majhi, T.~Padmanabhan, {Thermal nature of a generic null
  surface}, Phys. Rev. D 104~(12) (2021) 124080.
\newblock \href {http://arxiv.org/abs/2110.12665} {\path{arXiv:2110.12665}},
  \href {https://doi.org/10.1103/PhysRevD.104.124080}
  {\path{doi:10.1103/PhysRevD.104.124080}}.

\bibitem{Dey:2020tkj}
S.~Dey, B.~R. Majhi, {Covariant approach to the thermodynamic structure of a
  generic null surface}, Phys. Rev. D 102~(12) (2020) 124044.
\newblock \href {http://arxiv.org/abs/2009.08221} {\path{arXiv:2009.08221}},
  \href {https://doi.org/10.1103/PhysRevD.102.124044}
  {\path{doi:10.1103/PhysRevD.102.124044}}.

\bibitem{Dey:2021rke}
S.~Dey, K.~Bhattacharya, B.~R. Majhi, {Thermodynamic structure of a generic
  null surface and the zeroth law in scalar-tensor theory}, Phys. Rev. D
  104~(12) (2021) 124038.
\newblock \href {http://arxiv.org/abs/2105.07787} {\path{arXiv:2105.07787}},
  \href {https://doi.org/10.1103/PhysRevD.104.124038}
  {\path{doi:10.1103/PhysRevD.104.124038}}.

\bibitem{sakharov1967vacuum}
A.~D. Sakharov, Vacuum quantum fluctuations in curved space and the theory of
  gravitation, in: Doklady Akademii Nauk, Vol. 177, Russian Academy of
  Sciences, 1967, pp. 70--71.

\bibitem{Padmanabhan:2009kr}
T.~Padmanabhan, {Equipartition of energy in the horizon degrees of freedom and
  the emergence of gravity}, Mod. Phys. Lett. A 25 (2010) 1129--1136.
\newblock \href {http://arxiv.org/abs/0912.3165} {\path{arXiv:0912.3165}},
  \href {https://doi.org/10.1142/S021773231003313X}
  {\path{doi:10.1142/S021773231003313X}}.

\bibitem{Padmanabhan:2010xh}
T.~Padmanabhan, {Surface Density of Spacetime Degrees of Freedom from
  Equipartition Law in theories of Gravity}, Phys. Rev. D 81 (2010) 124040.
\newblock \href {http://arxiv.org/abs/1003.5665} {\path{arXiv:1003.5665}},
  \href {https://doi.org/10.1103/PhysRevD.81.124040}
  {\path{doi:10.1103/PhysRevD.81.124040}}.

\bibitem{Padmanabhan:2013nxa}
T.~Padmanabhan, {General Relativity from a Thermodynamic Perspective}, Gen.
  Rel. Grav. 46 (2014) 1673.
\newblock \href {http://arxiv.org/abs/1312.3253} {\path{arXiv:1312.3253}},
  \href {https://doi.org/10.1007/s10714-014-1673-7}
  {\path{doi:10.1007/s10714-014-1673-7}}.

\bibitem{Chakraborty:2014rga}
S.~Chakraborty, T.~Padmanabhan, {Evolution of Spacetime arises due to the
  departure from Holographic Equipartition in all Lanczos-Lovelock Theories of
  Gravity}, Phys. Rev. D 90~(12) (2014) 124017.
\newblock \href {http://arxiv.org/abs/1408.4679} {\path{arXiv:1408.4679}},
  \href {https://doi.org/10.1103/PhysRevD.90.124017}
  {\path{doi:10.1103/PhysRevD.90.124017}}.

\bibitem{Parattu:2013gwa}
K.~Parattu, B.~R. Majhi, T.~Padmanabhan, {Structure of the gravitational action
  and its relation with horizon thermodynamics and emergent gravity paradigm},
  Phys. Rev. D 87~(12) (2013) 124011.
\newblock \href {http://arxiv.org/abs/1303.1535} {\path{arXiv:1303.1535}},
  \href {https://doi.org/10.1103/PhysRevD.87.124011}
  {\path{doi:10.1103/PhysRevD.87.124011}}.

\bibitem{Chakraborty:2014joa}
S.~Chakraborty, T.~Padmanabhan, {Geometrical variables with direct
  thermodynamic significance in Lanczos-Lovelock gravity}, Phys. Rev. D 90~(8)
  (2014) 084021.
\newblock \href {http://arxiv.org/abs/1408.4791} {\path{arXiv:1408.4791}},
  \href {https://doi.org/10.1103/PhysRevD.90.084021}
  {\path{doi:10.1103/PhysRevD.90.084021}}.

\bibitem{ASENS_1923_3_40__325_0}
E.~Cartan, \href{http://www.numdam.org/articles/10.24033/asens.751/}{Sur les
  vari\'et\'es \`a connexion affine et la th\'eorie de la relativit\'e
  g\'en\'eralis\'ee (premi\`ere partie)}, Annales scientifiques de l'\'Ecole
  Normale Sup\'erieure 3e s{\'e}rie, 40 (1923) 325--412.
\newblock \href {https://doi.org/10.24033/asens.751}
  {\path{doi:10.24033/asens.751}}.
\newline\urlprefix\url{http://www.numdam.org/articles/10.24033/asens.751/}

\bibitem{Hehl:1976kj}
F.~W. Hehl, P.~Von Der~Heyde, G.~D. Kerlick, J.~M. Nester, {General Relativity
  with Spin and Torsion: Foundations and Prospects}, Rev. Mod. Phys. 48 (1976)
  393--416.
\newblock \href {https://doi.org/10.1103/RevModPhys.48.393}
  {\path{doi:10.1103/RevModPhys.48.393}}.

\bibitem{kibble1961lorentz}
T.~W. Kibble, Lorentz invariance and the gravitational field, Journal of
  mathematical physics 2~(2) (1961) 212--221.

\bibitem{sciama1962}
D.~W. Sciama, Recent developments in general relativity, Pergamon Press and
  PWN, Warsaw (1962) 415.

\bibitem{sciama1964physical}
D.~W. Sciama, The physical structure of general relativity, Reviews of Modern
  Physics 36~(1) (1964) 463.

\bibitem{sciama1964erratum}
D.~Sciama, Erratum: The physical structure of general relativity, Reviews of
  Modern Physics 36~(4) (1964) 1103.

\bibitem{hehl1971does}
F.~W. Hehl, How does one measure torsion of space-time?, Physics Letters A
  36~(3) (1971) 225--226.

\bibitem{de1986introduction}
V.~De~Sabbata, M.~Gasperini, Introduction to gravitation, World Scientific
  Publishing Company, 1986.

\bibitem{de1994spin}
V.~De~Sabbata, C.~Sivaram, Spin and torsion in gravitation, world scientific,
  1994.

\bibitem{kleinert1989gauge}
H.~Kleinert, \href{https://books.google.co.in/books?id=PgJRygEACAAJ}{Gauge
  Fields in Condensed Matter}, no. v. 2 in Gauge Fields in Condensed Matter,
  World Scientific, 1989.
\newline\urlprefix\url{https://books.google.co.in/books?id=PgJRygEACAAJ}

\bibitem{Poplawski:2009fb}
N.~J. Poplawski, {Classical Physics: Spacetime and Fields} (11 2009).
\newblock \href {http://arxiv.org/abs/0911.0334} {\path{arXiv:0911.0334}}.

\bibitem{Shapiro:2001rz}
I.~L. Shapiro, {Physical aspects of the space-time torsion}, Phys. Rept. 357
  (2002) 113.
\newblock \href {http://arxiv.org/abs/hep-th/0103093}
  {\path{arXiv:hep-th/0103093}}, \href
  {https://doi.org/10.1016/S0370-1573(01)00030-8}
  {\path{doi:10.1016/S0370-1573(01)00030-8}}.

\bibitem{Obukhov:2006gea}
Y.~N. Obukhov, {Poincare gauge gravity: Selected topics}, Int. J. Geom. Meth.
  Mod. Phys. 03 (2006) 95--138.
\newblock \href {http://arxiv.org/abs/gr-qc/0601090}
  {\path{arXiv:gr-qc/0601090}}, \href
  {https://doi.org/10.1142/S021988780600103X}
  {\path{doi:10.1142/S021988780600103X}}.

\bibitem{PhysRevD.104.084073}
S.~Hensh, S.~Liberati,
  \href{https://link.aps.org/doi/10.1103/PhysRevD.104.084073}{Raychaudhuri
  equations and gravitational collapse in einstein-cartan theory}, Phys. Rev. D
  104 (2021) 084073.
\newblock \href {https://doi.org/10.1103/PhysRevD.104.084073}
  {\path{doi:10.1103/PhysRevD.104.084073}}.
\newline\urlprefix\url{https://link.aps.org/doi/10.1103/PhysRevD.104.084073}

\bibitem{PhysRevD.96.024021}
P.~Luz, V.~Vitagliano,
  \href{https://link.aps.org/doi/10.1103/PhysRevD.96.024021}{Raychaudhuri
  equation in spacetimes with torsion}, Phys. Rev. D 96 (2017) 024021.
\newblock \href {https://doi.org/10.1103/PhysRevD.96.024021}
  {\path{doi:10.1103/PhysRevD.96.024021}}.
\newline\urlprefix\url{https://link.aps.org/doi/10.1103/PhysRevD.96.024021}

\bibitem{PhysRevD.58.044021}
D.~Hochberg, M.~Visser,
  \href{https://link.aps.org/doi/10.1103/PhysRevD.58.044021}{Dynamic wormholes,
  antitrapped surfaces, and energy conditions}, Phys. Rev. D 58 (1998) 044021.
\newblock \href {https://doi.org/10.1103/PhysRevD.58.044021}
  {\path{doi:10.1103/PhysRevD.58.044021}}.
\newline\urlprefix\url{https://link.aps.org/doi/10.1103/PhysRevD.58.044021}

\bibitem{Gourgoulhon:2005ng}
E.~Gourgoulhon, J.~L. Jaramillo, {A 3+1 perspective on null hypersurfaces and
  isolated horizons}, Phys. Rept. 423 (2006) 159--294.
\newblock \href {http://arxiv.org/abs/gr-qc/0503113}
  {\path{arXiv:gr-qc/0503113}}, \href
  {https://doi.org/10.1016/j.physrep.2005.10.005}
  {\path{doi:10.1016/j.physrep.2005.10.005}}.

\bibitem{padmanabhan2002classical}
T.~Padmanabhan, Classical and quantum thermodynamics of horizons in spherically
  symmetric spacetimes, Classical and Quantum Gravity 19~(21) (2002) 5387.

\bibitem{Pilling:2002dz}
T.~G. Pilling, {Gauge torsion gravity, string theory, and antisymmetric tensor
  interactions}, Other thesis (2002).

\bibitem{carter1997extended}
B.~Carter, Extended tensorial curvature analysis for embeddings and foliations,
  Contemporary Mathematics 203 (1997) 207--220.

\bibitem{wald2010general}
R.~M. Wald, General relativity, University of Chicago press, 2010.

\bibitem{thorne1986black}
K.~S. Thorne, K.~S. Thorne, R.~H. Price, D.~A. MacDonald, Black holes: the
  membrane paradigm, Yale university press, 1986.

\bibitem{Chakraborty:2018qew}
S.~Chakraborty, R.~Dey, {Noether Current, Black Hole Entropy and Spacetime
  Torsion}, Phys. Lett. B 786 (2018) 432--441.
\newblock \href {http://arxiv.org/abs/1806.05840} {\path{arXiv:1806.05840}},
  \href {https://doi.org/10.1016/j.physletb.2018.10.027}
  {\path{doi:10.1016/j.physletb.2018.10.027}}.

\bibitem{Kothawala:2010bf}
D.~Kothawala, {The thermodynamic structure of Einstein tensor}, Phys. Rev. D 83
  (2011) 024026.
\newblock \href {http://arxiv.org/abs/1010.2207} {\path{arXiv:1010.2207}},
  \href {https://doi.org/10.1103/PhysRevD.83.024026}
  {\path{doi:10.1103/PhysRevD.83.024026}}.

\bibitem{Fabbri:2006xq}
L.~Fabbri, {On a completely antisymmetric Cartan torsion tensor}, Annales Fond.
  Broglie 32 (2007) 215--228.
\newblock \href {http://arxiv.org/abs/gr-qc/0608090}
  {\path{arXiv:gr-qc/0608090}}.

\bibitem{SenGupta:2001cs}
S.~SenGupta, S.~Sur, {Spherically symmetric solutions of gravitational field
  equations in Kalb-Ramond background}, Phys. Lett. B 521 (2001) 350--356.
\newblock \href {http://arxiv.org/abs/gr-qc/0102095}
  {\path{arXiv:gr-qc/0102095}}, \href
  {https://doi.org/10.1016/S0370-2693(01)01238-2}
  {\path{doi:10.1016/S0370-2693(01)01238-2}}.

\bibitem{SDBRM}
S.~Dey, B.~R. Majhi, {Work in progress}.

\bibitem{blagojevic2001gravitation}
M.~Blagojevic, Gravitation and gauge symmetries, CRC Press, 2001.

\bibitem{Blagojevic:2003cg}
M.~Blagojevic, {Three lectures on Poincare gauge theory}, SFIN A 1 (2003)
  147--172.
\newblock \href {http://arxiv.org/abs/gr-qc/0302040}
  {\path{arXiv:gr-qc/0302040}}.

\bibitem{aldrovandi2013teleparallel}
R.~Aldrovandi, J.~G. Pereira, Teleparallel gravity, Teleparallel Gravity: An
  Introduction 173 (2013).
\newblock \href {https://doi.org/https://doi.org/10.1007/978-94-007-5143-9}
  {\path{doi:https://doi.org/10.1007/978-94-007-5143-9}}.

\bibitem{Boersma:1994pc}
S.~Boersma, T.~Dray, {Parametric manifolds. 1. Extrinsic approach}, J. Math.
  Phys. 36 (1995) 1378--1393.
\newblock \href {http://arxiv.org/abs/gr-qc/9407011}
  {\path{arXiv:gr-qc/9407011}}, \href {https://doi.org/10.1063/1.531127}
  {\path{doi:10.1063/1.531127}}.

\end{thebibliography}

\end{document}